\documentclass{article}
\pdfoutput=1

\usepackage{ifpdf}
\usepackage{longtable}
\usepackage{natbib}
\usepackage{booktabs}
\usepackage{epubtk}
\usepackage{graphicx}
\usepackage{textcomp}
\usepackage{xspace}

\newcommand{\kms}{km~s\super{-1}\xspace}
\newcommand{\ms}{m~s\super{-1}\xspace}
\newcommand{\mum}{\textmu\hspace{-0.2pt}m\xspace}
\newcommand{\Bf}{\textit{Bf}\xspace}
\newcommand{\B}{\textit{B}\xspace}
\newcommand{\f}{\textit{f}\xspace}
\newcommand{\Ro}{\textit{Ro}\xspace}
\newcommand{\R}{\textit{R}\xspace}

\begin{document}

\title{Observations of Cool-Star Magnetic Fields}

\author{\epubtkAuthorData{Ansgar Reiners}{%
Georg-August-Universit\"at\\
Institut f\"ur Astrophysik \\
Friedrich-Hund-Platz 1\\
37077 G\"ottingen}{%
Ansgar.Reiners@phys.uni-goettingen.de}{%
http://www.astro.physik.uni-goettingen.de/~areiners}%
}

\date{}
\maketitle

\begin{abstract}
Cool stars like the Sun harbor convection zones capable of producing
substantial surface magnetic fields leading to stellar magnetic
activity. The influence of stellar parameters like rotation, radius,
and age on cool-star magnetism, and the importance of the shear layer
between a radiative core and the convective envelope for the
generation of magnetic fields are keys for our understanding of
low-mass stellar dynamos, the solar dynamo, and also for other
large-scale and planetary dynamos. Our observational picture of
cool-star magnetic fields has improved tremendously over the last
years. Sophisticated methods were developed to search for the subtle
effects of magnetism, which are difficult to detect particularly in
cool stars. With an emphasis on the assumptions and capabilities of
modern methods used to measure magnetism in cool stars, I review the
different techniques available for magnetic field measurements. I
collect the analyses on cool-star magnetic fields and try to compare
results from different methods, and I review empirical evidence that
led to our current picture of magnetic fields and their generation in
cool stars and brown dwarfs.
\end{abstract}

\epubtkKeywords{stars: activity, magnetic fields}

\newpage
\tableofcontents 


\newpage

\section{Introduction}

One of the reasons why the physics of magnetic fields are so
attractive but poorly understood probably is that magnetic fields are
invisible. Both, the detection of magnetic fields and the
interpretation of field measurements connect a variety of research
fields because magnetic effects are manifold and measurement processes
involve a number of sophisticated techniques and unknowns. In the
stellar context, magnetic fields are believed to be the reason why
young stars can accrete material from their surrounding disk, they
rule the evolution of angular momentum, and the stellar dynamo
converts kinetic into thermal energy that appears in the many facets
of stellar activity.

The observation of stellar magnetic fields is difficult because they
are not directly visible, but also because we have only a very limited
idea about the nature of the fields that may exist in stars other than
the Sun. Because no measurement technique is capable of capturing
the entire complexity of a stellar magnetic field, observations always
only reveal that part of a magnetic field the observing strategy is
specialized for -- and in most cases it is not entirely clear what
that is.

Our imagination of magnetic fields in cool stars rests on observations
of the star we can observe in most detail -- the
Sun. Figure~\ref{fig:sun} shows an image of the solar surface from SOHO
together with a magnetogram taken at the same time during solar
maximum in 2001. Groups of cool spots appear where the magnetogram
reveals regions of high fields. Interestingly, the fields appear in
groups consisting of at least two areas in close proximity and of
opposite polarity \citep[for an overview of the solar magnetic field,
see, e.g., ][]{2006RPPh...69..563S}. Obviously, such groups, if they
exist, cannot be resolved in other stars where generally we can only
observe the light integrated from the whole projected stellar surface.

\epubtkImage{MDI_2001-10-19.jpg}{%
  \begin{figure}[htbp]
    \centerline{
      \includegraphics[width=7cm]{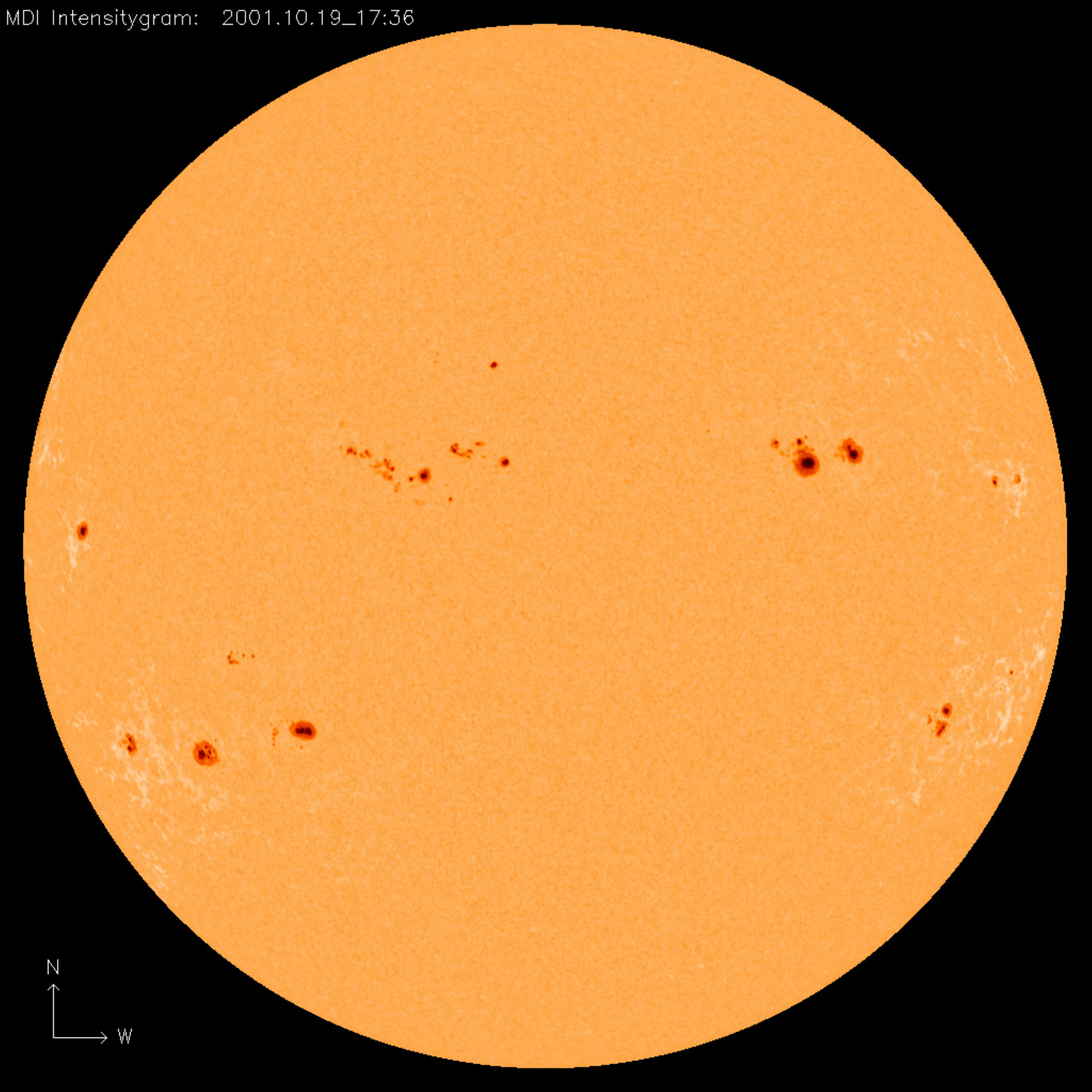}\qquad
      \includegraphics[width=7cm]{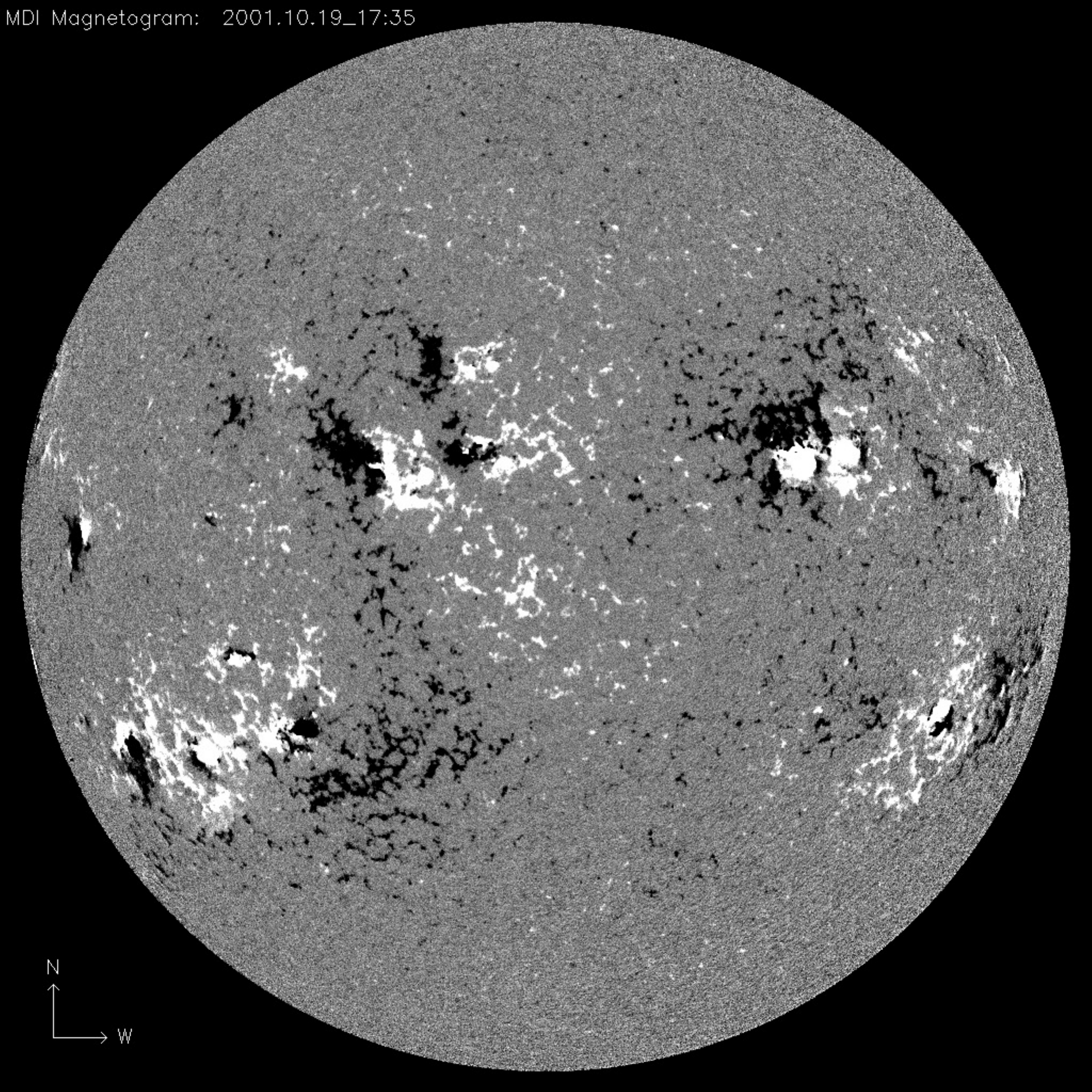}
    }
    \caption{Images of the Sun taken with the SOHO (ESA \& NASA)
      satellite. Cool sunspots appear as dark areas in the visible light image
      (left). They correspond to magnetic areas seen as black or white
      areas in the magnetogram (right; black and white show magnetic
      areas of different polarity). The magnetic field is arranged in spot
      groups of opposite polarity.}
    \label{fig:sun}
  \end{figure}
}

Although it serves as reference for cool-star magnetism, the solar magnetic
field is not at all easy to understand in all its details. The mean unsigned
magnetic flux density on the solar surface is often reported to be on the
order of 10~G using Zeeman splitting diagnostics. However,
\citet{2004Natur.430..326T} reported average flux densities one order of
magnitude higher employing a more sophisticated three-dimensional radiative
transfer approach taking into account the consequences of the Hanle
effect. The solar magnetic field is not the subject of this review, but the
example shows how confusing even the magnetic field of the Sun can be if
reduced to a single number. The reason for this is the wide range in strengths
and scales that are probed using different methods.

Fortunately, not all stars have average magnetic flux densities as low
as the solar one, and we absolutely can go out and look for fields
that are stronger or have a more obvious observational signature than
the solar field. Nevertheless, one has to keep in mind that all
observations can only reveal the type of field they are sensitive to,
and it is often more difficult to find out what that means than to
actually carry out the observation. This article reviews the existing
measurements of magnetic fields in cool stars. I define these to be
stars with efficient convection in their surface layers, i.e., stars
later than spectral type early F. Since F-type stars tend to be fast
rotators, which makes a magnetic field detection even more difficult,
a review on magnetic fields in cool stars essentially narrows down to
stars cooler than the Sun. Magnetic field measurements are available
for late-type dwarfs and also for some giants. One of the main
motivations for investigating stellar magnetic fields is to understand
the solar dynamo by assuming that the same mechanism works in other
stars but runs with a different set of parameters. By studying
magnetic fields in a sample of stars with different temperature,
convective velocities, and rotation rates, one can hope to shed light
on the fundamental mechanisms of a presumably universal cool-star
dynamo mechanism.

A particularly interesting class of stars are cool stars of spectral
type M. Covering the mass spectrum between $\sim 0.6$ and
$0.1\,M_{\odot}$, M~dwarfs are the most frequent type of stars,
which makes them very interesting by themselves. Furthermore, within
this mass range, the stars can have very different physical properties
rendering them very attractive targets for comparative studies. The
transition from partly convective (sun-like) to fully convective stars
happens in the M~dwarf regime, probably around spectral type
M3/M4. This area is in the center of interest for dynamo theory
because the tachocline is believed to be the place where at least one
important part of the solar dynamo is located. Furthermore,
atmospheres of M~dwarfs can be very different and both molecules and
dust gain importance as the temperature drops toward late spectral
types. It is important to understand how this influences magnetic field
generation, and how the coupling between magnetic fields and stellar
atmospheres changes. Towards even cooler objects, brown dwarfs are
objects with masses below $0.08\,M_{\odot}$ that are described as
failed stars because they do not burn hydrogen in their core. Although
they are not considered stars, their physical properties are very
similar to low-mass stars, especially close to the surface. I will,
therefore, include them in the discussion of stellar magnetic fields.

\newpage

\section{Methodology of Magnetic Field Measurements}
\label{sect:methodology}

Magnetic fields are not directly visible. Their effects on the
observable world are manifold, but all we can hope to measure are the
consequences the presence of a magnetic field has on any properties
that are accessible to observation. A particularly useful indicator of
stellar magnetic activity, for example, is the non-thermal emission
generated through magnetic heating \citep[for a review,
see][]{2008LRSP....5....2H}. Non-thermal emission is an example for
indicators of magnetism that I will call \emph{indirect} in the
following. Indirect indicators require an additional mechanism to
provide evidence for magnetic fields, and it is often difficult to
entirely rule out alternative mechanisms as a source for its
presence. For example, non-thermal emission may be generated by
acoustic heating mechanisms so that the detectability cannot be
translated into a magnetic field strength without further knowledge
\citep{1996SSRv...75..453N}. Nevertheless, there is ample evidence
that indirect indicators like chromospheric Ca or coronal X-ray
emission are reliable tracers of magnetic flux density at least in
sun-like stars \citep{1989ApJ...337..964S, 2003ApJ...598.1387P}. For
the following, an observable is called \emph{direct} if its detection
or amplitude necessarily implies the presence of a magnetic field.

The most successfully employed mechanism for direct detection of
stellar magnetic fields is the Zeeman effect
\citep{1897ApJ.....5..332Z}. Different approaches to use the Zeeman
effect for magnetic field measurements and resulting field
determinations are discussed and build the main part of the following
chapters. Another mechanism that appears similar to the Zeeman effect
and has been used in solar magnetic field measurements is the Hanle
effect \citep{1924ZPhy...30...93H}. The Hanle effect
describes how selective level population can be modified by a magnetic
field \citep{2004ASSL..307.....L, 2006spse.conf...77T}. It can be used
to measure tangled, very small-scale fields of up to a few hundred
Gauss but requires a very detailed understanding of atomic level
population and three-dimensional scattering processes
\citep{2004Natur.430..326T}. This level of detail cannot yet be
achieved in stellar observations and the Hanle effect could so far not
be used to detect magnetic fields in stars other than the Sun.

Observations of magnetically induced emission, i.e., indirect magnetic
field diagnostics, provide a wealth of information on stellar magnetic
activity that is often easier accessible than direct field
measurements. For reviews on observations of coronal emission,
chromospheric emission, and starspots the reader is referred to the
reviews by \citet{2002ARAA..40..217G, 2004AARv..12...71G}, 
\citet{2005LRSP....2....8B}, and \citet{2008LRSP....5....2H}. In this
article, the main consequences of indirect observations for our
picture of stellar magnetism will be discussed only briefly in
Section~\ref{sect:indirect}.

\subsection{Zeeman effect}
\label{Sect:ZeemanEffect}

\subsubsection{Absorption lines in a magnetic field}

In this section, I will give a brief introduction on the basics of the
Zeeman effect. More comprehensive discussions of the Zeeman effect and
equations to calculate Zeeman splitting in stellar atomic absorption
lines can be found, e.g., in \citet{1963tas..book.....C}, 
\citet{1969tzm..book.....B}, \citet{1988ApJ...324..441S}, 
\citet{1992AARv...4...35L}, \citet{2005LNP...664..183M}, and 
\citet{2009ARAA..47..333D}.

An atomic or molecular absorption or emission line is excited if
electrons make a transition from one energy level to another. The
energy of each level is altered in the presence of a magnetic field
according to the vector product between the spin ($S$) and orbital
angular momenta ($L$) of the electron, and the magnetic field vector
(so-called LS coupling). Each energy level with total angular momentum
quantum number $J$ splits into (2$J$ + 1) states of energy with
different magnetic quantum numbers $M$. The difference between
subsequent states of energy is proportional to $B g$ with $B$ the
magnetic field and $g$ the Land\'e factor, the latter being a function
of the energy level's orbital and spin angular momentum quantum
numbers,
\begin{equation}
  \label{Eq:Landetrans}
  g_i = \frac{3}{2} + \frac{S_i(S_i+1)-L_i(L_i+1)}{2J_i(J_i+1)}.
\end{equation} 

A dipole transition between two energy levels must obey the selection
rule $\Delta M = -1, 0, +1$, hence there are generally three groups of
transitions between two energy levels. Spectral lines with $\Delta M =
0$ are called $\pi$ components, spectral lines with $\Delta M = -1$ or
$+1$ are called $\sigma_{\mathrm{blue}}$ and $\sigma_{\mathrm{red}}$-components,
respectively. Since orbital and spin angular momentum quantum numbers
can be different between the two energy levels, the Land\'e-factors of
both levels can be different, and transitions between energy levels
are not only a function of $M$ but depend on the Land\'e-factors of
the energy states. The result is that each component consists of a
group of transitions.

An often used quantity in the characterization of Zeeman splitting is
the so-called effective Land\'e-factor, which is the average
displacement of the group of $\sigma$-components with respect to line
center. The effective Land\'e factor $g$ of a transition is a
combination of Land\'e values of the two energy levels involved
\citep{1969tzm..book.....B},
\begin{equation}
  \label{Eq:Landeeff}
  g = \frac{1}{2}(g_u + g_l) + \frac{1}{4}(J_u - J_l)(g_u - g_l)(J_u + J_l +1).
\end{equation}
In principle, the effective Land\'e factor can be calculated from the
energy level's individual Land\'e factors and
Equation~(\ref{Eq:Landeeff}). For many transitions, however, LS-coupling is
a poor approximation of the real situation leading to large errors in
the calculation of individual $g_i$ values. In such cases, it can be
more appropriate to measure $g_i$ in laboratory experiments
\citep[e.g.,][]{1975JPCRD...4..353R} and use Equation~(\ref{Eq:Landeeff}) to
obtain more useful empirical effective Land\'e factors
\citep{1982SoPh...77..285L, 1985AA...148..123S}.

In summary, in the presence of a magnetic field, the transition
energies of the $\sigma$-components are shifted according to the
sensitivity of the transition (the \emph{Land\'e}-factor $g$) and the
strength of the magnetic field $B$. The energy perturbation depends
on the energy level's quantum numbers, condensed in $g$, and the
magnetic field, $B$. If we measure the energy shift in terms of
wavelength shift $\Delta \lambda$ or as Doppler displacement $\Delta
v$, the perturbation becomes a function of the initial wavelength of
the transition, $\lambda_0$. The wavelength displacement of the
$\sigma$ components is
\begin{equation}
  \label{Eq:Zeemanl}
  \Delta \lambda = 46.67 \, g \lambda_0^2 B ,
\end{equation}
with $\Delta \lambda$ in m\AA, $\lambda_{0}$ in \mum, and $B$ in kG.
The average velocity displacement of the spectral line components can
then be written as
\begin{equation}
  \label{Eq:Zeemanv}
  \Delta v = 1.4 \, \lambda_0 g B,
\end{equation}
with $B$ in kG, $\lambda_{0}$ in \mum, and $\Delta v$ in
\kms. The typical Zeeman velocity displacement of a spectral
line at visual wavelengths in the presence of a kG-field is on the
order of 1~\kms, which is somewhat smaller than the typical
resolving power of a high-resolution spectrograph and the intrinsic
line-width of stellar absorption lines. The Doppler displacement is
proportional to the wavelength $\lambda_{0}$, which facilitates the
detection of Zeeman splitting at infrared wavelengths compared to
measurements in the visual.

\subsubsection{Polarization of Zeeman components}

The three groups of Zeeman components, $\sigma_{\mathrm{blue}}$,
$\sigma_{\mathrm{red}}$, and $\pi$, are characterized by different magnetic
moments, which means that the three Zeeman components have distinct
polarization states. Furthermore, the actual intensity and
polarization seen by an observer depends on the angle between the line
of sight and the magnetic field at the atom. Figure~\ref{fig:scheme}
shows a simplified scheme of the splitting (left panel) and of the
different observable polarization states (right) of the $\pi$ and
$\sigma$ components: The $\pi$ component is not shifted in energy, it
is always linearly polarized but is not observed if the line of sight
is parallel to the magnetic field vector. The $\sigma$ components, on
the other hand, are shifted according to the formulae above. They can
be observed linearly or circularly polarized depending on the
observer's view. If the line of sight is parallel (\emph{longitudinal}
field) to the magnetic field vector, both components are circularly
polarized but in opposite directions. If the line of sight is
perpendicular (\emph{transverse} field) to the magnetic field, the
$\sigma$-components are linearly polarized in the direction
perpendicular to the polarization of the $\pi$-component.

It is important to realize that measurements of longitudinal and
transerve fields as seen in circular and linear polarization, and also
field measurements from unpolarized light are usually not identical to
the real surface magnetic field because of the measuring principles
discussed here. For the following, I will speak of longitudinal fields
if magnetic fields are derived from circular polarization. This
includes results from Stokes~V magnetic maps, which combine
observational information of longitudinal fields visible at different
epochs.

\epubtkImage{Bild2-Bild1.png}{%
  \begin{figure}[htbp]
    \centerline{
      (a)\parbox[c]{7cm}{\vspace{0pt}\includegraphics[width=7cm]{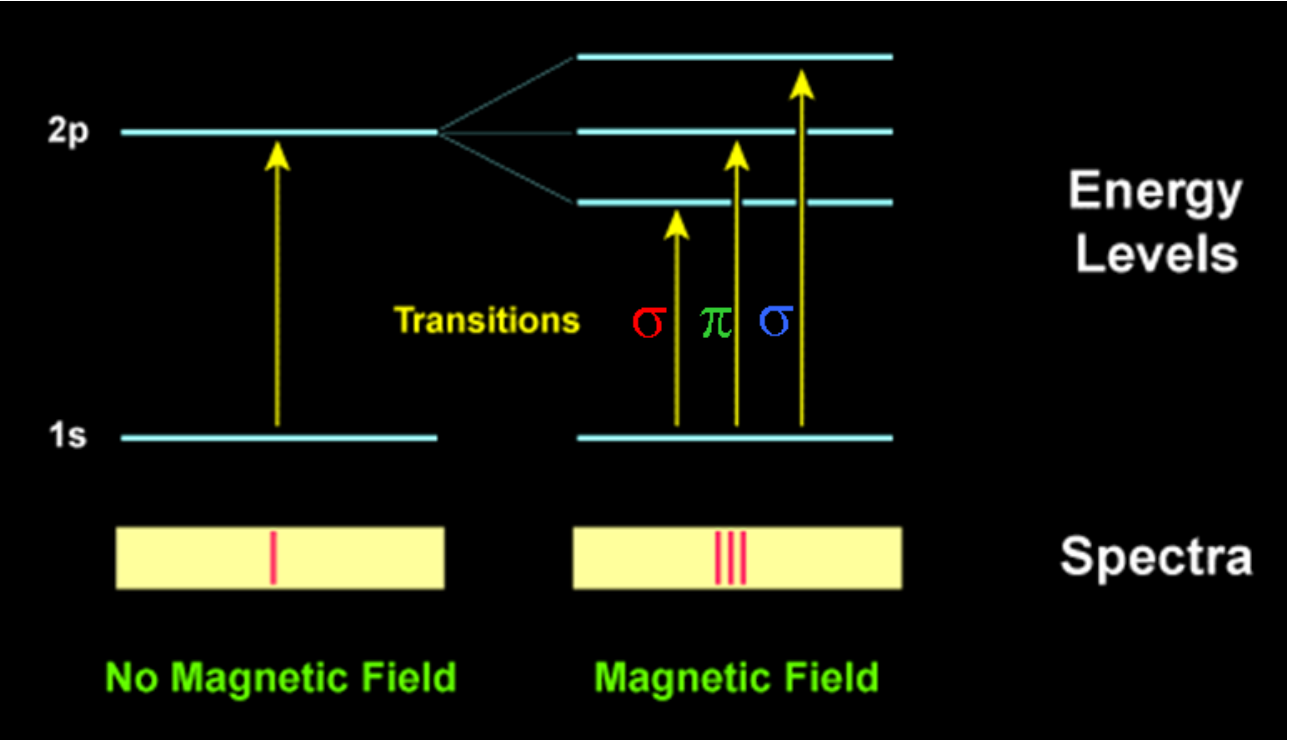}}\qquad
      (b)\parbox[c]{5cm}{\vspace{0pt}\includegraphics[width=5cm]{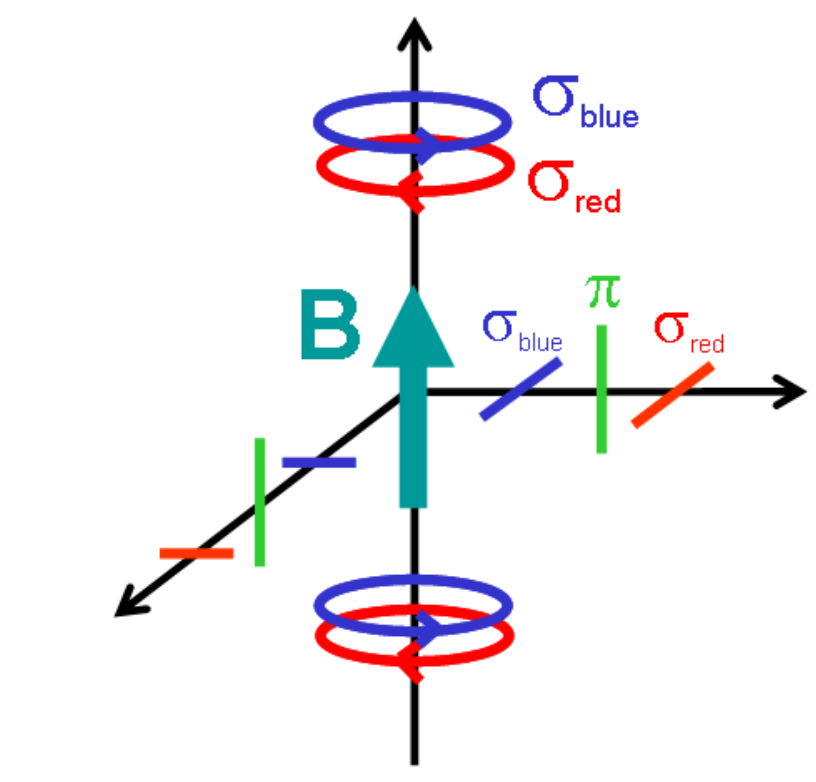}}}
    \caption{Schematic view of Zeeman splitting. (a) The
  upper level in the example is split into three levels
  producing three spectral lines that are separated. (b) Polarization
  of the $\pi$ and $\sigma$ components.}
\label{fig:scheme}
\end{figure}}

\subsubsection{The Stokes vectors}

For the characterization of a magnetic field, the measurement of
intensity in different polarization states can be of great
advantage. This is immediately clear from 
Figure~\ref{fig:scheme}b since the different Zeeman components are
polarized in a characteristic fashion. A commonly used system are the
Stokes components I, Q, U, and V \citep{1851TCaPS...9..399S} defined in
the following sense:\\

\centerline{\includegraphics{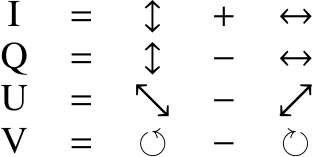}}
\ \\
Stokes~I is just the integrated (unpolarized) light. Stokes~Q and U
measure the two directions of linear polarization, and Stokes~V
measures circular polarization. Note that Stokes~Q and U are the
differences between two linearly polarized beams with perpendicular
directions of polarization. The components measured in Stokes~U are
rotated by 45\textdegree\ with respect to the components measured in
Stokes~Q; directions of Q and U are not defined in an absolute sense
but require the definition of a frame of reference, in which
polarization is measured. For a very readable introduction to Stokes
vectors and alternative forms the reader is referred to
\citet{1996aspo.book.....T}.

Stokes vectors are useful in astronomy because perpendicular circular
and linear polarization states can be measured with relatively
straightforward instrumentation. The representation of astronomical
polarization measurements is usually done in terms of Stokes
vectors. The problem we are concerned with, however, is under what
circumstances the magnetic field of a star can be recovered from
measurements of the Stokes vectors. If solar magnetic fields are a
good example for other stellar magnetic fields, we can expect that
they often show up in groups of different polarity. This is a problem
to measurements in Stokes V because equal amounts of polarized light
with opposite polarization simply cancel out and become invisible. In
Stokes Q and U, on the other hand, regions with magnetic field vectors
pointing into directions perpendicular to each other will
cancel. Thus, magnetism on a star with one magnetic region on the
eastern limb and another one of an identical field strength and
geometry on the northern limb will show no linear polarization and be
invisible to Stokes Q and U measurements. 

In reality, stellar magnetic fields can be expected to be very
complicated structures with a continuum of field strengths and
orientations. Therefore, we can in general not expect to resolve
classical splitting patterns with three groups of Zeeman components,
because magnetic field strengths and, thus, the velocity displacement of
the $\sigma$-components will be continuous. Finally, the visible
surface of a star is not a flat disk but one half of a sphere and even
the academic case of a completely radial (non-potential) field has
vector components that would be observed under viewing angles between
0\textdegree\ and 90\textdegree. It is, therefore, a formidable task to
measure magnetic fields and their geometries in stars that are
spatially entirely unresolved.

For a first estimate of the signals that may be expected from Stokes
measurements in sun-like stars, we can take a look at sunspot data. In
the center of a sunspot with $B$~=~2200~G, the polarization of
spectral lines at around 630~nm is on the order of 10\% in Stokes Q
and U, and about 20\% in Stokes V if measured at very high spectral resolving
power (\R~=~200\,000) \citep{2000RvGeo..38....1L}. Observations of
spatially unresolved sun-like stars will obviously not reach that
level because of canceling effects and lower spectral
resolution. \citet{2002AA...381..736P} calculated the four Stokes
parameters in profiles at $\lambda$~=~615~nm for a star with a dipolar
magnetic field configuration with a field strength of 8~kG at a
spectral resolving power of \R~=~100\,000. This model star is
probably not very similar to a low-mass star, but the example nicely
shows what level of polarization can be expected in an extreme case of
a very strong and organized magnetic field. The signal from
disk-integrated observations reaches maximum values of 0.5\% in Stokes
Q and U, and 5\% in Stokes~V. Thus, if magnetic fields are to be
detected in all four Stokes parameters, extremely high data quality is
required. At visual wavelengths, linear polarization observations must
have signal-to-noise ratios of roughly 1000 while circular
polarization perhaps is relaxed by a factor 10, approximately. For
similar magnetic field measurements in sun-like stars, the
requirements are likely higher by at least one order of magnitude.

A series of very high quality stellar measurements in all four Stokes
vectors of magnetic Ap and Bp stars was presented by
\citet{2000MNRAS.313..823W}. Again, no comparable measurements are
available for sun-like stars, but hotter stars with very strong fields
may serve as guideline for our expectations of polarization in cool
star observations. \citet{2000MNRAS.313..823W} show that in magnetic
Ap and Bp stars, circular polarization detected in strong,
magnetically sensitive lines is typically around
1\,--\,2~\texttimes~10\super{-2} while linear polarization is a factor
10\,--\,20 lower. Typical fields in cool stars are probably much
weaker so that polarization can be expected to be a lot weaker than
this, too. Recently, \citet{2011ApJ...732L..19K} presented the first
detection of linearly polarized spectra in cool stars. In an active
K-dwarf, they detected circular polarization at a level of
5~\texttimes~10\super{-5}, and linear polarization of roughly a factor
of 10 weaker.

In order to demonstrate the visibility of magnetic fields in the
Stokes parameters and the canceling effects of spot groups with
opposite polarization (as observed on the Sun),
Figures~\ref{fig:StokesHori} and \ref{fig:StokesTang} show some basic
simulations of line profiles from magnetic regions in the Stokes
parameters. The left panel visualizes the ``geometry'', which is
actually not a geometry of some real stellar magnetic field, but
nothing else than two areas of radial magnetic fields put on a flat
surface, where the spherical shape of a star has not been taken into
account in this example. It should be emphasized that on a real,
spherical star, cancellation effects would not be as obvious as in
this simplistic example and some net flux would usually remain. This
is even more important in the case of rotating stars where opposite
polarization of magnetic regions with different (local) radial
velocities would not lead to complete flux cancellation.

The magnetic regions in this toy model as shown in
Figure~\ref{fig:StokesHori} are observed with the field direction
perpendicular to the line of sight, i.e., observations of the
transverse field. Figure~\ref{fig:StokesTang} shows the same cases for
observations of magnetic regions observed with the field direction
parallel to the line of sight, i.e., the longitudinal field. In the
first case of transverse field orientation, no circular polarization
is visible at all. For longitudinal field observations, linear
polarization is invisible. Spectral resolving power is set to
\R~=~100\,000, the model line is a fictitious Fe line at rest
wavelength 600~nm, thermally broadened according to a temperature of
4000~K. Broadening due to turbulence and rotation are not taken into
account.

\epubtkImage{StokesPicHori.png}{%
\begin{figure}[htb]
  \centerline{\includegraphics[width=\textwidth]{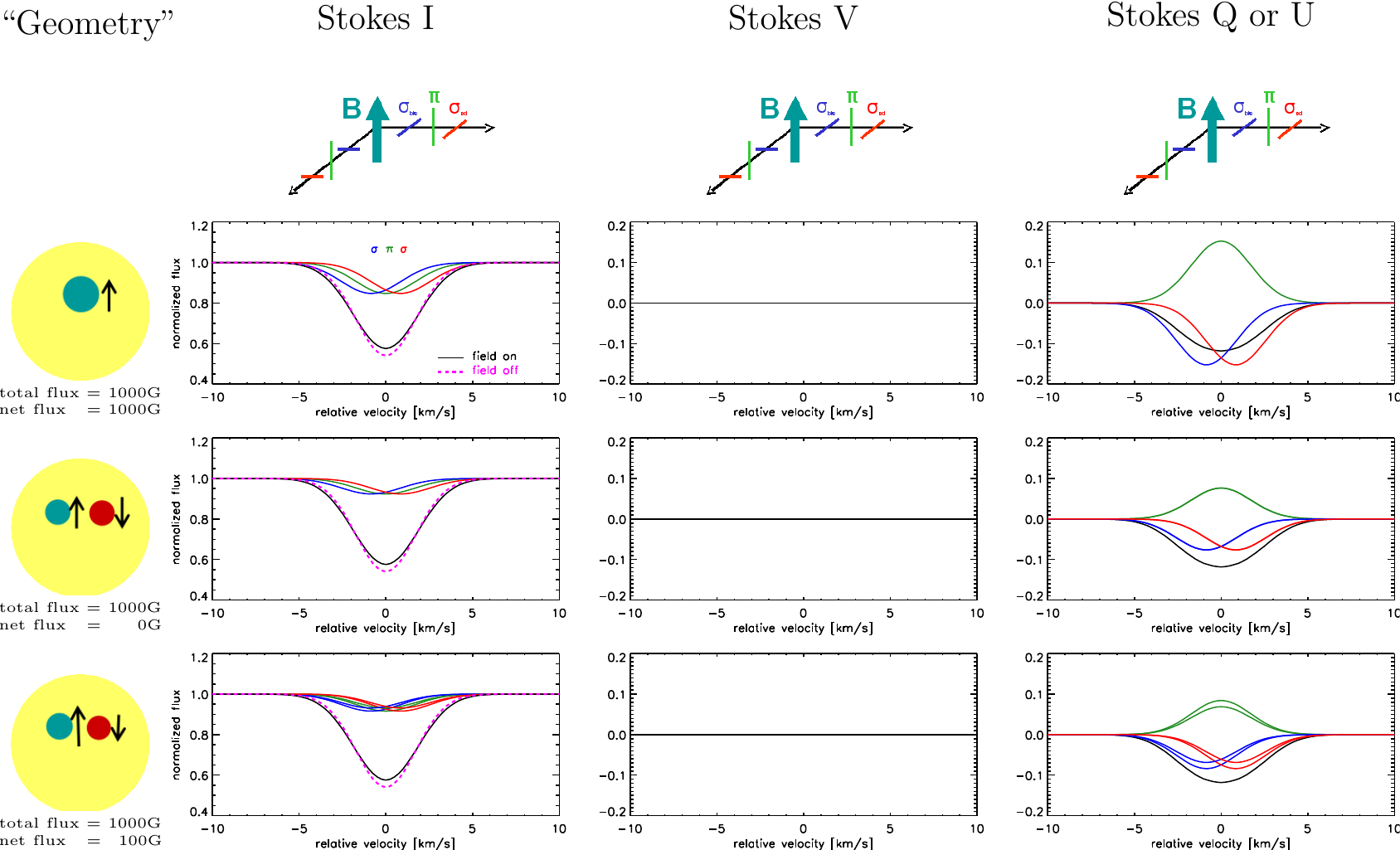}}
  \caption{Three examples of simplified field geometries and their
    signals in Stokes~I, V, and Q or U if the field is perpendicular
    to the line of sight (transverse field). Blue, green, and red
    lines show the line profiles of individual Zeeman components
    $\sigma_{\mathrm{blue}}$, $\pi$, and $\sigma_{\mathrm{red}}$,
    respectively. The black line is the sum of the three, that means
    the line that will be observable. In the Stokes~I panel, the
    magenta line shows how the line would appear with zero magnetic
    field. Rotational Doppler effects are ignored in these examples.}
  \label{fig:StokesHori}
\end{figure}}

\epubtkImage{StokesPicTang.png}{%
\begin{figure}[htb]
  \centerline{\includegraphics[width=\textwidth]{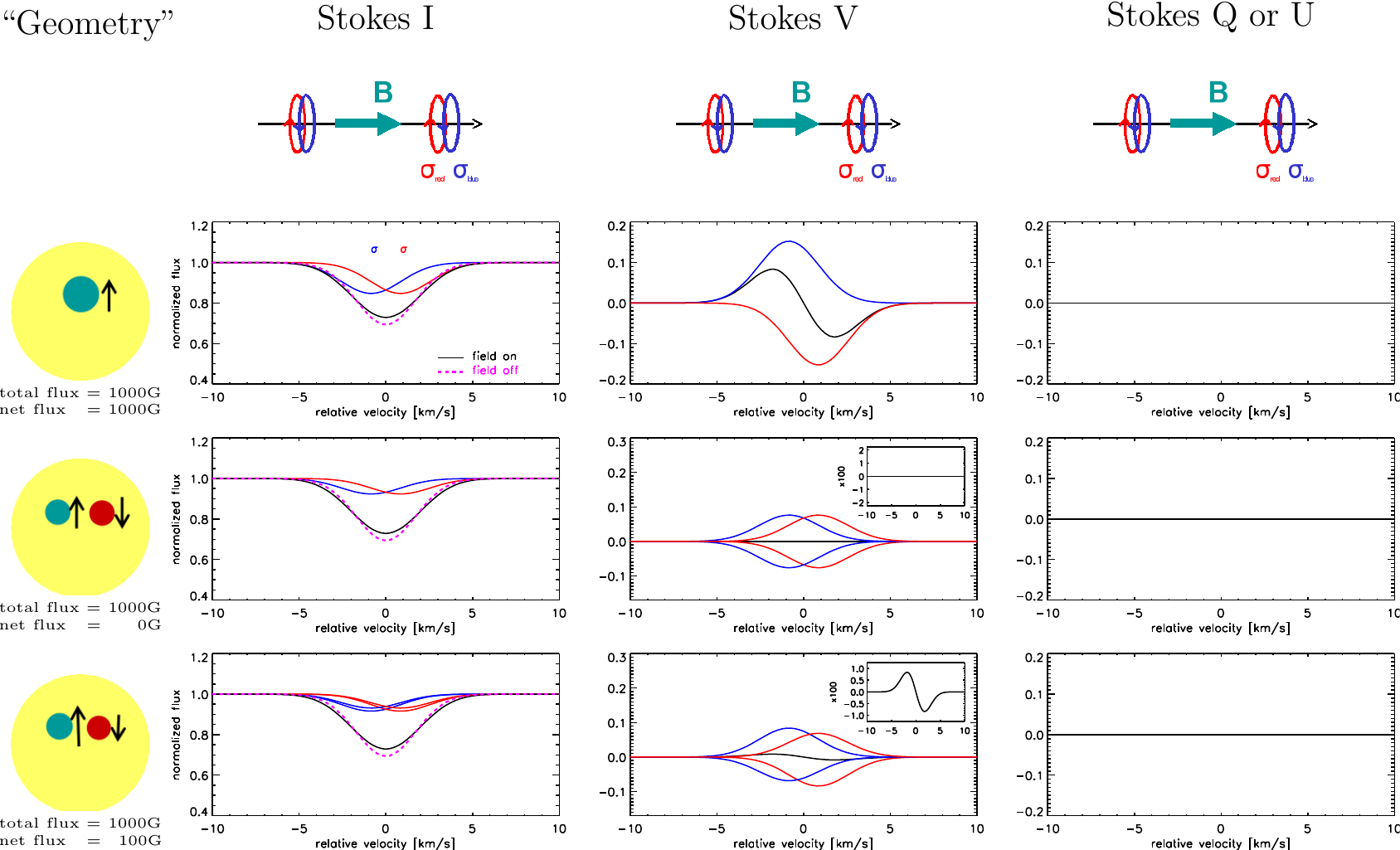}}
  \caption{Three examples of simplified field geometries and their
    signals in Stokes~I, V, and Q or U if the field is tangential to
    the line of sight (longitudinal field). Colored lines like in
    Figure~\ref{fig:StokesHori}. Rotational Doppler effects are
    ignored in these examples.}
  \label{fig:StokesTang}
\end{figure}}

The first example in the top row of both figures is a simple magnetic
field region with only one polarity; total field strength and signed
``net'' field both are 1000~G in this example. Stokes~I exhibits very
little broadening that is difficult to detect. There is a difference
between the two directions of observation since for the longitudinal
field (Figure~\ref{fig:StokesTang}), the $\pi$ component does not
appear. Linear polarization of the transverse field
(Figure~\ref{fig:StokesHori}) and circular polarization of the
longitudinal field (Figure~\ref{fig:StokesTang}) are on the order of
10\%. Note that the direction of the Stokes~V signal indicates the
orientation of the magnetic field vector. In the second row, two
magnetic regions, each with only half the size as in the first example are
observed. Both regions have the same absolute field strength and area,
but opposite polarity. The Stokes~I signal is identical to the first
example (the individual components are weaker but there are twice as
many). The same is true for the linear polarization signal in Stokes Q
and U because a shift in the polarization direction of 180\textdegree\
has the same signal as the original one. The signal in Stokes~V,
however, entirely vanishes (for all observing angles) because the net
(signed) field of this configuration is exactly zero; any field
strength in this canceling configuration is invisible to Stokes~V. The
last row in Figures~\ref{fig:StokesHori} and \ref{fig:StokesTang} show
the case of two magnetic areas with slightly different sizes; the
total field is still 1000~G, but here the net field is 100~G. Again,
Stokes~I, and Q and U are the same as in the examples above. Because
of the non-vanishing net field, the amplitude in Stokes~V is now
different from zero at ca. 1\%. In Figure~\ref{fig:StokesHori}, linear
polarization always provides a strong signal because opposite magnetic
field polarities do not cancel out. In a situation with two spots
located at relative polarization of 90\textdegree\ to each other,
linear polarization would completely cancel, too. As mentioned above
and will be discussed again in Section~\ref{sect:DI}, Doppler shifts on
a rotating star add valuable signal to polarimetric
measurements. Since real stars have usually non-zero rotation, most
cool stars show non-zero magnetic features visible in polarimetric
measurements.

\subsubsection{Reconstruction of stellar magnetic fields from Stokes vectors}

The few examples shown in Figures~\ref{fig:StokesHori} and
\ref{fig:StokesTang} demonstrate the principal sensitivity of the four
Stokes vectors to magnetic fields and their configurations. In
spatially resolved regions on the solar surface, measurements of
polarization provide relatively well-defined information on the
magnetic field (at least if compared to the case in other stars). In
other stars, however, we do not quite know what kinds of fields to
expect. The average flux density on the Sun is only on the order of a
few G and remains undetectable in observations of integrated solar
light. Slowly rotating stars of a comparable activity level probably
have fields as weak as the solar one. On the other hand, the magnetic
geometry of more rapidly rotating and, hence, more active stars is
entirely unknown and may not be very similar to the solar case.

A major difficulty in measuring stellar Zeeman splitting is the small
value of $\Delta v$ compared to other broadening agents like intrinsic
temperature and pressure broadening, and rotational broadening. In a
kG-magnetic field, typical splitting at optical wavelengths is of the
order 1~\kms, which is well below intrinsic line-widths of
several \kms and also below the spectral resolving power of
typical high-resolution spectrographs. Thus, individual components of
a spectral line can normally not be resolved even if the star only had
one well-defined magnetic field component. Real stars, however, can be
expected to harbor a magnetic field distribution that is much more
complex than this. Thus, even if spectral lines were intrinsically
very narrow and spectral resolving power infinitely high, we would
expect the Zeeman-broadened lines to look smeared out since in our
observations we integrate over all magnetic field components on the
entire visible hemisphere.

Stellar activity manifests itself in magnetic regions that can be
darker than the quiet photosphere (e.g., spots) or brighter (e.g.,
faculae). The contribution of a surface region to an observed spectral
line depends on its intensity contrast and local opacity while average
field densities in active regions like spots or faculae are known to
be systematically different from each other. This implies that regions
of different field strenghts are systematically weighted in their
contribution to the observed Zeeman pattern, and that the choice of
diagnostic is very important for the field density measured.

Another point that becomes immediately clear is that the geometric
interpretation of Zeeman splitting on an unresolved stellar disk can
be arbitrarily complex, no matter if polarized or unpolarized light is
used. In addition to the ambiguity between magnetic field strength and
the fraction of the star being occupied with magnetic fields (which
includes our ignorance about the number and distribution of magnetic
components), the signature of a magnetic field region in stellar
spectra depends on the angle between the magnetic field lines and the
line of sight. In reality, a continuous distribution of angles can be
expected because field lines are probably bent on the stellar surface,
and because the stellar surface is spherical. As a result, even
geometrically relatively simple field distributions will lead to
highly complex splitting patterns. If the star is rotating at
significant speed, as most active stars probably do, that pattern
again depends a lot on the time a star is observed. This, in turn, can
be utilized to reconstruct the geometry of the magnetic field by
observing the variation of the observed spectra with rotation.

There are two basically quite different ways to gather information
about stellar magnetic fields:
\begin{enumerate}
\item Measure the integrated scalar, unsigned magnetic field
  (Stokes~I).
\item Measure the magnetic vector field.
\end{enumerate}
The most promising way, clearly, to obtain information about the
magnetic field is to determine simultaneously the integrated field
\emph{and} its vector components. Observationally, however, there are
important differences between measurements in Stokes~I (integrated
flux measurements) and measurements in polarized light, so that in
practice both parts are often done separately.

\subsubsection*{Integrated field measurements}

The value of the integrated magnetic field strength can be derived
from observations in Stokes~I. Such observations can be carried out
with every high-resolution spectrograph and do not require
polarization optics. Stokes~I measurements are sensitive to the entire
magnetic field on the star, independent of field geometry and
canceling effects. A simultaneous measurement of Stokes~I is, therefore,
always helpful in order to determine the fraction of a magnetic field
that may be invisible to polarized light measurements.

Unfortunately, in a measurement of Zeeman splitting in Stokes~I one
faces the difficulty to disentangle the effect of Zeeman broadening
from all other broadening agents. This requires precise knowledge of
the spectral line appearance in the absence of a magnetic field. This
task requires extremely good knowledge about spectral line formation,
velocity fields, and the temperature distribution on the
star. Signatures of cool spots or differential rotation, for example,
can be very similar to Zeeman splitting patterns in integrated
starlight. The amplitude of Zeeman spitting due to a strong magnetic
field (e.g., 1000~G) is very subtle in sun-like stars observed at
visual wavelengths because intrinsic line width, surface velocity, and
typical instrumental resolution are of the same order as Zeeman
broadening. This implies that the detection of magnetic fields lower
than $\sim$~1~kG is extremely difficult at visual wavelengths (see
Section~\ref{sect:results_I}). Thus, stellar Stokes~I measurements are
typically not sensitive to magnetic fields lower than a few hundred
Gauss. The degeneracy between Zeeman splitting and other broadening
agents is lifted at longer wavelengths, hence infrared observations
have much higher sensitivity to magnetic fields. Unfortunately, only very few
high-resolution infrared spectrographs exist today but more and more
measurements are being reported (Section~\ref{sect:results_I}).

The Zeeman splitting pattern in surface-integrated starlight is the
sum of Stokes~I patterns from the entire stellar surface. The
absorption line from a star is very different from a sunspot
observation in which individual components from relatively
well-defined magnetic regions can be visible. The line broadening
pattern in Stokes~I depends on the magnetic field strength of the
individual components, the strongest fields are visible in the
components responsible for the widest line wings. The fractional area
of the surface filled with magnetic fields (filling factor) and the
weight of individual surface features in the final line profile are
parameters that are hidden in the line profile shape and are
degenerate with respect to each other. The information on the field
distribution and the contribution of individual magnetic areas is,
therefore, very limited in observations of Stokes~I alone. Another
limitation of Stokes~I measurements became visible in observations of
the solar magnetic field using the Hanle effect (see above). These
measurements revealed that the Sun harbors a field that is not of
10~G but more of 100~G strength. It is unclear whether a similar
difference (either in absolute or relative units) would also appear if
stars with much higher field strengths are observed, but it clearly
shows that Stokes~I measurements have difficulties capturing the
entire magnetic flux but can mainly provide a lower limit.

\subsubsection*{Reconstruction of the magnetic vector field}

Observations in Stokes~V, Q, or U are sensitive to the magnetic field
vector, not only to the unsigned field. This provides information
about the direction of the magnetic field that is not accessible to
Stokes~I measurements. The signal of a non-polarized spectral line is
zero in Stokes~V, Q, and U. This means that the problem of
disentangling Zeeman splitting from other line broadening mechanisms
does not exist, and the method is much more sensitive to small field
values (1~G and below). A problem is, however, that the signal seen in
polarized light is only the ``net'' magnetic field; regions of
opposite polarity cancel out in Stokes~V and magnetic fields at
90\textdegree\ orientation cancel out in Stokes~Q and U. Therefore,
depending on what observing technique is used, an arbitrary large
magnetic field may be hidden on the stellar surface without any signal
in Stokes~V or Stokes~Q and U alone. The problem is more severe for
circular polarization because the $\pi$ components are not detected
here.

It has been shown that the magnetic field distribution of a star can
be reconstructed in great parts from simultaneous observations of all
four Stokes parameters \citep{2002AA...388..868K,
  2010AA...524A...5K}. Successful reconstruction requires that the
star is observed over an entire rotation period for two reasons: 1) to
reconstruct the surface field hemisphere, the star needs to be seen
from different sides (note that if the star is seen under high
inclination angles, the invisible part close to the hidden pole
always remains undetectable); 2) at different phases, the angles
between the magnetic field lines and the line of sight vary with the
result that field components that may have canceled when observed at
disk center, can become visible when observed close to the limb. The
spatial resolution of magnetic field reconstructions depends on the
frequency of observations during stellar rotation and on intrinsic
line broadening (all Stokes components are subject to line
broadening). Typically, a resolution element has a size of ten or
several ten degrees on the stellar surface. \citet{2002AA...388..868K}
showed that using only a subset of Stokes vectors leads to ambiguities
that should be interpreted with great caution. Unfortunately,
measurements of linear polarization are extremely challenging in cool
stars because of the low polarization signal so that typically only
Stokes~I and (sometimes) Stokes~V are available (see
Section~\ref{sect:VMapsDwarfs}). Zeeman broadening in Stokes~I is very
subtle at least at visual wavelengths where most available
spectrographs operate, and Stokes~V, Q, and U measurements are both
difficult to acquire and exhibiting subtle Zeeman signals. The
observational difficulties obtaining all four Stokes components led to
the practice that in cool stars in the past usually either Stokes~I or
Stokes~V alone were investigated.

\subsubsection{Field, flux, and filling factor}

In general, a stellar surface may be covered with a homogenous field
of one particular field strength, or it can be covered with several
magnetic areas of different field strength. One example is a surface
of which 50\% is covered with a field of strength \B. If the other
50\% of the surface has no magnetic field, the average field is
\Bf~=~\B/2 with filling factor \f~=~0.5. An important consequence of
the fact that individual Zeeman-components are usually not resolved is the
degeneracy between magnetic field \B and filling factor \f. A strong
magnetic field covering a small portion of the star looks similar to a
weaker field covering a larger portion of the star. An often used way
around this ambiguity is to specify the value \Bf, i.e., the product
of the magnetic field and the filling factor; if more than one
magnetic component is considered, \Bf is the weighted sum over all
components. Products of \B with some power of \f, for example
\textit{Bf}\super{0.5} or \textit{Bf}\super{0.8} are often considered because they seem to be
better defined by observations \citep[see][]{1984ApJ...277..640G,
  1988ApJ...324..441S, 1995ApJ...439..939V}. One important point to
observe is that \Bf is often called the ``flux'' -- because it is the
product of a magnetic field and an area -- but it has the unit of a
magnetic field. In fact, the term flux is very misleading since: 1)
with \f specifying a relative fraction of the stellar surface, \Bf
is really the average \emph{flux density} that is identical to the
average unsigned magnetic field on the visible stellar surface, i.e.,
$Bf \equiv \langle B \rangle$; and 2) the total magnetic flux of two stars with
the same values of \Bf can be extremely different according to their
radii because the actual flux is proportional to the radius squared,
$\mathcal{F} \propto Bfr^2$. As a consequence, the value \Bf will be
much lower in a young, contracting star compared to an older (smaller)
one if flux is conserved.

A related source of confusion is the difference between the signed
magnetic field (or flux), and the unsigned values or the square of the
fields (used to calculate magnetic energy). With Stokes~I, both
polarities produce the same signal and the total unsigned flux is
measured. This implies that Stokes~I carries only partial information
about field geometry, but it also means that Stokes~I always probes
the entire magnetic flux of the star (see above). On the other hand,
Stokes~V can provide information on the sign of the magnetic fields,
but this comes with the serious caveat that opposite magnetic fields
cancel out and can become invisible to the Stokes~V signal. Thus,
results on \Bf from Stokes~V measurements can be much lower than
Stokes~I measurements.

\subsubsection{Equivalent widths}

Shifting of the $\sigma$-components to either side of the line center
leads to broadening of the spectral line and, in general, to a
flattening of the line core (see Figure~\ref{fig:StokesHori}). An
interesting effect can be used to measure magnetic fields if lines 
that are saturated are used , i.e., lines that have equivalent widths
smaller than the sum of the individual $\pi$- and
$\sigma$-components. If such a saturated line is split in the presence
of a magnetic field, the core depth of the line will remain at
approximately constant level while the line grows wider (see
Figure~\ref{fig:Leroy}). As a result, the equivalent width of a
saturated, magnetically sensitive line will grow with magnetic field
strength.

\epubtkImage{Leroy.png}{%
\begin{figure}[htbp]
  \centerline{\includegraphics[width=0.6\textwidth]{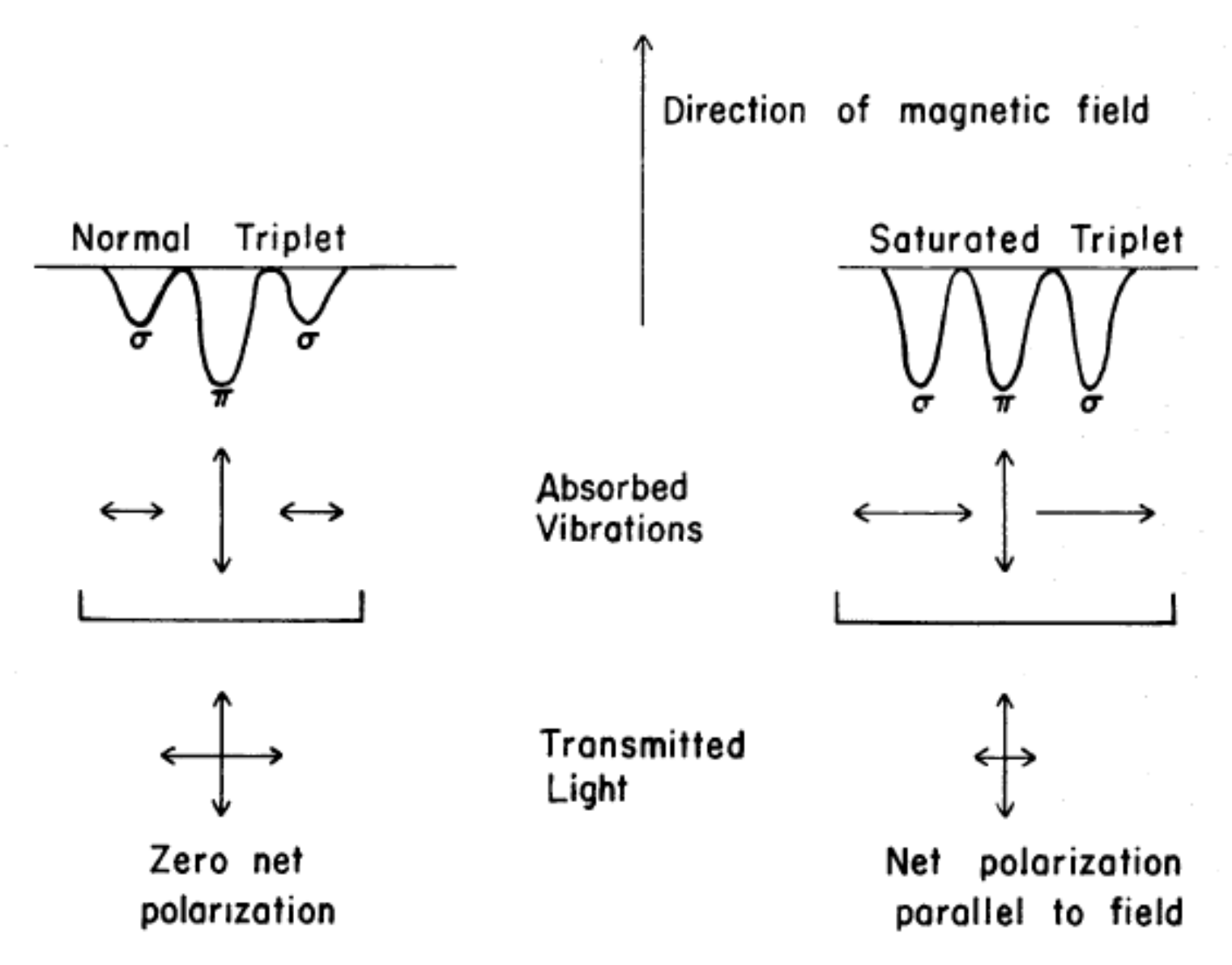}}
  \caption{The net polarization of a weak line is
    zero, and its equivalent width remains constant if a magnetic
    field is applied. In contrast, the net polarization of a saturated
    line in a transverse magnetic field is nonzero, and the equivalent
    width of a saturated line becomes larger in a magnetic
    field (from \citealp{1976ApJ...204..818M}, after 
    \citealp{1962AnAp...25..127L}; reproduced by permission of the AAS).}
  \label{fig:Leroy}
\end{figure}}

\cite{1992ApJ...390..622B} introduced a method to detect cool star
magnetic fields searching for enhanced equivalent widths of
Zeeman-sensitive absorption lines. As in other work searching for
Zeeman splitting in Stokes~I observations, they carefully modeled
polarized line transfer and compared the appearance of Zeeman
sensitive to Zeeman insensitive lines. The advantage of the equivalent
width method is that equivalent widths are more easily measured than
the subtle differences in line shape, in other words, information from
several spectral bins within one spectral line is extracted into one
number that can be measured more accurately. Nevertheless, the method
cannot lift degeneracies between magnetic field strength (times
filling factor) and other features like starspots or uncertainties in
the model atmosphere; the equivalent width method can only make
existing differences in the lines easier detectable.

The variation of line equivalent widths can be monitored over time. If
one assumes that variations occur because of varying visible magnetic
field strength, spectroscopic time series can be used to obtain
information about the surface distribution of co-rotating magnetic
regions (see also next section). This method was used for example by
\citet{1992ASPC...26..255S, 1994ASPC...64..661S} for Stokes~I magnetic
surface imaging.

\subsubsection{Doppler Imaging}
\label{sect:DI}

In addition to measuring the average magnetic field on a star, signed
or unsigned, the Doppler shift of individual features carries
information about the geometry of the stellar surface. Doppler Imaging
exploits the correspondence between wavelength position across a
rotationally broadened spectral line and spatial position across the
stellar disk to reconstruct surface maps of rotating stars
\citep{1983PASP...95..565V}; the method goes back to work by
\citet{1958IAUS....6..209D}, \citet{1974ApJ...192..409F}, and
\citet{1977SvAL....3..147G}. Spatial resolution of the maps depends on the
rotation velocity of the star and the sampling frequency at which
spectra are taken, among other factors. It has been used very
successfully to reconstruct temperature maps of cool stars \citep[see,
e.g.,][]{2002AN....323..309S} and abundance maps of hotter stars
\citep[e.g.,][]{2004AA...424..935K}. Zeeman Doppler Imaging (ZDI)
follows the same approach but investigating polarized light
\citep{1989AA...225..456S}. As the star is observed at different
phases, the magnetic field vectors are observed under different
projection angles leading to characteristic signatures in polarized
light; field components that may be invisible at one phase can have
large Stokes parameters at other phases.

\epubtkImage{img90.png}{%
\begin{figure}[htbp]
  \centerline{\includegraphics[width=0.9\textwidth]{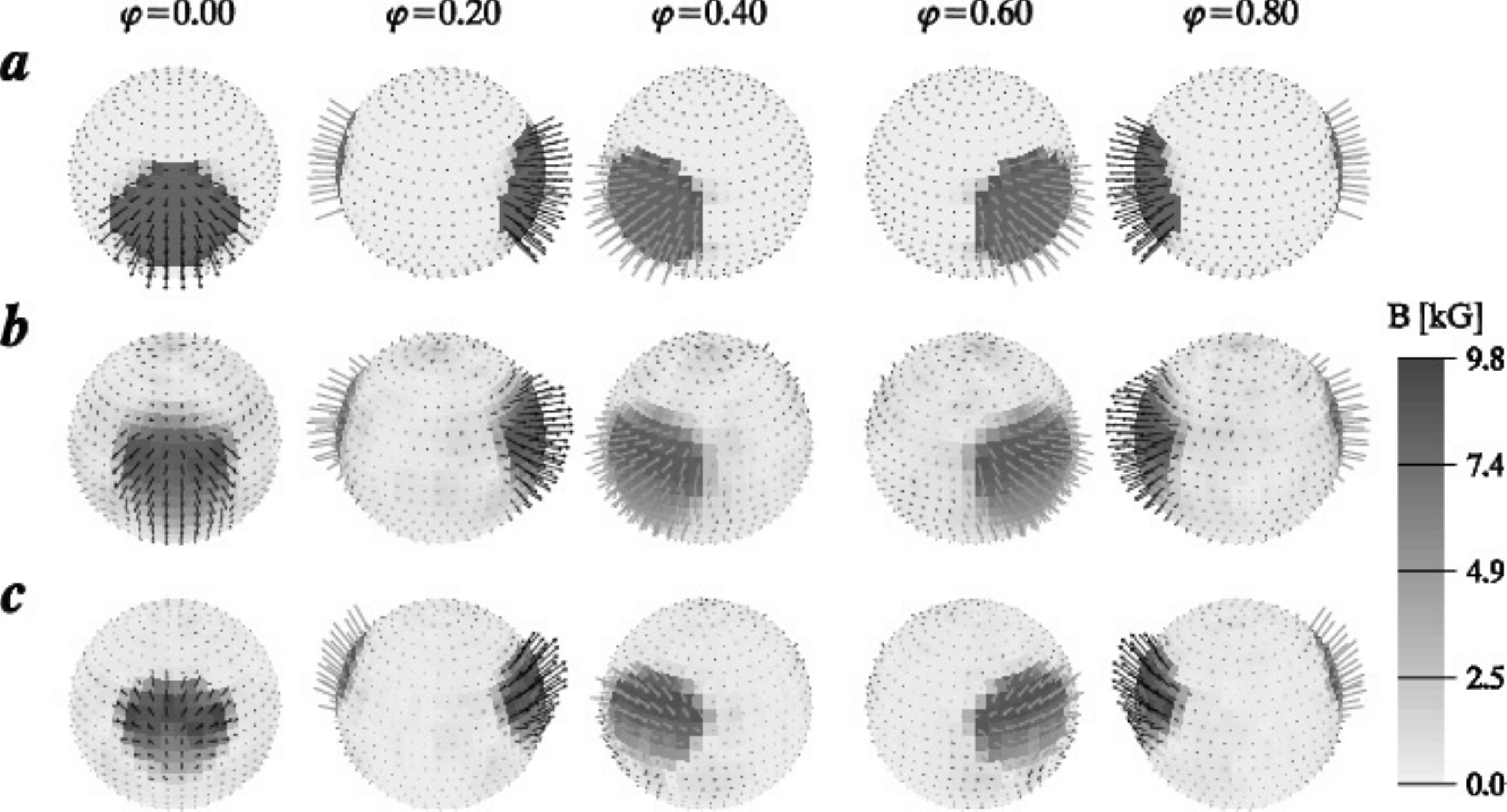}}
  \caption{(a) Surface image consisting of two magnetic spots with 8~kG 
    radial field of opposite polarity, and (b) reconstructions involving 
    all four Stokes parameters and (c) involving only Stokes~I and V 
    \citep[from][]{2002AA...388..868K}.}
  \label{fig:Piskunov}
\end{figure}}

\epubtkImage{Donati01_05-09.png}{%
\begin{figure}[htbp]
  \centerline{
    \includegraphics[height=0.9\textwidth]{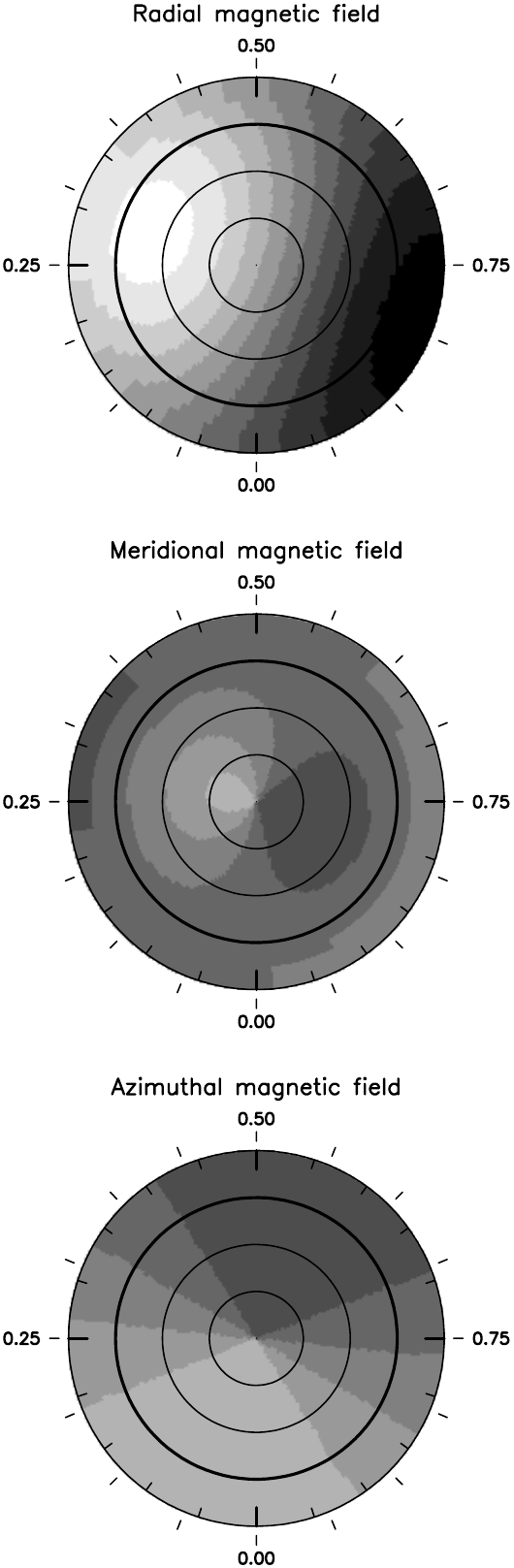}
    \includegraphics[height=0.9\textwidth]{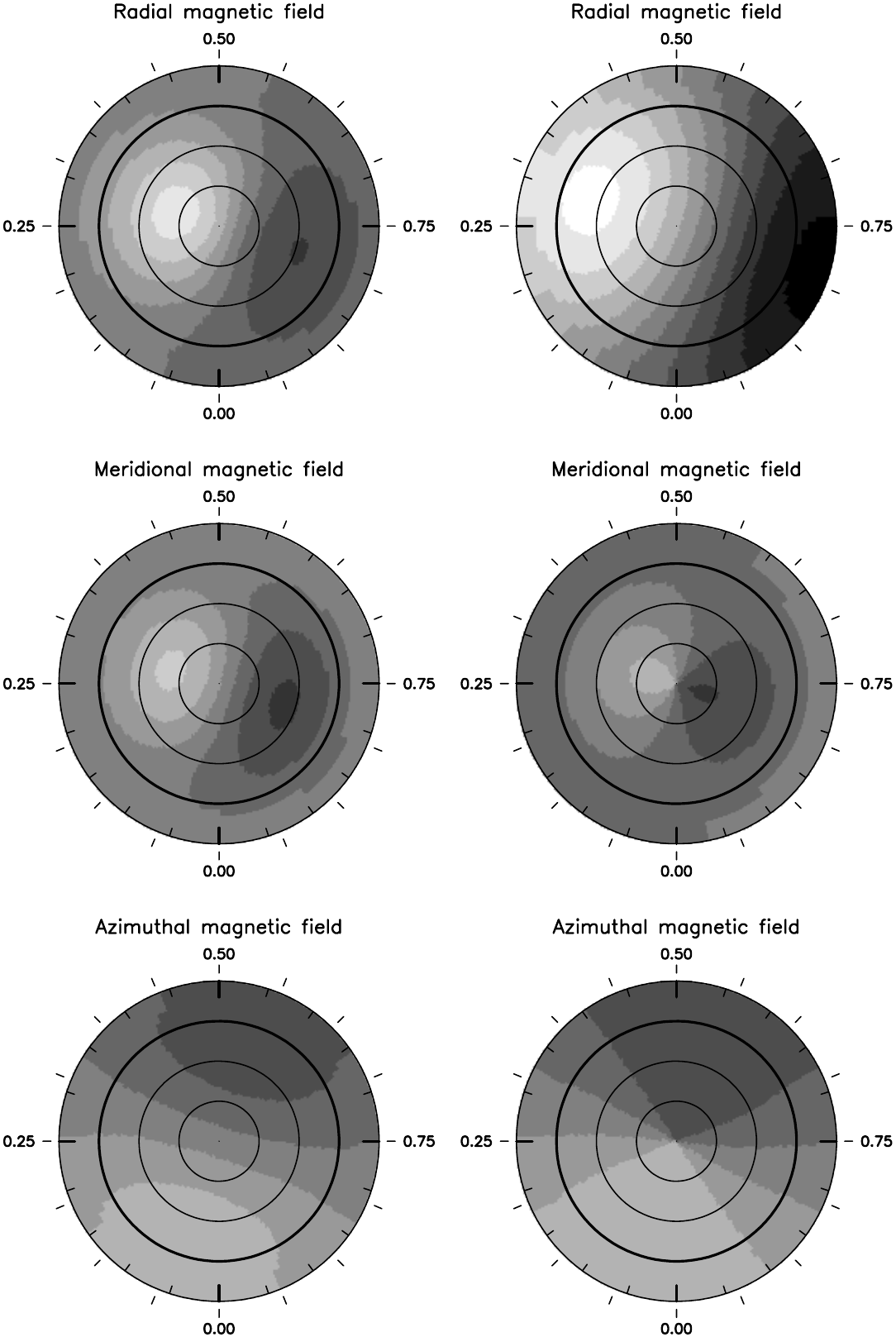}
  }
  \caption{Field reconstructions shown in a flattened polar projection
    with parallels drawn as concentric circles every 30\textdegree\ down
    to a latitude of --30\textdegree. Bold circle and central dot denote
    equator and visible pole, respectivey. Black and white code field
    intensities of 1000~G and --1000~G. Reconstructions of a
    synthetic dipole field (\emph{left panel}) are shown assuming
    unconstrained field structure (\emph{center panel}) and linear
    combination of force-free fields (\emph{right panel})
    \citep[from][]{2001LNP...573..207D}.}
  \label{fig:Donati}
\end{figure}}

\epubtkImage{DonatiBrown_4-6.png}{%
\begin{figure}[htb]
  \centerline{
    \includegraphics[width=0.46\textwidth]{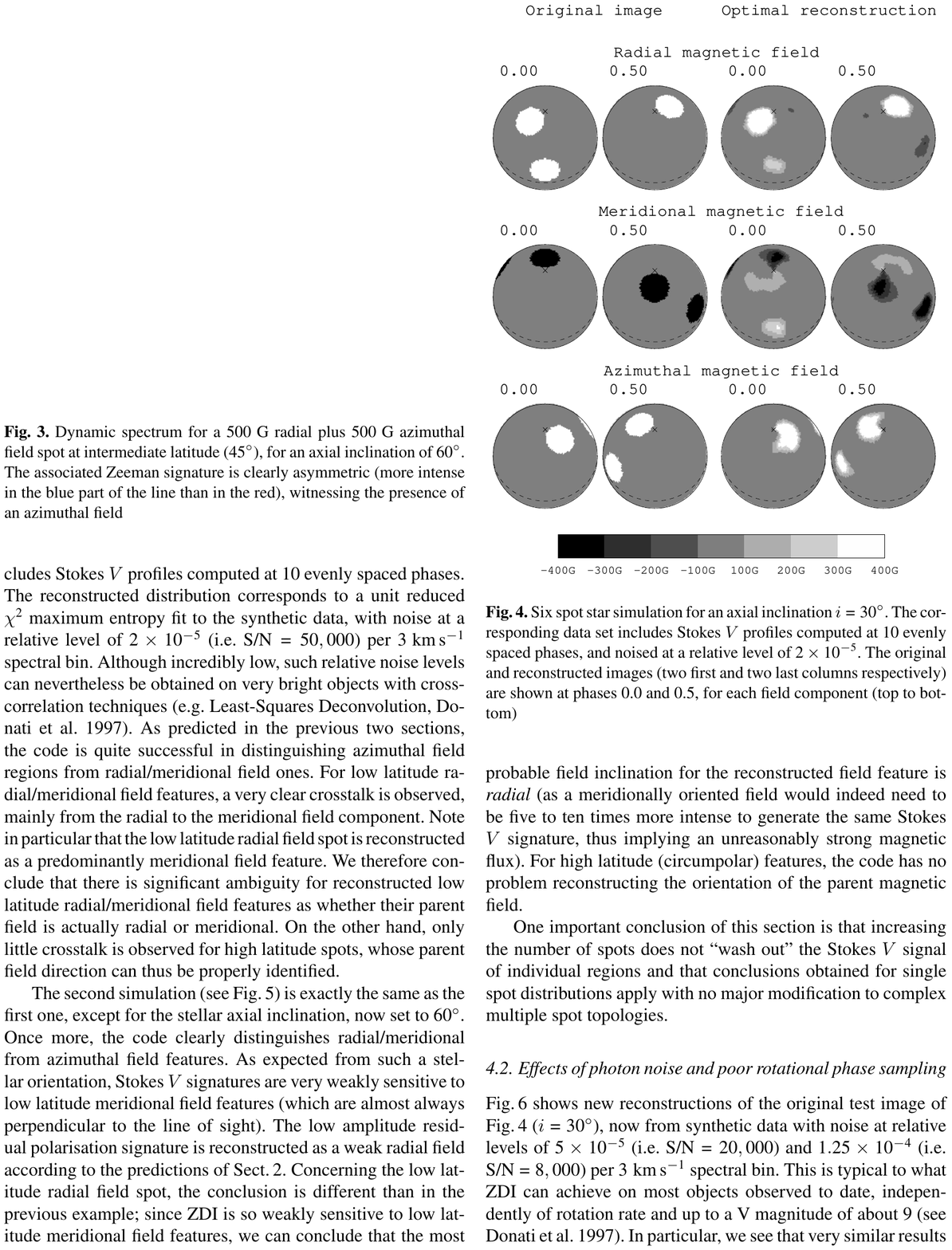}\qquad
    \includegraphics[width=0.46\textwidth]{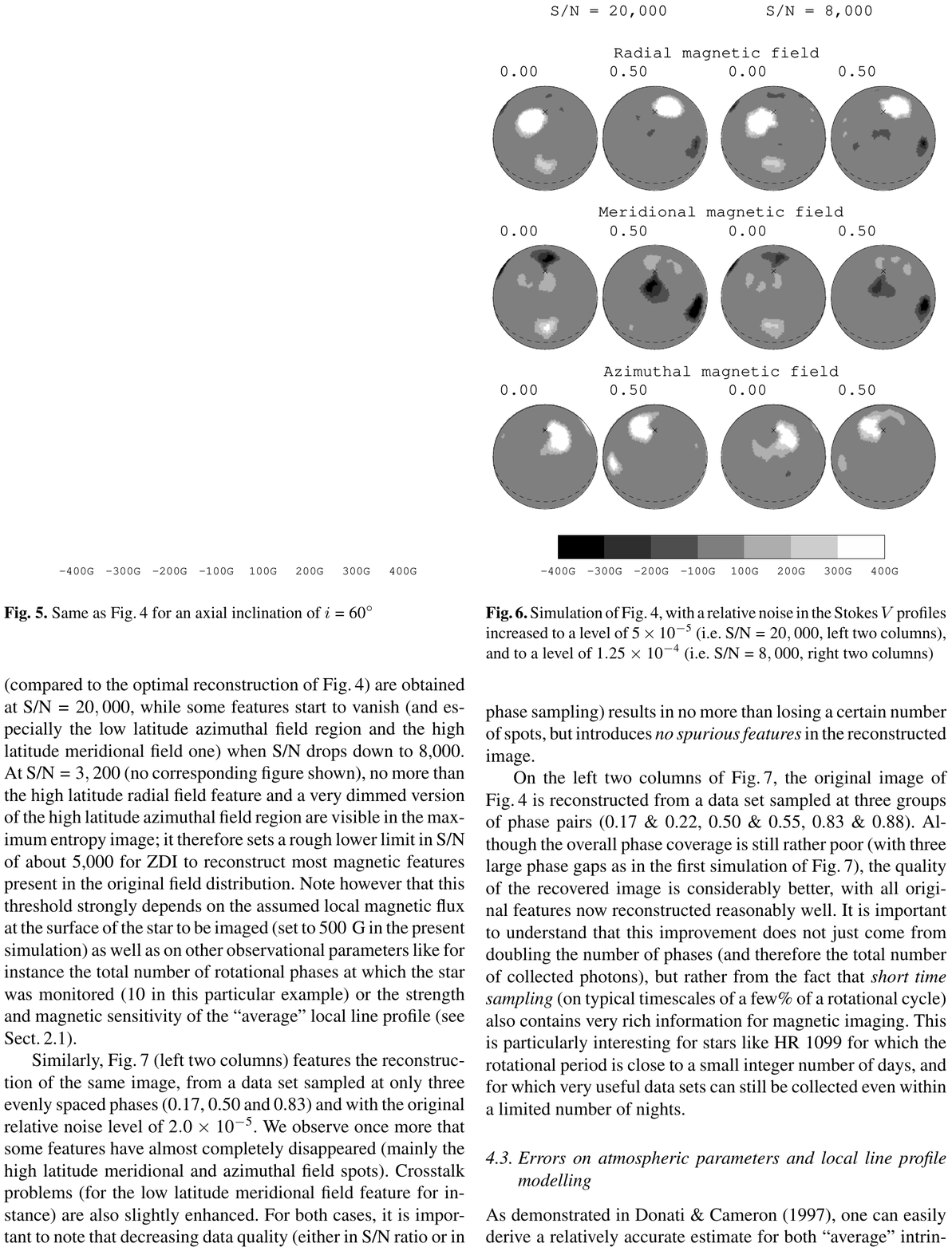}
  }
  \caption{Six spot star simulations for a star observed under an
    inclination angle of \textit{i}~=~30\textdegree. The data set includes
    Stokes~V profiles at 10 evenly spaced phases. The original images
    is shown in the left two columns. The next two columns show the
    optimal reconstruction followed by reconstructions with noise
    levels increased to 5~\texttimes~10\super{-5} (S/N~=~20\,000) and
    1.25~\texttimes~10\super{-4} (S/N~=~8000)
    \citep[reprinted with permission
      from][\copyright\ ESO]{1997AA...326.1135D}.}
  \label{fig:DonatiBrown}
\end{figure}}

Two fundamental issues for Doppler Imaging techniques are that DI
assumes the field not to be evolving, and that temperatures of
magnetic regions are not generally known. The assumption of
non-evolving fields is questionable given the high level of activity
and rate of flaring of these stars, but we have only little
information on characteristic timescales and evolution patters. Also,
temperatures of stellar active regions are poorly known in stars other
than the Sun, but regions of higher (lower) temperature add more
(less) flux to the observed spectra than the quiet stellar
photosphere.

The approaches to construct Doppler Images can be very different. It
has been shown that relatively simple magnetic geometries can be
reconstructed using all four Stokes parameters simultaneously and
calculating magnetic radiative transfer. An example from
\citet{2002AA...381..736P} is shown in Figure~\ref{fig:Piskunov},
another one from \citet{2001LNP...573..207D} is reproduced in
Figure~\ref{fig:Donati}, and a third example from
\citet{1997AA...326.1135D} is shown in
Figure~\ref{fig:DonatiBrown}. There is an extensive literature on the
applicability of ZDI that goes far beyond the scope of this
review. For detailed information, the reader is referred to
\citet{2009ARAA..47..333D}, \citet{2002AA...388..868K}, and
\citet{2001LNP...573..207D} and references therein.  As a few
examples, Figure~\ref{fig:Piskunov} shows a reconstruction of a star
with two magnetic spots \citep{2002AA...388..868K},
Figure~\ref{fig:Donati} show reconstructions of a large-scale dipolar
configurations \citep{2001LNP...573..207D} using different assumptions
on the field structure, and Figure~\ref{fig:DonatiBrown} shows a
configuration with two relatively large spots
\citep{1997AA...326.1135D}.

In cool stars, no Zeeman Doppler Image from all four Stokes parameters
exists today, but may become achievable with high-resolution
spectro-polarimeters like PEPSI \citep{2004AN....325..278S}. Because
the signal in Stokes~I is extremely weak at visual wavelengths and for
magnetic fields much weaker than several kG (as used for example in
Figure~\ref{fig:Piskunov}), even using only Stokes~I and V together is
usually not an option in cool stars (see also next section). Effects
of using Stokes~I and V, or Stokes~V alone are shown in the examples in
Figure~\ref{fig:Piskunov} and \ref{fig:DonatiBrown}. Neglecting
Stokes~Q and U leads to an underestimate of the area covered by the
magnetic spots at low latitudes and to strong crosstalk from the
radial to the meridional field map while no crosstalk appears from the
radial to the azimuthal maps
\citep{2002AA...388..868K}. \citet{1997AA...326.1135D}, using examples
with two large spots, show that imaging in Stokes~V suffers
essentially from crosstalk between low-latitude radial and meridional
field features at low inclinations, but otherwise reasonably well
recovers the input field structure. They also demonstrate how
reconstructions deteriorate when data quality is lower
(Figure~\ref{fig:DonatiBrown}).  Another example addressing the
crosstalk issue is given by \citet{2001LNP...573..207D} using examples
of a large magnetic spot and dipolar magnetic field configurations.

Obviously, ZDI is a powerful method that can be used to recover useful
information on stellar magnetic field configurations. While it is
undisputable that pure large-scale fields are more easily observable
than small-scale field components, and that crucial information about
the large-scale surface magnetic field can be recovered, it is not
entirely clear what part of a more complex field geometry is
reconstructed under realistic conditions in low-mass stars (including
cool spots and hot emission regions, small spot groups, and temporal
evolution). A very practical limitation for the Doppler Imaging
technique in cool stars is that extremely high signal-to-noise ratios
are required in polarized light in order to measure the subtle
signatures of net polarization. Simply integrating over long times in
order to collect enough photons is not applicable because individual
exposures for Doppler Imaging must be kept short enough so that
adequate spatial resolution can be achieved. One way out is to use
bigger telescopes, another is to cleverly co-add the information
contained in the many spectral lines that all contain similar
information from the star; this can be done with a technique called
Least Squares Deconvolution.

\subsubsection{Least Squares Deconvolution}
\label{sect:LSD}

The basic idea of Doppler Imaging is to translate line profile
variations into a map of the stellar surface. The information of the
surface itself is contained in every spectral line, but each line is
sampled with relatively high noise in the spectroscopic data. If one
assumes that line formation is similar in all lines, the full spectrum
can be described as a convolution between a broadening function
characteristic of the stellar surface at a given rotational velocity,
and the spectrum of the star as it would look if the star was not
rotating. Least Squares Deconvolution (LSD, developed by 
\citealp{1989AA...225..456S} and \citealp{1997MNRAS.291..658D})
is the inverse process:
assuming a non-broadened intrinsic spectrum of the star, one searches
for the broadening function that must be convolved with this intrinsic
function so that the result of the convolution provides the best match
to the observed data. \citet{1997MNRAS.291..658D} treat the observed
spectrum as the convolution of the broadening function with a set of
weighted ``delta'' functions located at the wavelengths taken from a
spectral line list. \citet{2003AA...412..813R} used a similar approach
but iteratively optimizing the weights of individual lines so that the
fit to the spectrum is improved.

In its simplest incarnation, LSD can provide the broadening function
that is inherent in all spectral lines, and using many lines can boost
the signal-to-noise ratio of the derived broadening function with
respect to individual lines. Furthermore, line blending can be treated
very effectively. LSD can provide an accurate measure of the
broadening profile inherent to all spectral lines if one makes the
assumption that the broadened template spectrum captures all
differences between the lines used
\citep[e.g.][]{2003AA...398..647R}. This implies that lines are not
allowed to follow different broadening patterns or line formation
processes \citep{2010AA...522A..57S}. As a consequence, lines with
different Land\'e factors following different broadening patters
cannot be used to derive a broadening profile that can be interpreted
as the broadening profile inherent in each line. If the broadening
patterns of individual spectral lines differ, however, LSD can still
be used to determine an average broadening function from many
lines. As an approximation for Zeeman broadening, average Land\'e $g$
values are sometimes assumed to derive an average Zeeman broadening
profile in Stokes~I \citep[e.g.,][]{2008MNRAS.390..567M}. The
interpretability of these signatures is limited
\citep{2010AA...522A..57S} but can still allow a useful mapping of the
stellar surface.

For polarized light, \citet{1997MNRAS.291..658D} show an elegant way
how LSD can be used to extract mean broadening profiles from circular
polarization in Stokes~V data, and \citet{2000MNRAS.313..823W} extend
this formalism to linear polarization. A crucial step is to apply the
so-called weak-field approximation \citep[see][]{1956PASJ....8..108U,
  1994StenfloBook}: if Zeeman splitting is much smaller than the
Doppler width of spectral lines, the following equations hold for
every line $i$:

\begin{eqnarray}
  \label{Eq:WeakField}
  V (v) & \propto & g_i B \frac{\partial I(v)}{\partial v} , \\
  Q (v) & \propto & g_i^2 B^2 \frac{\partial^2 I(v)}{\partial v^2} ,
\end{eqnarray}
with $V$ and $Q$ the Stokes parameters, $g_i$ the Land\'e factor for
line $i$, $B$ the magnetic field, and $v$ the Doppler velocity. Thus,
under the weak-field assumption, polarized spectra can be written as a
convolution between an average line profile ($\frac{\partial
  I(v)}{\partial v}$ or $\frac{\partial^2 I(v)}{\partial v^2}$) and a
line list in which each line is weighted by its Land\'e factor. The
amplitude of the deconvolved broadening function in is proportional to
$B$ and $B^2$ in $V$ and $Q$, respectively. In the weak-field
approximation \citep[together with the weak-line
approximation;][]{2010AA...522A..57S}, all line profiles have identical
shape and only differ in intensity, which allows the use of a linear
multi-line approach like LSD, which makes interpretation of the
derived profile relatively straightforward. If fields are strong
enough so that Stokes~V splitting patterns significantly differ in
shape between different lines, or if several lines are saturated, the
meaning of the derived function becomes less obvious. Several other
methods that overcome these limitations like Principal Component
Analysis \cite[PCA;][]{2008AA...486..637M} or Zeeman Component
Decomposition \cite[ZCD;][]{2010AA...522A..57S} were developed during
the last years.

As was mentioned several times already, polarization signals from
integrated observations of cool stars are so small that usually they
cannot be detected in individual spectral lines with current
instrumentation. If the weak-field approximation is used, it is
difficult to assess how the reconstruction of magnetic fields is
affected, in particular together with ZDI. \citet{1997AA...326.1135D}
point out that the weak field approximation is in principle no longer
valid for field strengths above 1.2~kG, but the authors claim that in
special cases the weak field approximation can adequately describe
Stokes~V profiles up to 5~kG \citep[see
  also][]{1997MNRAS.291....1D}. In summary, it appears not obvious
that algorithms applying the weak field approximation are sensitive to
(and can correctly interpret) the signatures of fields much larger than
1~kG. A potential consequence could be that they are not only
insensitive to \emph{average} fields above kG-strength, but would also
systematically miss spatially small magnetic components with fields of
this strength, as for example large spots similar to the largest
sunspots.

\subsection{Broad band polarization}
\label{sect:broadband}

Obtaining high resolution spectra of cool stars, in particular of very
faint M~stars and cooler objects, is challenging because the required
signal-to-noise ratios are difficult to reach. It is, therefore, very
desirable to develop a method to measure magnetic field properties
from low-resolution spectroscopy or even
photometry. \citet{1962AnAp...25..127L} proposed that broad band
linear polarization can be caused by differences between saturation of
$\pi$ and $\sigma$ components (Figure~\ref{fig:Leroy}). Based on the
observation of linear polarization in different filters by
\citet{1976ApJ...204L..47K}, \citet{1976ApJ...204..818M} present
evidence for a magnetic field of 10~kG strength on the bright spotted
dwarf BY~Dra. This scenario, however, was ruled out by several later
measurements. \citet{1991ApJ...374..319H} and
\citet{1993ApJ...404..739S} modeled broad band linear polarization in
cool stars including polarization from scatter in the stellar
atmosphere. They show that broad band polarization probably dominates
over Rayleigh and Thomson scattering. Measurements of linear
polarization in cool stars were reported, e.g., by
\citet{1981AA...101..223T} and \citet{2003ARep...47..430A}, but the
correspondence to magnetic regions is not entirely
clear. \citet{2003ARep...47..430A} show linear polarization depending
on wavelength with higher degrees of polarization at short
wavelengths. This dependence is expected if the signal comes from the
magnetic surface of the star, but in some cases the detected
polarization strongly exceeds the maximum level expected. Thus, a
supplementary source of polarization is suggested, which is proposed
to be most likely the remnant of a circumstellar disk. If such a disk
is required, however, polarization due to magnetism and polarization
from the disk are difficult to disentangle. Other potential sources of
broadband linear polarization include light source anisotropy
\citep{1999AA...347..919A} and stellar flares \citep{1994AA...286..194S}.

\epubtkImage{Bagnulo_3a-3b.png}{%
\begin{figure}[htbp]
  \centerline{
    \includegraphics[width=0.5\textwidth]{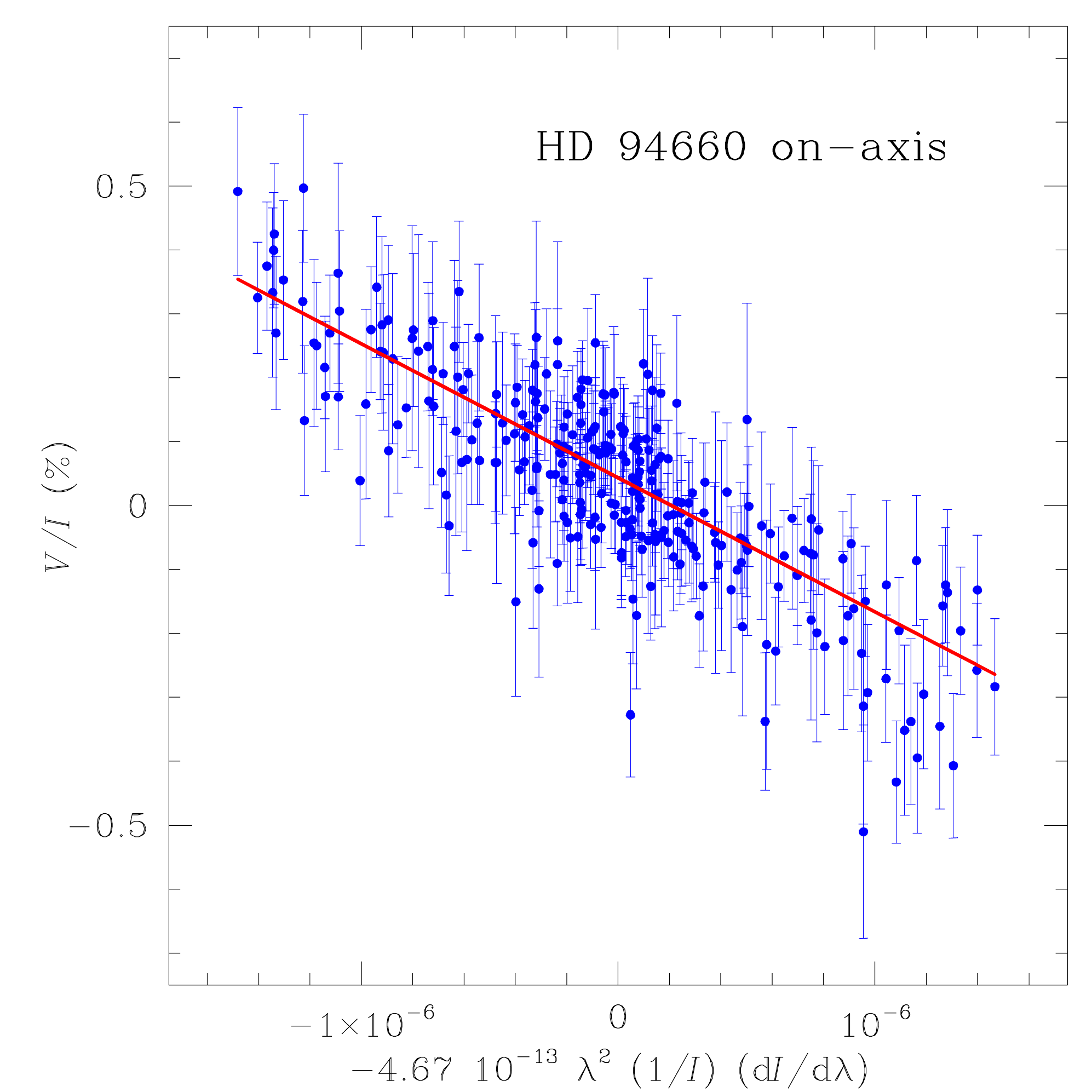}
    \includegraphics[width=0.5\textwidth]{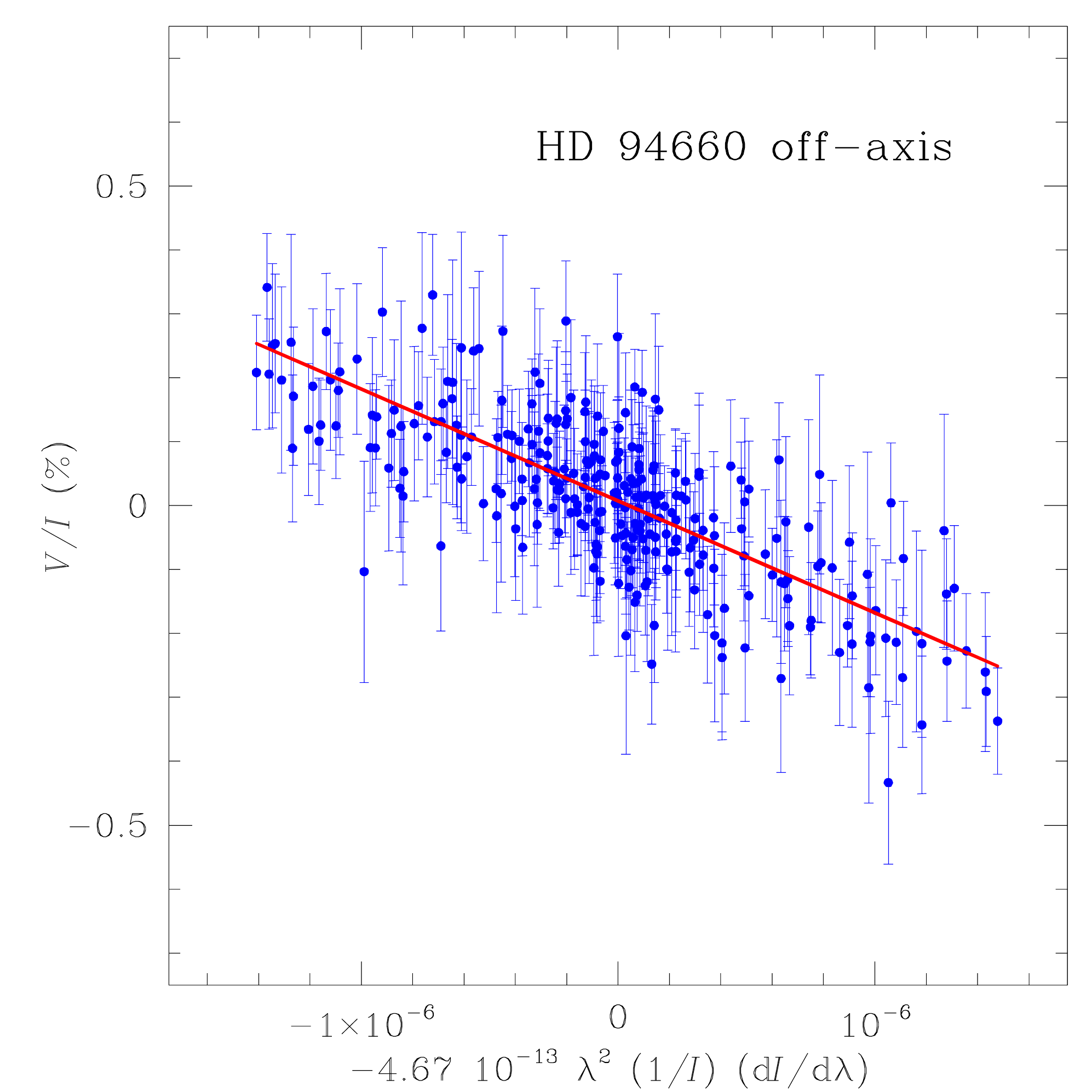}
  }    
  \caption{Stokes~V plotted vs.\ the derivative
    of Stokes~I (weak field approximation). The linear relation shows
    that the star has a magnetic field, and from the slope of the
    relation magnetic field strengths on the order of 2~kG are
    derived for the two configurations \citep[from][]{2002AA...389..191B}.}
  \label{fig:Bagnulo}
\end{figure}}

\citet{2002AA...389..191B} demonstrated a method they used to
successfully measure magnetic fields in hot stars from low spectral
resolution. Assuming that the weak-field approximation holds, they
plot circular polarization $V$ against the derivative of intensity,
$dI/d\lambda$ (see Equation~(\ref{Eq:WeakField})). If the weak-field
approximation holds and the intensity derivative is proportional to
Stokes~V, the longitudinal magnetic field can be determined from the
slope of their relation (Figure~\ref{fig:Bagnulo}). The method is
fairly straightforward in hot stars with well-separated hydrogen lines
that all have very similar Land\'e factors. The method has not yet
been applied successfully to cool stars (understood here as stars of
spectral type F and later) and it is not clear whether it would work
given the large number of blended lines with very different Land\'e
factors. \citet{2009AA...498..543K} applied the method to RR~Lyr
stars that are technically similar to cool stars (regarding their
outer convection zones) but have spectra very different to later
spectral types. Nevertheless, this may be a promising method to
determine longitudinal net field strengths in cool stars that are not
observable at very high spectral resolution.

\subsection{Indirect diagnostics}
\label{sect:indirect}

We know from the Sun that magnetic regions lead to enhanced emission
both in the solar chromosphere and in the corona. Chromospheric and
coronal emission can be observed in tracers like Ca\,{\sc ii}
emission in H \& K lines or the Ca triplet, in H$\alpha$, in UV, X-ray,
or radio emission. If we assume that other stars obey the same
relations between magnetic fields and emission processes, we can
determine their magnetic fields from observations of these
tracers. For most of the indirect tracers, the determination of
magnetic field requires: 1) that magnetic fields and their
configuration in other stars are not too different from the solar
field; and 2) that we correctly identify the mechanism coupling
magnetic fields to observable emission. In this review, I will not
give a detailed discussion of the results from indirect diagnostics,
but rather introduce the general ideas and refer to the original
literature.

\subsubsection*{Intensity contrast}

Figure~\ref{fig:sun} gives a clear example of the correspondence
between surface brightness and magnetic flux density on the
Sun. Relations between these two values were provided, e.g., by
\citet{2002AA...388.1036O}. They determine contrasts of active region
faculae and the network as a function of heliocentric angle and
magnetogram signal. Although this information is not available in
spatially unresolved observations of other stars, it can be very
helpful for the analysis of stellar variability from high quality
photometric data, e.g., from the CoRoT or \emph{Kepler} satellites.

\subsubsection*{Chromospheric emission}

The correspondence between chromospheric Ca\,{\sc ii} K emission and
magnetic fields for the solar surface was investigated by
\citet{1989ApJ...337..964S}. Figure~\ref{fig:Schrijver} shows that a
close relation exists between the field strength and Ca\,{\sc ii}
emission in solar surface observations:
\begin{equation}
  \frac{I_c - 0.13}{I_W} = 0.008\, \langle fB \rangle^{0.6}, 
\end{equation}
with $I_C$ the core intensity and $I_W$ the intensity in the wings
\citep[see][]{1989ApJ...337..964S}. \citet{1990AA...234..315S} found
a relation between C\,{\sc iv} and magnetic flux density of the
form
\begin{equation}
  F_{\mathrm{C\,{\sc iv}}} \propto \langle fB \rangle^{0.7}. 
\end{equation}
A similar correspondence was also observed in M~stars but using
H$\alpha$ \citep{2007ApJ...656.1121R,
  2010ApJ...710..924R}. Disk-integrated measurements of stellar
chromospheric activity can therefore trace changes in surface activity
induced by, e.g., rotation or magnetic cycles \citep[see,
e.g.,][]{1995ApJ...438..269B, 2008LRSP....5....2H}.

\epubtkImage{Schrijver.png}{%
\begin{figure}[htbp]
  \centerline{\includegraphics[width=0.6\textwidth]{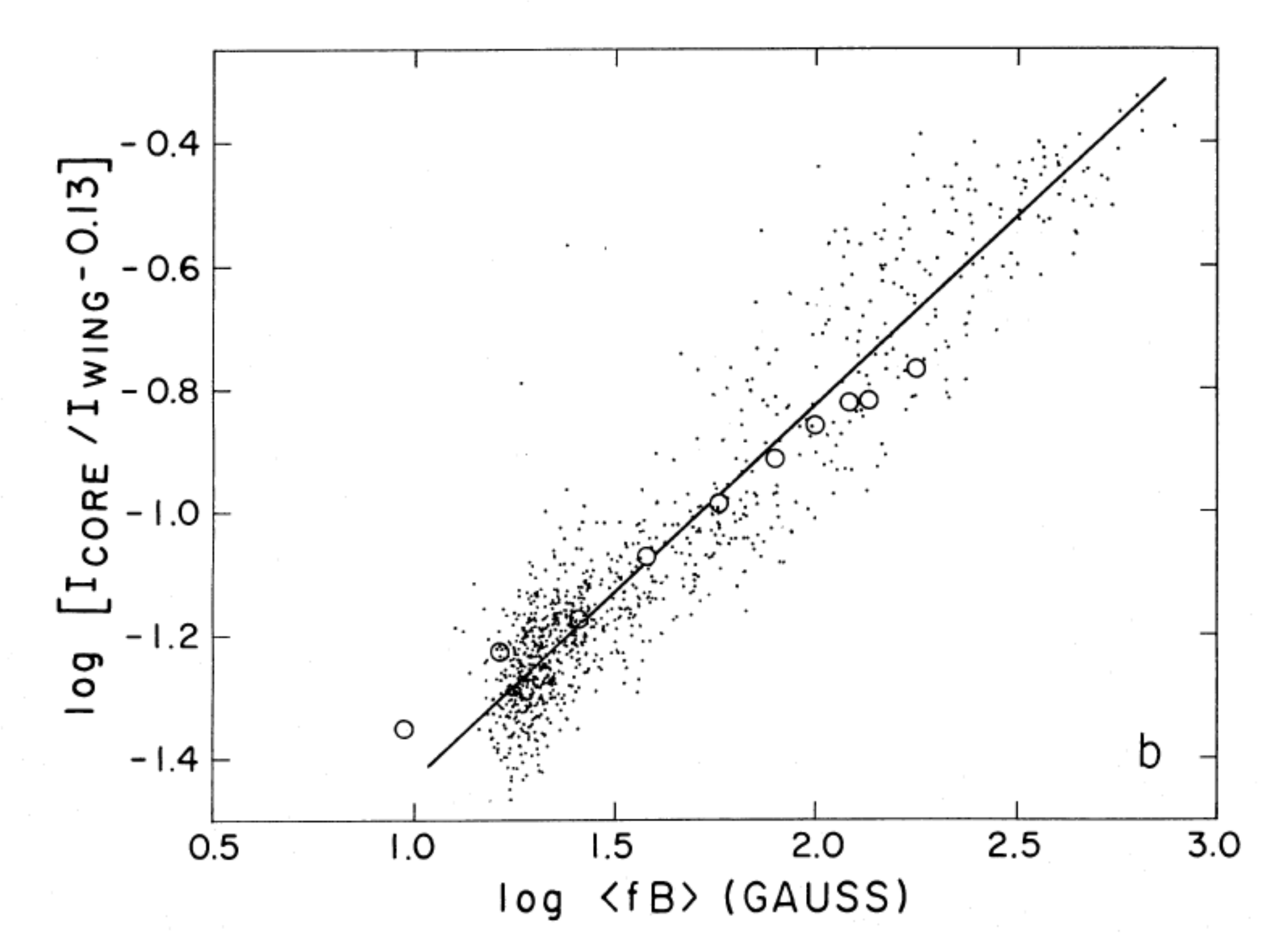}}
  \caption{Ca\,{\sc ii}~K core wing intensity
    ratio vs.\ absolute value of the magnetic flux density from
    resolved solar surface observations after degradation of the
    resolution to 14.4"~\texttimes~14.4" 
    \citep[from][reproduced by permission of the AAS]{1989ApJ...337..964S}.}
  \label{fig:Schrijver}
\end{figure}}

\subsubsection*{X-ray emission}

X-ray observations are available for the Sun and many stars. A close
relation between magnetic flux and X-ray spectral radiance was shown
by \citet{2003ApJ...598.1387P} \citep[see
also][]{2004AARv..12...71G}. The relation holds for solar quiet
regions, active regions, and disk-integrated measurements of very
active stars and covers more than ten orders of magnitude in both
parameters (see Figure~\ref{fig:Pevtsov}). The relation is approximated by 
\begin{equation}
  L_X \propto \Phi^{1.15},
\end{equation}
with $L_X$ the X-ray spectral radiance and $\Phi$ the magnetic
flux. The relation is similar to the one found by
\citet{2001ASPC..223..292S} for cool stars,
\begin{equation}
  F_X \propto \Phi^{0.95}.
\end{equation}

\epubtkImage{Pevtsov.png}{%
\begin{figure}[htbp]
  \centerline{\includegraphics[width=0.6\textwidth]{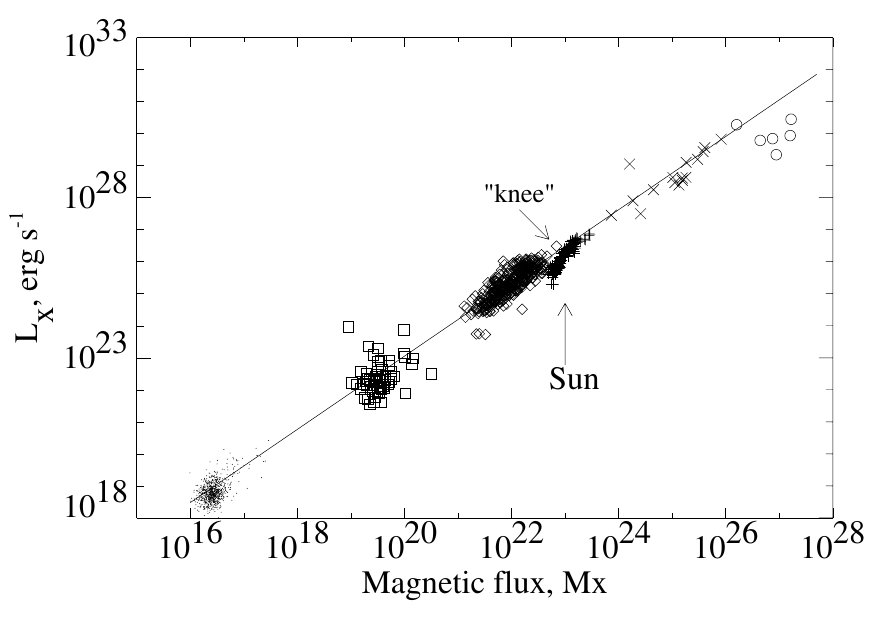}}
  \caption{\label{fig:Pevtsov} X-ray spectral radiance vs.\ total
    unsigned magnetic flux for solar and stellar
    observations. \emph{Dots:} Quiet Sun. \emph{Squares:} X-ray bright
    points. \emph{Diamonds:} Solar active regions. \emph{Pluses:}
    Solar disk averages. \emph{Crosses:} G, K, and M
    dwarfs. \emph{Circles:} T Tauri stars. \emph{Solid line:}
    Power-law approximation $L_X \propto
    \Phi^{1.15}$
    \citep[from][reproduced by permission of the AAS]{2003ApJ...598.1387P}.}
\end{figure}}

\subsubsection*{Radio emission}
\label{sect:radio}

Emission of radiation at radio wavelengths is indicative of ionized
atmospheres that in many cases are related to stellar magnetic
activity. Radio emission can be generated by different processes
leading to characteristic signatures of radio emission; for an
overview see \citet{2002ARAA..40..217G}. The close correlation
between radio and X-ray emission \citep[the G\"udel--Benz
relation;][]{1994AA...285..621B} shows that radio and X-ray emission
are generated by the same or at least correlated processes. The
relation holds for quiescent and active emission of the Sun and a wide
variety of stars. In very low-mass stars or brown dwarfs, however,
\citet{2005ApJ...627..960B} showed that this relation is violated with objects
that are overluminous at radio wavelengths.

Depending on the emission process, or on the question whether the
emitting electrons are relativistic or not, radio emission has
characteristic properties that can allow the determination of magnetic
fields \citep[see][]{2002ARAA..40..217G}. The gyrofrequency, or
cyclotron frequency, in a magnetic field is
\begin{equation}
  \nu_c = \frac{eB}{2 \pi m_e c} \approx 2.8 \times 10^6\,B,
\end{equation}
with the magnetic field strength $B$ in Gauss and $\nu_c$ in
Hz. Gyrosynchrotron emission from a power-law electron distribution is
proportional to $\gamma^{-\delta}$ (with $\gamma$ the Lorentz factor)
and shows polarization characteristic for the magnetic field. For a chosen
angle between the line of sight and the magnetic field, $\theta =
\pi/3$, the polarization $p$ can be written
\citep{1985ARAA..23..169D}
\begin{equation}
  p \approx 10^{3.35 + 0.035 \delta} (\nu/B)^{-0.51}.
\end{equation}
In principle, this equation can be used to determine the magnetic
field strength from the fractional polarization of radio emission, but
it rests on several assumptions and is very sensitive to the geometry
of the emitting regions, which is not known in spatially unresolved
stars.

\enlargethispage{\baselineskip}
Coherent emission, in particular electron cyclotron maser emission,
can be a reliable tracer of the magnetic field strength because it is
emitted mostly at the fundamental and the second harmonic of
$\nu_c$. Detection of radio emission at a given frequency indicates
the presence of magnetic field corresponding to that frequency, for
example the detection of 8.5~GHz radio emission indicates a field of
strength $B \ge 3\mathrm{\ kG}$ \citep[see, e.g.,][]{2008ApJ...684..644H}.

\newpage

\section{Magnetic Field Measurements in Cool Stars}

\subsection{Average magnetic fields from integrated light}
\label{sect:results_I}

\bigskip

\mbox{
  \parbox[t]{.35\textwidth}{\ }
  \parbox[t]{.6\textwidth}{\emph{As more physical effects were included in the modeling, 
      more sources of line broadening were treated and the magnetic parameters decreased.\\ 
      But one must ask: will $fB \rightarrow 0$ eventually?}\\ \cite{1996mpsa.conf..367S}}}

\bigskip

The history of magnetic field measurements in cool and, in particular,
in sun-like stars, is not easily followed. The fundamental paradigm of
magnetic fields leading to chromospheric and coronal emission, as
observed on the Sun, has motivated clear expectations on the presence
and properties of magnetic fields. The relation between rotation and
activity, hence presumably also between rotation and magnetic flux,
and the difficulty to detect Zeeman signatures in rotationally
broadened spectral lines causes great practical difficulty, especially
in sun-like stars. In low-mass (M-type) and pre-main sequence stars,
the relation between activity and rotation is presumably more
observer-friendly, facilitating the detectability of Zeeman
broadening. I will, therefore, distinguish between magnetic field
observations in sun-like stars, low-mass stars, and pre-main sequence
stars.

\subsubsection{Sun-like stars}

The general difficulties detecting the subtle effects of Zeeman
broadening in a spectral line from the spatially unresolved stellar
disk were discussed in Section~\ref{Sect:ZeemanEffect}. A promising
way to overcome the problem of degeneracies between Zeeman broadening
and other broadening agents is to compare spectral lines with
different Zeeman sensitivities in the same spectrum. An enhanced width
(or equivalent width) of the magnetically sensitive lines often is
good indication for the presence of a magnetic field. This strategy
was successfully applied to Ap stars with fields of 1~kG-strength by
\citet{1971ApJ...164..309P}. \citet{1980ApJ...240..567V} used a
multichannel photoelectric Zeeman analyzer mainly to measure
polarization in sun-like stars, but also presents comparison
between the widths (FWHM) of magnetically sensitive lines at
6173~\AA\ and two nearby, magnetically less sensitive lines. Four
stars were analyzed with this method finding no evidence for magnetic
fields. \citet{1980ApJ...240..567V} concludes that this rules out the
presence of non-coherent longitudinal fields in excess of 1000\,--\,1500~G
and covering the entire surface, which is similar to $Bf \le 1500$~G.

\citet{1980ApJ...239..961R} introduced a new method based on the
comparison between magnetically sensitive and insensitive
lines. Realizing that the increase in line width for fields less than
several kG is very small at optical wavelengths, he suggested to
employ a Fourier transform technique to easily separate the broadening
effects due to magnetism from other broadening effects. The underlying
principle is the very same as if one is comparing line shapes or line
widths directly in the wavelength regime (instead of Fourier
regime). However, the Fourier transform technique is able to cleanly
separate the different broadening effects at least in principle, and
thus could ideally separate magnetic broadening from other
effects. The main limitation of Zeeman broadening measurements at
optical wavelengths, however, cannot be overcome by this method: it is
still necessary to precisely measure a magnetically non-broadened line
in order to use it as a template for the (potentially stronger)
broadening observed in a magnetically sensitive line. Both lines must
be of very similar nature in terms of formation height and temperature
response. It is, therefore, not surprising that the limitations
discussed by \citet{1980ApJ...239..961R} are essentially identical to
the limitations arising when line widths are compared
directly. Consequently, the Fourier technique was not applied to a
great many spectra, but the paper became a benchmark for line
comparison techniques in general because it thoroughly discusses the
requirements and limitations of this technique.

What followed was a series of attempts trying to measure magnetic
fields in more or less active sun-like stars. Driven by detections of
chromospheric and coronal activity, active stars with relatively low
rotational broadening ($v\,\sin{i}$) were observed in order to search
for the effects of Zeeman broadening. Highest obtainable data quality
at this time was typically on the order of $R \sim$~50\,000\,--\,70\,000 and
SNR~$\sim$~100\,--\,200. A remarkable conclusion from the magnetic field
observations taken during this time was pointed out by
\citet{1985PASP...97..719G}. Investigating the reports on magnetic
field measurements, he finds that for G- and K-dwarfs, the product
between the magnetic field strength \B, and the areal coverage factor
\f, i.e., the average magnetic field strength \Bf, \emph{``is a
  constant independent of physical parameters such as spectral type
  and rotational velocity''}. Realizing that this is rather unlikely,
he concludes that \emph{``either we have systematic misconceptions
  involved in our Zeeman-broadening analysis or else we have before us
  a remarkable magnetic conservation condition''}. The value of this
``magnetic constant'' is roughly \Bf~=~500~G. According to
Equation~(\ref{Eq:Zeemanv}), this means an extra-broadening of
700~\ms for a magnetically sensitive line ($g = 2.5$) over an
insensitive line ($g = 1.0$) at red optical wavelengths (670~nm);
this is typically between 10\% and 20\% of a resolution element.

This example demonstrates that searching for the subtle effects of a
several hundred Gauss magnetic field is close to the theoretical
detectability of the Zeeman effect, and that it is extremely difficult
to judge whether differences between lines of different magnetic
sensitivities are really due to magnetism. Consequently, the Zeeman
analysis methods were criticized by many authors \citep[see
e.g.,][]{1988ApJ...324..441S} centering on two flaws: 1) incomplete
treatment of radiative transfer, and 2) lack of correction for line
blends. \citet{1988ApJ...324..441S} presents a set of improved methods
for the analysis of magnetic fields in cool stars. Main ingredients
are radiative transfer effects, treatment of exact Zeeman patterns,
and improved correction for line blends. Following up on this
improvement, \citet{1990ApJ...360..650B} went one step further
introducing a two-component analysis by applying their more detailed
line-transfer analysis to the (more realistic) situation in which the
magnetic component of the stellar atmosphere is not identical to the
non-magnetic component. The authors also point out that the derived
magnetic flux still could be in error by a factor of 2 because
atmospheres from one-dimensional calculations are used for a
multi-component analysis (neglecting gradients and differences in
atmospheric structure); misestimates of abundance, turbulence, and
subsequently magnetic field can be quite severe. A detailed parameter
study estimating the accuracy of magnetic field analysis methods in
detailed radiative transfer calculations with embedded fluxtubes is
given by \citet{1992ASPC...26..259S} and \citet{1994ASPC...64..474S}.

Obviously, a straightforward way to improve magnetic field
measurements is to observe at longer wavelengths (see
Equation~(\ref{Eq:Zeemanv})). Useful lines are found for example at
1.56~\mum (Fe\,{\sc i}) and 2.22~\mum (Ti\,{\sc i}), i.e.,
at wavelengths a factor of 3\,--\,4 longer than typical red/optical
observations. First suitable instrumentation at such long wavelengths
became available in the early-1990s. The first detailed analysis of a
high-resolution infrared spectrum in a sun-like star (for earlier work
on M~stars, see Section~\ref{Sect:Mstars}) was performed by
\citet{1995ApJ...439..939V}. These authors used a high-resolution
(\textit{R}~=~103\,000), high SNR (100\,--\,200) spectrum (taken during
several hours of exposure) to determine the magnetic field of
$\epsilon$\,Eri, and upper limits on the order of 100~G in two other
early K-dwarfs. $\epsilon$\,Eri has been subject to magnetic field
investigations many times earlier at optical
wavelengths. \citet{1995ApJ...439..939V} also show a compilation of
reports on magnetic field measurements in this star published between
1984 and their work in 1995. Interestingly, average magnetic fields of
$\epsilon$\,Eri decreased over time starting at $\sim$~800~G in 1984
and reaching 130~G in 1995. Possible interpretations of this result
are that the field in $\epsilon$\,Eri is variable, or that
observations reporting lower field strengths (predominantly near-IR
measurements) probe a different part of the stellar
atmosphere. \citet{1995ApJ...439..939V} discuss possible scenarios
reaching the conclusion that probably optical investigations have
overestimated the magnetic flux of $\epsilon$\,Eri.

A critical compilation of magnetic field measurements obtained between
the paper of \citet{1980ApJ...239..961R} and 1996 was attempted by
\citet{1996mpsa.conf..367S}. The selection process leading to a
condensed sample of ``improved'' field measurements was described as
follows: \emph{``I have therefore compiled a carefully selected sample
  of magnetic measurements from analyses which treat radiative
  transfer effects and use disk-integration in their models. In
  addition, I (ruthlessly!) neglect results from low S/N IR data,
  measurements using Fe\,{\scriptsize I} 8468~\AA\ in K dwarfs, Zeeman/magnetic
  Doppler imaging results, and curve-of-growth analyses''} \citep[for
the reasons why some techniques were neglected,
see][]{1996mpsa.conf..367S}. A similar, upgraded collection of Zeeman
analyses carried out in the period 1996\,--\,2001 was given by
\citet{2001ASPC..223..292S}.

For this review, I have tried in Table~\ref{tab:StokesIsunlike} to
compile magnetic field measurements available for sun-like
stars. Following \citet{1996mpsa.conf..367S}, I include only those
measurements that rely on relatively high data-quality and analysis
techniques. Since apparently not very many magnetic field measurements
were reported in sun-like stars after 2001,
Table~\ref{tab:StokesIsunlike} does not contain many results in
addition to the compilations by \citet{1996mpsa.conf..367S,
  2001ASPC..223..292S}. However, in the light of the results reported
by \citet{1995ApJ...439..939V}, I distinguish between work done at
optical wavelengths and work done at infrared wavelengths, the former
probably being more prone to overestimating the magnetic field.

\epubtkImage{59Vir_B3_Comp.png}{%
\begin{figure}[htbp]
  \centerline{\includegraphics[width=\textwidth]{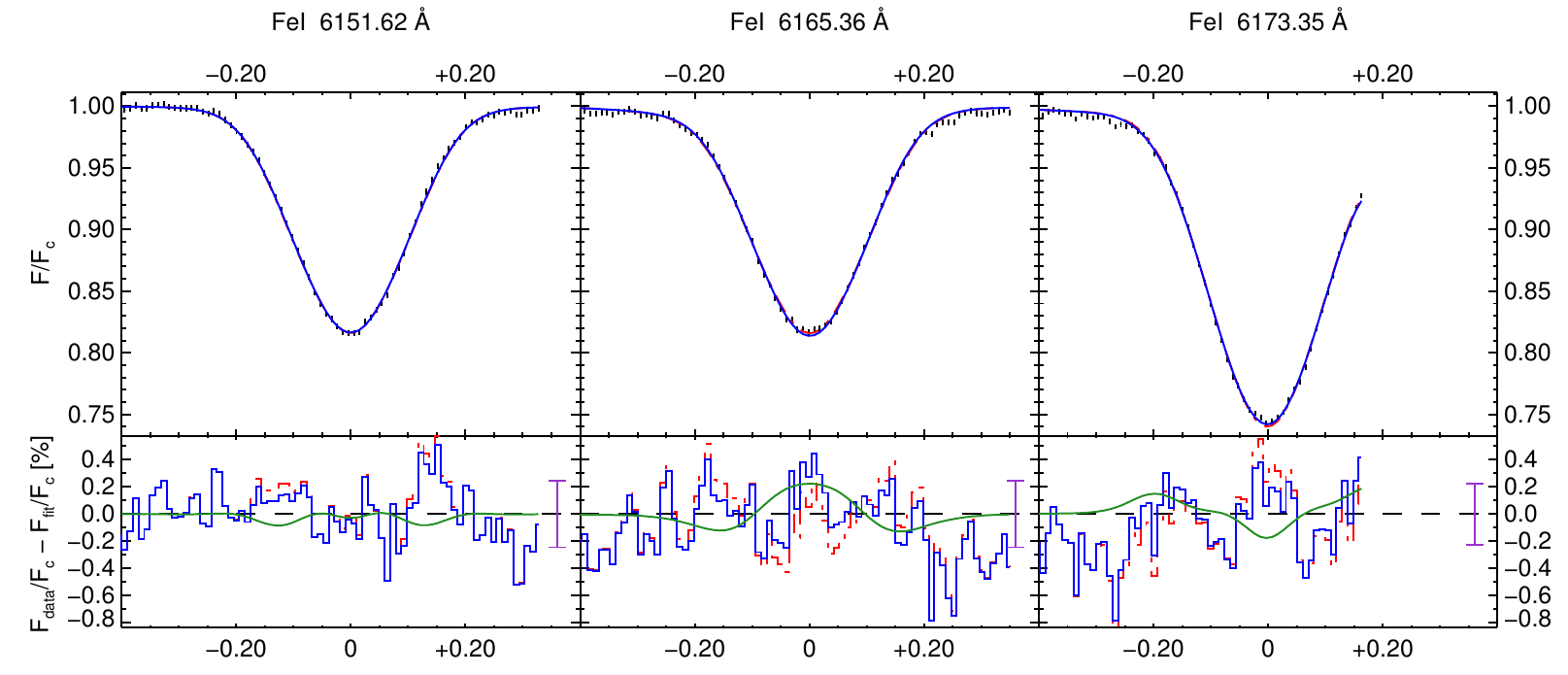}}
  \vspace{5mm}
  \centerline{\includegraphics[width=\textwidth]{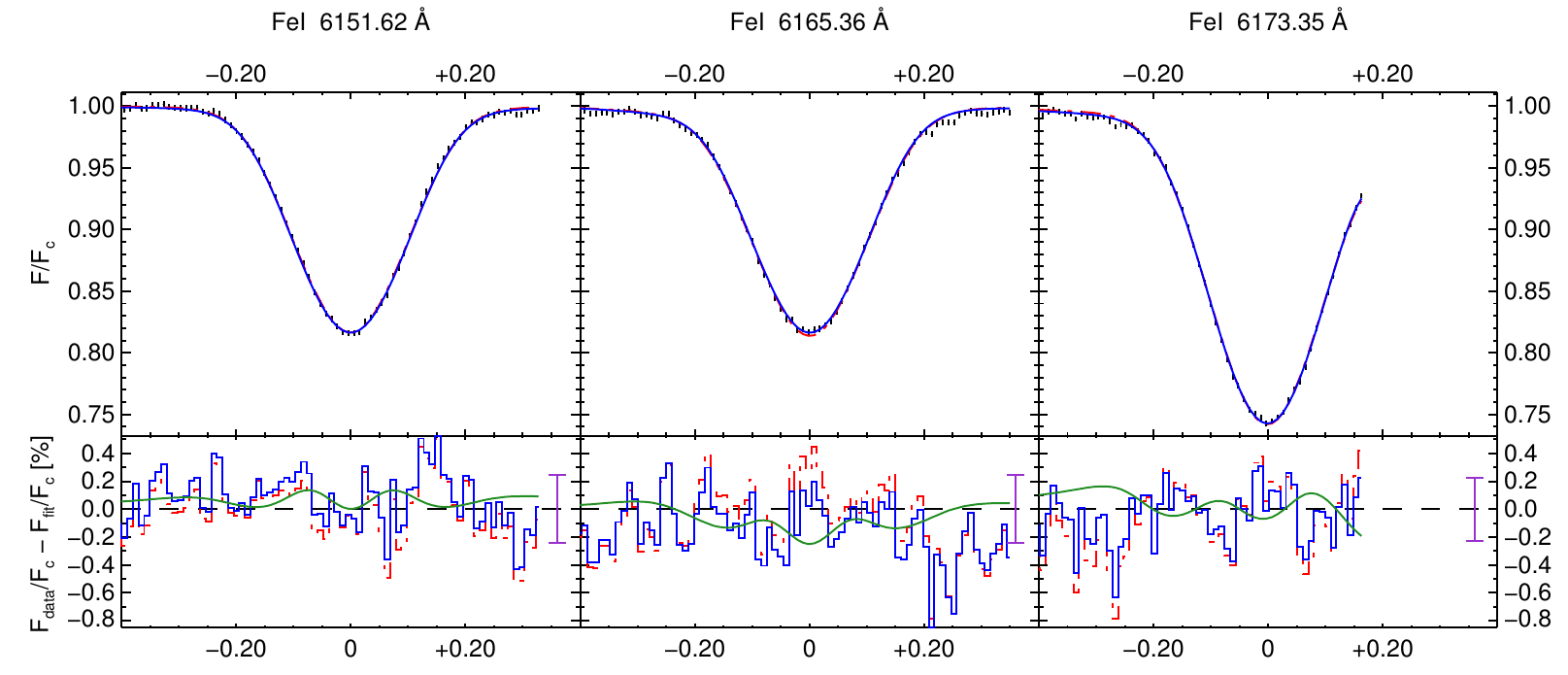}}
  \caption{CES spectra of 59~Vir with
    uncertainties overplotted by best-fit solutions. Solid blue lines
    represent the overall best-fit solutions, dash-dotted red lines
    are other solutions shown for comparison. Residuals drawn below
    the fits visualize differences between measured and calculated
    line profiles, scaled by factor 100. The purple error bar to the
    right shows $\sigma_{i}$. Green lines indicate the difference
    between overall best-fit and the comparison model, i.e., the change
    in line shape due the presence of magnetic flux (other fit
    parameters vary freely). \emph{Top:} Model with identical
    temperature for magnetic and non-magnetic regions; best-fit:
    \Bf~=~500~G (solid blue), comparison: \Bf~=~0~G (dash-dotted
    red). \emph{Bottom:} Best fit for model with different
    temperatures for magnetic and non-magnetic regions; \Bf~=~120~G
    (blue solid), comparison: solution from upper panel (\Bf~=~500~G,
    same temperatures, red dashed line)
    \citep[from][]{2010AA...522A..81A}.}
  \label{fig:AndersonSpectra}
\end{figure}}

A critical re-investigation of the detectability of magnetic fields in
high-quality optical spectra was carried out by
\citet{2010AA...522A..81A}. The data material used for this work is
of much higher quality than most magnetic field investigations before,
and the data therefore allows a critical view on the published results
and some of the limitations of the method. \citet{2010AA...522A..81A}
used optical spectra around the Fe\,{\sc i} line at 6173~\AA\
observed at a spectral resolving power of \R~=~220\,000 and
SNR~$\sim$~400. The analysis is carried out for a one-component
model with the same atmosphere for the magnetic and the non-magnetic
parts of the stellar surface, and also for a two-component model
employing different atmospheres for the two components. The results
are reproduced in Figures~\ref{fig:AndersonSpectra} and
\ref{fig:AndersonResults}. For the active G0 star 59~Vir, the authors
find a magnetic field with \Bf~$\approx$~420~G for the one-component
case. For the two-component analysis, they cannot exclude a zero-field
solution reporting an upper limit of
300~G. Figure~\ref{fig:AndersonSpectra} shows how subtle the
differences between solutions with different magnetic field strengths
are if all other relevant parameters are allowed to vary freely (there
is currently no way to constrain these parameters at the level
required). Figure~\ref{fig:AndersonResults} demonstrates the relation
between magnetic field strength \B and filling factor \f in case of
a one-component atmosphere (left panel). The two-component models
shown in the center and right panels, however, can lift the \Bf
degeneracy but manage to reproduce the spectra even without the
presence of a significant magnetic field. In other words, at optical
wavelengths, the signal of temperature spots on the surface of a cool
star can dominate the influence of the magnetic field through Zeeman
broadening. Unfortunately, we have so far no clear empirical evidence
for the relation between temperature and magnetic field strength on
stellar surfaces other than on the Sun.

A look at Table~\ref{tab:StokesIsunlike} reveals that infrared
measurements are only available in six sun-likes stars, all of them
are of spectral type K. Two of the six data points are actually
non-detections, and three were reported in conference summaries in
which, unfortunately, no comprehensive presentation of the data and
its analysis is given.

\epubtkImage{59Vir_B3_chi.png}{%
\begin{figure}[htbp]
  \def\epsfsize#1#2{.32\textwidth}
  \centerline{
    \includegraphics[width=0.3\textwidth]{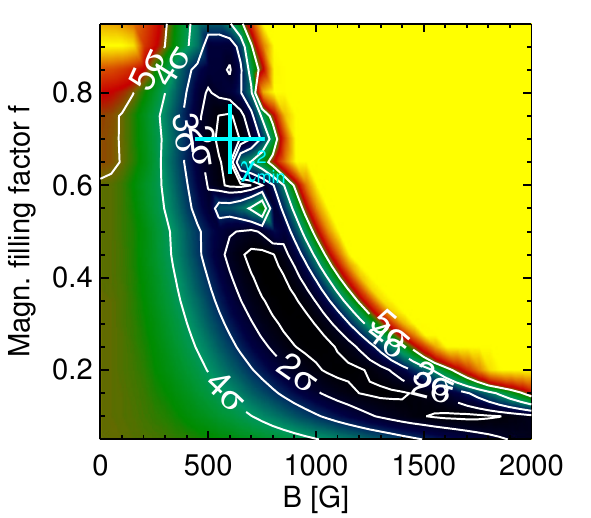}\quad
    \includegraphics[width=0.3\textwidth]{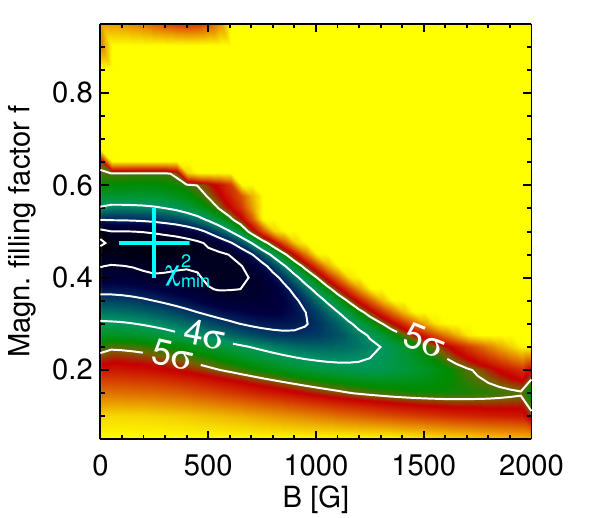}\quad
    \includegraphics[width=0.3\textwidth]{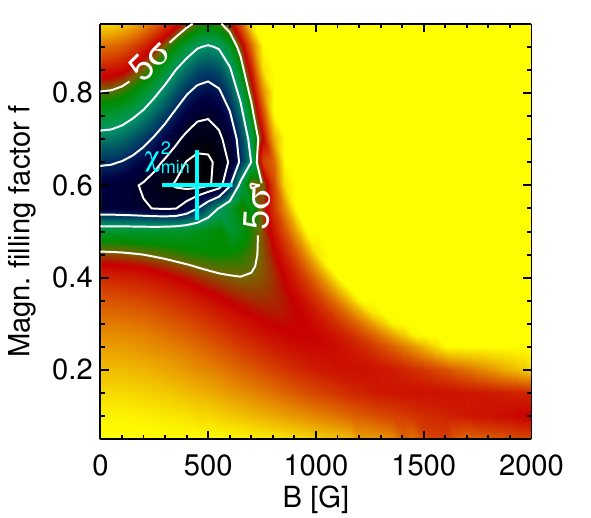}
  }
  \caption{$\chi^2$-maps for 59~Vir solutions (see
    Figure~\ref{fig:AndersonSpectra}). \emph{Left panel:} Solution
    using the same atmospheres for magnetic and non-magnetic regions;
    \emph{Center panel:} the same using cool magnetic regions;
    \emph{Right panel:} the same for warm magnetic regions
    \citep[from][]{2010AA...522A..81A}.}
  \label{fig:AndersonResults}
\end{figure}}

\subsubsection{M-type stars}
\label{Sect:Mstars}

Low-mass stars of spectral type M have radii of approximately half a
solar radius and less. If the stellar dynamo depends on the value of
the Rossby number, $Ro = P/\tau_{\mathrm{conv}}$, the magnetic field
strength expected in sun-like and low-mass stars is a function of
rotational period and convective overturn time. Values for the
convective overturn time are theoretically not well determined, but
$\tau_{\mathrm{conv}}$ is probably higher at lower masses
\citep[e.g.,][]{1996ApJ...457..340K}. Therefore, slower rotation is
sufficient to produce larger fields in less massive
stars. Furthermore, the smaller radii of less massive stars lead to
lower surface velocities hence less rotational broadening at a given
rotational period. Finally, less massive stars are also much cooler
thus exhibiting less temperature broadening in their spectral
lines. It is this combination of parameters that facilitates the
detection of Zeeman splitting in M-type stars in comparison to more
massive, sun-like stars; Zeeman broadening is more easily detected
because of generally narrower line widths \citep[see
also][]{2007AA...467..259R}.

\epubtkImage{JKV2000_1-2.png}{%
\begin{figure}[htbp]
  \centerline{
    \includegraphics[width=0.5\textwidth]{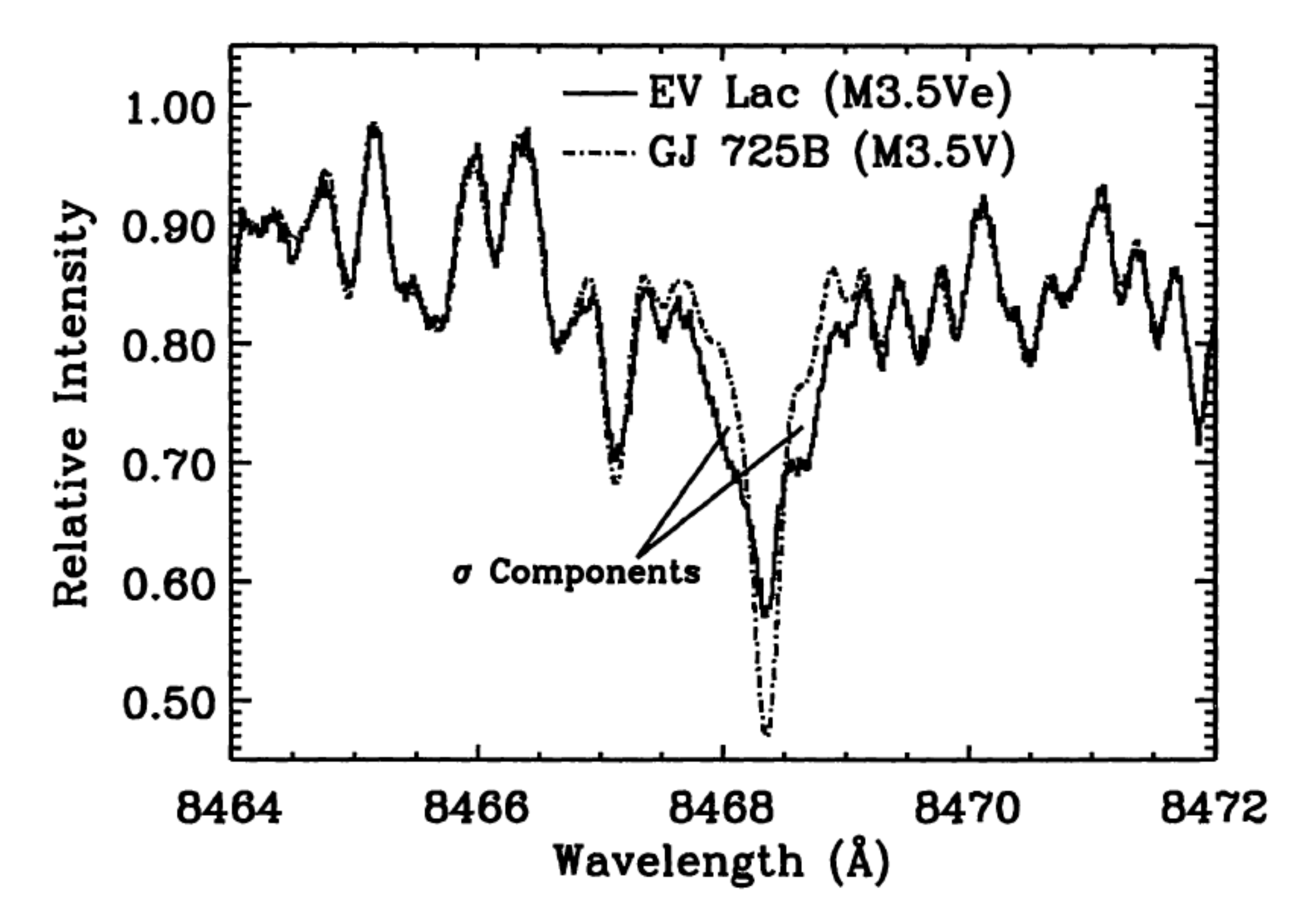}\quad
    \includegraphics[width=0.5\textwidth]{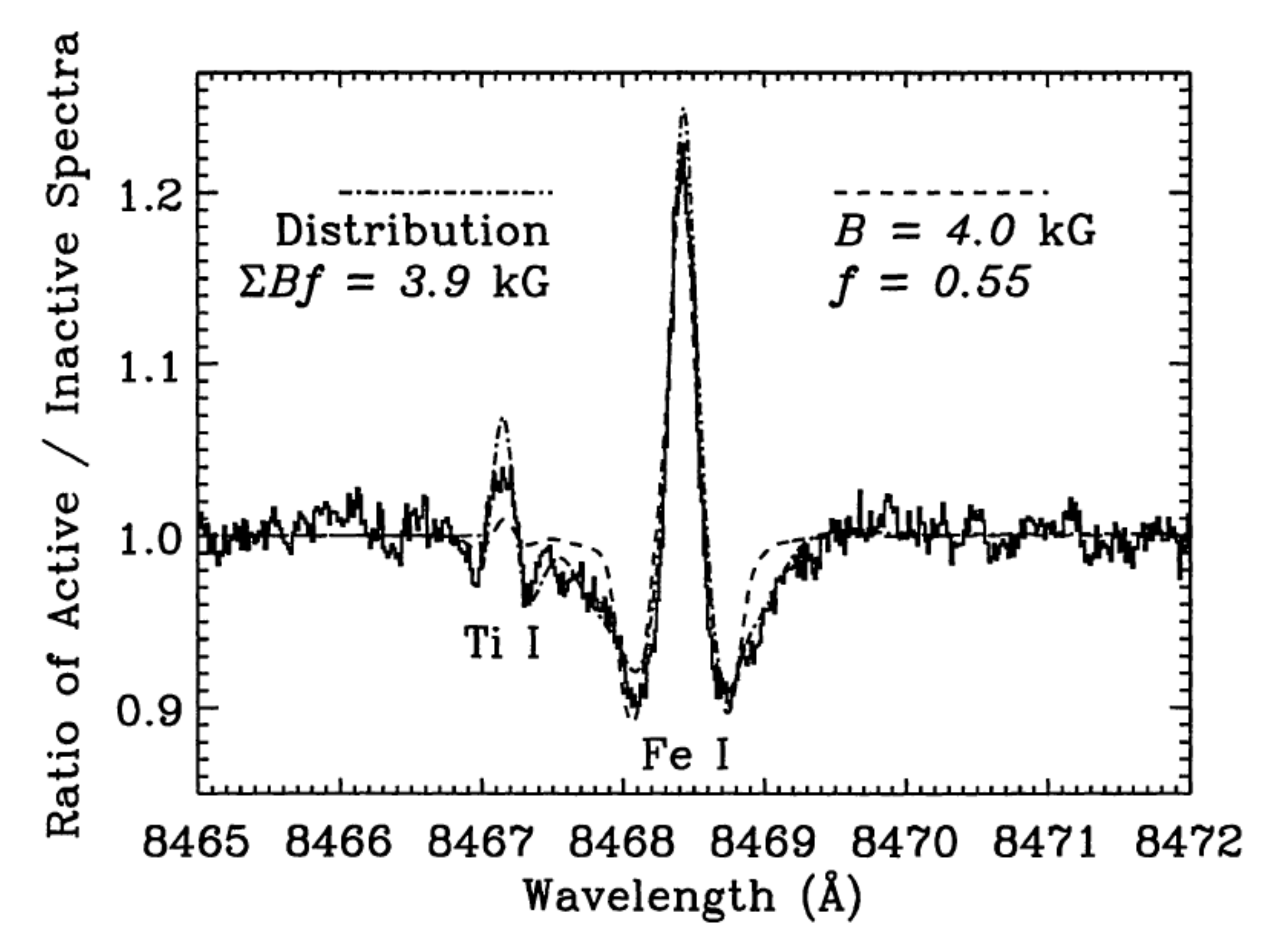}
  }
  \caption{Magnetic field measurements in active M
    dwarfs. \emph{Left:} Spectra of the flare star EV~Lac and the
    inactive star Gl~725B in the vicinity of the magnetically
    sensitive Fe\,{\sc i} line at 8468.4~\AA. \emph{Right:} The
    spectrum of EV~Lac divided by the inactive star Gl~725B (solid
    histogram). The dashed line shows a single field fit to the data
    (missing the line wings), the dashed-dot line show a fit allowing
    a distribution of magnetic fields
    \citep[from][]{2000ASPC..198..371J}.}
  \label{fig:JKV2000}
\end{figure}}

The first detection of Zeeman splitting in an M-type star, and also
the first detection of a photospheric magnetic field in cool stars at
all, was presented by \citet{1985ApJ...299L..47S}. They observed the
early-M flare star AD~Leo using a Fourier transform
spectrometer. After six hours of observation they had obtained a
spectrum with \R~=~45\,000 and SNR~$\approx$~25 around the
Ti\,{\sc i} lines at 2.22~\mum from which they measured an
average magnetic field strength of \Bf~=~2800~G. Similar data taken
with the same instrument was obtained in a few M-stars, and
\citet{1994IAUS..154..493S} presented a preliminary analysis of the
three M-type stars AU~Mic, AD~Leo, and EV~Lac.  Another benchmark was
the investigation of the Fe\,{\sc i} line at 8468~\AA\ in seven
early- to mid-M dwarfs by \citet{1996ApJ...459L..95J}. Substantial
magnetic fields were detected in two stars of the sample, EV~Lac and
Gl~729. A refined analysis of the two stars and AD~Leo and YZ~Cmi was
presented in \citet{2000ASPC..198..371J}. The latter work assumed a
distribution of magnetic fields on the stellar surface, which led to
significantly higher average field values compared to
\citet{1996ApJ...459L..95J}. The results from the 8468~\AA\ line were
comparable to the values from the 2.22~\mum line within 10\,--\,20\%.

A serious problem for the detection of Zeeman splitting in atomic spectral
lines of M-type stars is the appearance of molecular bands. For example, the
Fe\,{\sc i} line at 8468~\AA\ is embedded in a forest of TiO molecular
absorption lines, which makes the modeling of Zeeman splitting in this line a
delicate task. To overcome this problem, and the notorious difficulty to model
TiO absorption \citep[see][]{1998ApJ...498..851V}, \citet{1996ApJ...459L..95J}
modeled the ratio of the flux between an active and an inactive
star. Hopefully, our understanding of very cool atmospheres, molecular
chemistry, and molecular line formation will in the future allow a detailed
modeling of Zeeman splitting in the spectra of M~dwarfs \citep[see
also][]{2009AIPC.1094..124K, 2011arXiv1112.0141O, 2011MNRAS.418.2548S}.

At optical wavelengths, the main opacity contributors in M~dwarfs are
molecular bands from TiO and VO. Analysis of Zeeman broadening in
these bands, however, is difficult not only because of problems
getting the line formation right, but also because the lines are not
individually resolved. Nevertheless, for the detection of M~star
magnetic fields, it would be favorable to utilize molecular
absorption bands. A molecular band that appears to be extremely useful
for the analysis of M-star magnetic fields (and other purpose) is the
near-infrared band of molecular FeH. Its suitability for magnetic
analysis was shown by \citet{1999asus.book.....W}, and it was proposed
to be a useful diagnostic at low temperatures by
\citet{2001ASPC..223.1579V}. An observational problem of FeH is that
its most suitable band is located at around $\lambda$~=~1\mum, which
is too red for most CCDs and too blue for most astronomically used
infrared spectrographs. As a consequence, only very few
high-resolution spectrographs can provide spectra at this wavelength,
and efficiencies are typically ridiculously low. On the other hand, M
dwarfs emit much of their flux at near-infrared wavelengths so that in
comparison to optical measurements, the signal quality around 1\mum
is not much lower than around 700~nm if the spectra are obtained with
an optical/near-IR echelle spectrograph like HIRES (Keck observatory)
or UVES (ESO/VLT). \citet{2006ApJ...644..497R} developed a method to
semi-empirically determine the magnetic fields of M~dwarfs comparing
FeH spectra of the targets to spectra of two template stars; one with
no magnetic field and one with a known, strong magnetic field (Figure~\ref{fig:RB2007}).  This
method requires a known magnetic star to calibrate the Zeeman
splitting amplitude. The field strength of the target star is then
estimated by interpolation between the template spectra.

\epubtkImage{fg2b_online-h.png}{%
\begin{figure}[htbp]
  \centerline{\includegraphics[width=0.9\textwidth]{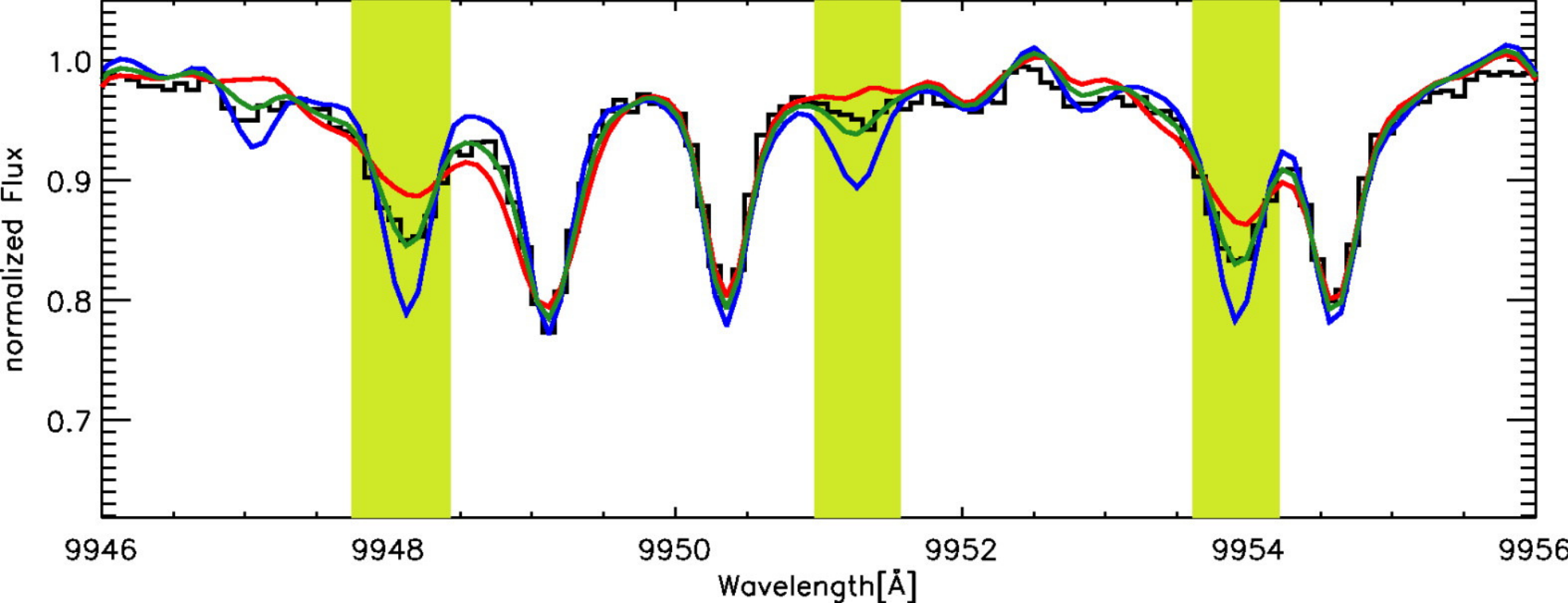}}
  \caption{Magnetic field measurement using the
    empirical method of \citet{2006ApJ...644..497R}. The black
    histogram shows the spectrum of Gl~729. The red and blue lines are
    scaled spectra of the active star EV~Lac and the inactive star
    GJ~1002, respectively. The green line is an interpolation between
    the red (\Bf~=~3.9~kG) and blue (0~kG) lines yielding a field
    strength of \Bf~=~2.2~kG for Gl~729
    \citep[from][reproduced by permission of the AAS]{2007ApJ...656.1121R}.}
\label{fig:RB2007}
\end{figure}}

The method of \citet{2006ApJ...644..497R} was first used in a sample
of 24 M~stars between spectral types M0 and M9
\citep{2007ApJ...656.1121R}. As reference, the field measurement of
EV~Lac measured by \citet{2000ASPC..198..371J} was used. Thus, all
magnetic field measurements are relative to this reference star
($\langle B \rangle$~=~3.9~kG), and magnetic fields higher than this value
cannot be quantified. Obviously, systematic uncertainties of the
measurements are quite large, typically several hundred Gauss, and
uncertainties probably grow towards very late spectral types where the
template spectra are less suited as a reference. Unfortunately, Zeeman
splitting of the FeH molecule is very complicated and could not
entirely be described at this point
\citep[see][]{2002AA...385..701B}. Meanwhile, progress has been made
using an empirical approach to understand FeH absorption and line
formation \citep{2009AA...508.1429W, 2010AA...523A..58W}, and to
model Zeeman splitting in FeH lines \citep{2009ASPC..405..527A,
  2010AA...523A..37S}. It was suggested that the fields determined
semi-empirically may be overestimated by some
$\sim$~20\%\epubtkFootnote{This could be due to an overestimate of the
  reference magnetic field measurement derived from the atomic line
  analysis.} \citep{2010AA...523A..37S}.

A (probably non-exhaustive) list of magnetic field measurements from
Stokes~I analysis in M~dwarfs is given in
Table~\ref{tab:StokesIMtype}, and I plot the distribution of field
strength as a function of spectral type in Figure~\ref{fig:MBFields}.
The field strengths of young, early-M and field mid- and late-M~dwarfs
are on the order of a few kG. This is the main results from Zeeman
analysis and consistently found using different indicators (at least
in mid-M~dwarfs). Compared to the Sun, the average magnetic field
hence is larger by two to three orders of magnitude, an observational
result that must have severe implications for our understanding of
low-mass stellar activity. It is not clear whether our picture of a
star with spots more or less distributed over the stellar surface is
actually valid in M~dwarfs. If, for example, 50\% of the surface of
a star with a mean magnetic field of 4~kG is covered with a ``quiet''
photosphere and low magnetic field, the other half of the star must
have a field strength as large as $\sim$~8~kG. The two components of
the stellar surface on such a star probably have very different
temperatures and properties, and the definition of effective
temperature must be considerably different from the temperature of the
``quiet'' photosphere.

\epubtkImage{StokesIMdwarfs.png}{%
\begin{figure}[htbp]
  \centerline{\includegraphics[width=0.8\textwidth]{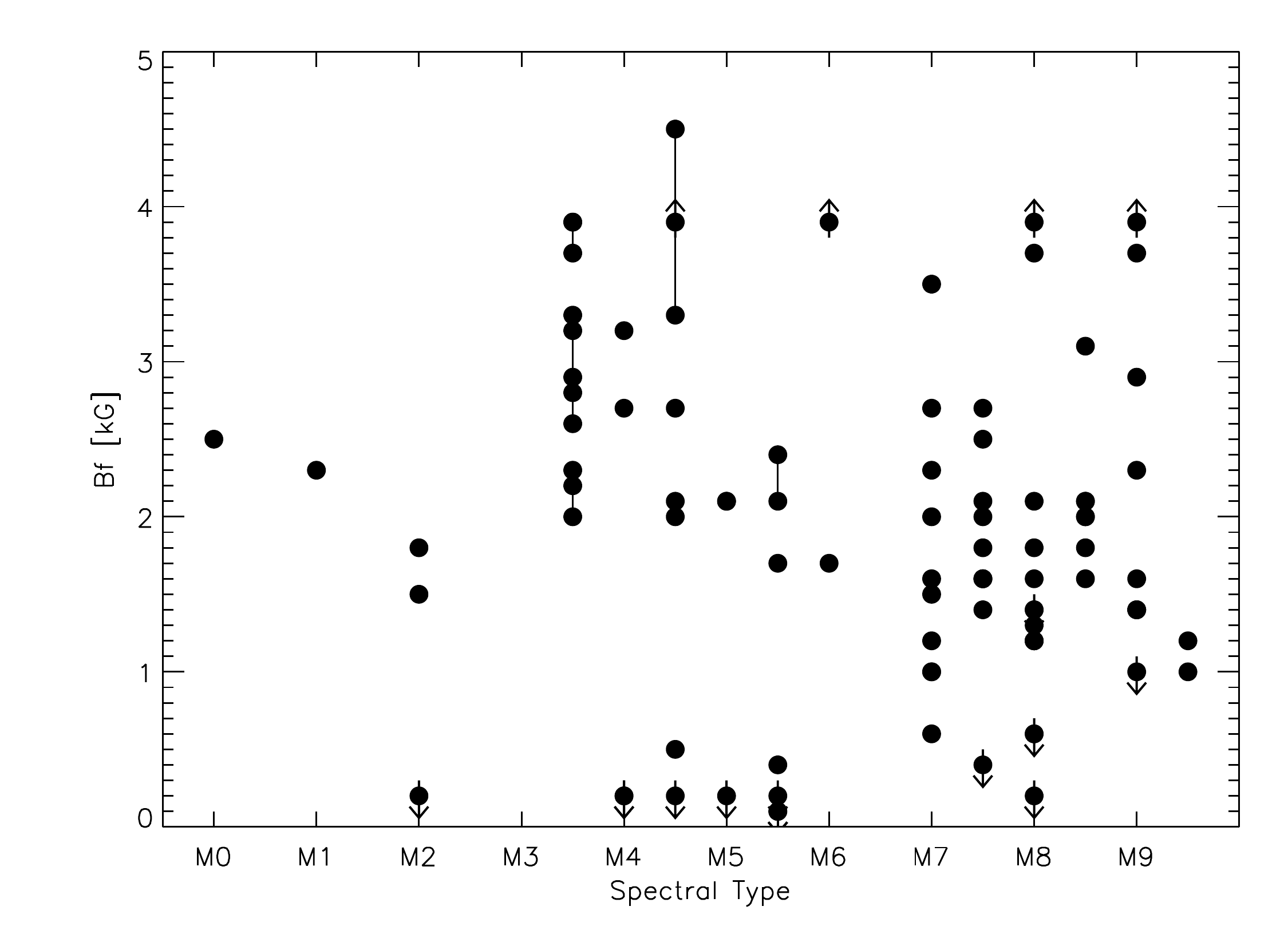}}
  \caption{Measurements of M~dwarf average magnetic fields from
    integrated light measurements. Data are given in
    Table~\ref{tab:StokesIMtype}. Limits are indicated by arrows, and
    multiple measurements of the same star are connected with vertical
    lines.}
  \label{fig:MBFields}
\end{figure}}

In early-M~dwarfs (M3 and earlier), magnetic fields were found in young stars
that are still rapidly rotating. Since old, early-M~dwarfs in the field are
generally slowly rotating and inactive there has been no search for magnetic
fields in any large sample of them. Typical field values can be expected to be
on the level of a few hundred Gauss and less, which is difficult to detect
with Stokes~I Zeeman measurements. Many mid- and late-M~stars are rapidly
rotating and fields of kG-strength are ubiquitously found among them.

\subsubsection{Pre-main sequence stars and young brown dwarfs}
\label{sect:results_young_I}

Magnetic fields of pre-main sequence stars are of particular interest
because accretion of circumstellar material onto the stellar surface
is believed to be controlled by the stellar magnetic field
\citep[e.g.,][]{2007prpl.conf..479B}. Evidence for accretion is
observed in pre-main sequence stars of very different mass including
young brown dwarfs. Field strengths predicted from several models of
magnetospheric accretion are on the order of several kG for T~Tauri
stars, and a few hundred Gauss for young brown dwarfs
\citep{1999ApJ...516..900J, 2009ApJ...697..373R}. On top of this, at
young ages, magnetic fields may be generated by a dynamo like in
older, sun-like stars (in contrast to fossil fields), but the dynamo
would probably operate similar to the one in low-mass M-type dwarfs
because pre-main sequence stars are still fully convective. On the
other hand, at ages of a few Myr, primordial fields may still be
present and not (yet) dissipated. Magnetic fields in pre-main sequence
stars may therefore carry important information about the star- and
planet-formation process.

First measurements of magnetic field strengths in T~Tauri stars were
attempted using the equivalent width method in (red) optical
absorption lines by \citet{1992ApJ...390..622B}, and
\citet{1999AA...341..768G} were following this strategy.
\citet{1999ApJ...516..900J} used infrared lines of Ti\,{\sc i} at
2.22~\mum to determine the magnetic field in BP~Tau. Obtaining
information on stellar parameters and rotation from optical lines and
magnetically insensitive CO lines, they were able to disentangle the
significant Zeeman broadening from other broadening agents. Similar
work on other T~Tauri stars using infrared spectra was done by
Johns-Krull \textit{et al.}, much of it is summarized in
\citet{2007ApJ...664..975J} where additional measurements of 14
T~Tauri star magnetic fields are presented. Another set of 14 magnetic
field measurements in very young T~Tauri stars in the Orion nebula
cluster are given in \citet{2011ApJ...729...83Y}. We will return to
the results from these substantial samples in
Sections~\ref{sect:Equipartition} and \ref{sect:Flux}. A summary of
magnetic field measurements in very young stars and brown dwarfs is
given in Table~\ref{tab:StokesIPreMain}, and are shown in
Figure~\ref{fig:PreMSBFields}.

\epubtkImage{StokesIPreMS.png}{%
\begin{figure}[htbp]
  \centerline{\includegraphics[width=0.8\textwidth]{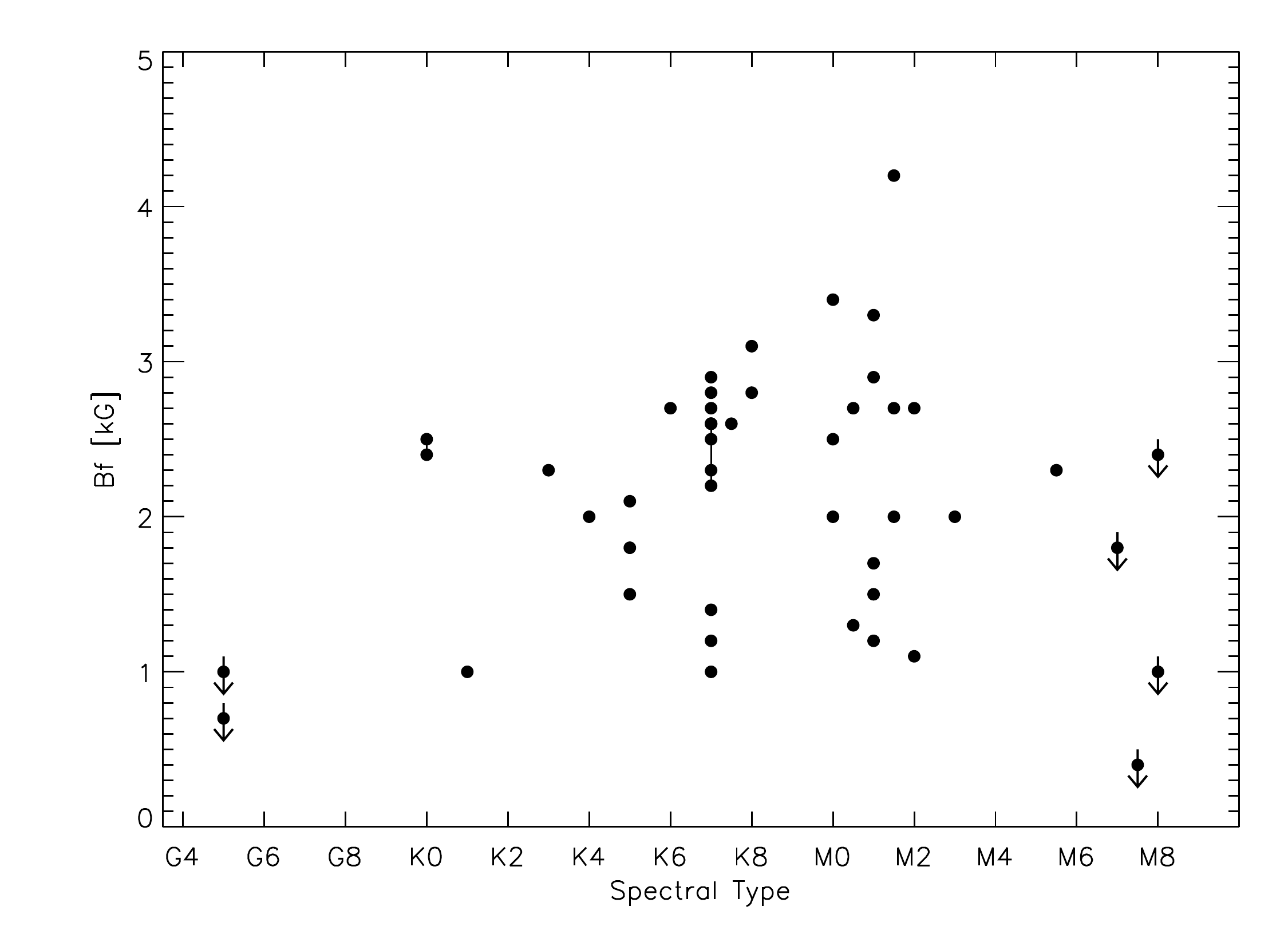}}
  \caption{Measurements of pre-main sequence and young brown dwarf
    magnetic fields from integrated light measurements. Data are given
    in Table~\ref{tab:StokesIPreMain}. Limits are indicated by arrows,
    and multiple measurements of the same star are connected with
    vertical lines.}
  \label{fig:PreMSBFields}
\end{figure}}

Using FeH measurements as laid out in Section~\ref{Sect:Mstars},
\citet{2009ApJ...697..373R} attempted to find evidence for kG-strength
magnetic fields in young brown dwarfs. Young brown dwarfs can be
expected to harbor substantial magnetic fields
\citep{2010AA...522A..13R}, and no fundamental difference is known to
exist in the parameters that are believed to be relevant for magnetic
flux generation between very-low mass stars and young brown dwarfs.
However, in contrast to pre-main sequence stars, and in contrast to
older brown dwarfs, none of the young brown dwarfs investigated by
\citet{2009ApJ...697..373R} exhibited a field above the detection
threshold that in all cases lay below the fields typically found among
the other groups.

An important property of the four young brown dwarfs investigated
for magnetic fields is that all of them show evidence for accretion
and, therefore, harbor a circumstellar disk. Magnetic field strengths
required for magnetospheric accretion in these objects are much lower
than in more massive, young stars, hence there is currently no
contradiction between the presence of accretion and the lack of
evidence for substantial fields. Observations of radio-emission,
however, indicate that fields of a few kG strength are in fact present
on some L-type field (old and non-accreting) brown dwarfs
\citep{2008ApJ...684..644H, 2009ApJ...695..310B}.  Direct measurement
of magnetism in non-accreting brown dwarfs, both young and old, are
required to further investigate whether the average fields are really
weaker in young brown dwarfs or in the presence of accretion.

It is an interesting question whether the non-detection of magnetic
fields in brown dwarfs is due to the presence of accretion disks
around the objects observed so far. If this is the case, there ought
to be some mechanism for the disk to regulate the magnetic field of
the central object, which is not easily understood. Alternatively,
large difference in radius may be of importance in this context
because the surface area of young brown dwarfs is about an order of
magnitude larger than the surface of old brown dwarfs. If magnetic
flux is approximately conserved during its evolution, the average
magnetic field would be an order of magnitude lower in young, large
brown dwarfs than in old, small, field brown dwarfs. We come back to
this point in Section~\ref{sect:Flux}.

\newpage

\begin{table}[htbp]
  \caption[Average magnetic fields from Stokes~I measurements in
    sun-like stars.]{Average magnetic fields from Stokes~I measurements in
    sun-like stars. Tables \ref{tab:StokesIsunlike} --
    \ref{tab:StokesVPreMS} are an attempt to collect information
    available on magnetic fields in cool stars. They are certainly
    incomplete to some extent simply because the author has overlooked
    many sources. The reader is encouraged to send references to
    papers that are missing so far and new work that appears in this
    field.}
  \label{tab:StokesIsunlike}
   \centering
   {\small
   \begin{tabular}{lcrl}
     \toprule
     \textbf{Star} & \textbf{SpType} & \textbf{\emph{Bf} [kG]} & \textbf{Reference}\\
   \midrule
   \multicolumn{4}{c}{IR data}\\
   \midrule
   $\sigma$ Dra   & K0  & $\le$~0.10 & \cite{1995ApJ...439..939V}\\
   40 Eri         & K1  & $\le$~0.10 & \cite{1995ApJ...439..939V}\\
   $\epsilon$ Eri & K2  & 0.13 & \cite{1995ApJ...439..939V}\\
   LQ Hya         & K2  & 2.45 & \cite{1996mpsa.conf..367S}\\
   $\xi$~Boo~B    & K4  & 0.46 & \cite{1994IAUS..154..493S}\\
   Gl~171.2A      & K5  & 1.40 & \cite{1996mpsa.conf..367S}\\
   \midrule
   \multicolumn{4}{c}{Optical data}\\
   \midrule
  HD 68456       & F6   & 1.00 & \cite{2010AA...522A..81A}\\
  59 Vir         & G0   & 0.19 & \cite{1994ASPC...64..438L} \citep[see][]{1996mpsa.conf..367S}\\
                 &      & 0.42 & \cite{2010AA...522A..81A} (one temperature)\\
                 &      & \textless~0.30 & \cite{2010AA...522A..81A} (cool spot solution)\\
  58 Eri	 & G1   & 0.20 & \cite{1997AA...318..429R}\\
  $\kappa$ Cet   & G5   & 0.36 & \cite{1992ASPC...27..197S}\\
  61 Vir         & G6   & \textless~0.15 & \cite{2010AA...522A..81A}\\
  $\xi$~Boo~A    & G8   & 0.48 & \cite{1988ApJ...330..274B}\\
                 &      & 0.35 & \cite{1989ApJ...345..480M}\\
                 &      & 0.34 & \cite{1994ASPC...64..438L} \citep[see][]{1996mpsa.conf..367S}\\
  70 Oph A       & K0   & 0.22 & \cite{1989ApJ...345..480M}\\
  40 Eri A	 & K1   & 0.06 & \cite{1997AA...318..429R}\\
  36 Oph B	 & K1   & 0.12 & \cite{1997AA...318..429R}\\
  $\epsilon$ Eri & K2   & 0.35 & \cite{1988ApJ...330..274B}\\
                 &      & 0.30 & \cite{1989ApJ...345..480M}\\
                 &      & 0.17 & \cite{1997AA...318..429R}\\
  HD 166620      & K2   & 0.23 & \cite{1988ApJ...330..274B}\\
  HD 17925       & K2   & 0.25 & \cite{1996mpsa.conf..367S}\\
  36 Oph A       & K2   & 0.20 & \cite{1989ApJ...345..480M}\\
  HR 222 	 & K2.5 & 0.19 & \cite{1989ApJ...345..480M}\\
  HR 5568        & K4   & 0.16 & \cite{1997AA...318..429R}\\
  EQ Vir         & K5   & 1.38 & \cite{1996mpsa.conf..367S}\\
  61 Cyg A       & K5   & 0.29 & \cite{1989ApJ...345..480M}\\
  $\epsilon$ Ind & K5   & 0.09 & \cite{1997AA...318..429R}\\
    \bottomrule
   \end{tabular}}
\end{table}

\clearpage

\ifpdf
\renewcommand{\tablename}{\normalsize Table}
{\small
\begin{longtable}[c]{llcrl}
\else
\begin{table}[htbp]
\fi
  \caption{Average magnetic fields from Stokes~I in M-dwarfs.}
  \label{tab:StokesIMtype}
\ifpdf\\\else
\centering
{\small
\begin{tabular}{llcrl}
\fi
  \toprule
  \textbf{Star} & \textbf{Other Name} & \textbf{SpType} & \textbf{\emph{Bf} [kG]} & \textbf{Reference}\\
  \midrule

    Gl 182             &           & M0.0 & 2.5     & \cite{2009AA...496..787R}\\
  Gl 803             &  AU Mic   & M1.0 & 2.3     & \cite{1994IAUS..154..493S}\\
  Gl 569A            &           & M2.0 & 1.8     & \cite{2009AA...496..787R}\\
  Gl 494             &  DT Vir   & M2.0 & 1.5     & \cite{1996mpsa.conf..367S}\\
  Gl 70	             &           & M2.0 & $<$~0.2 & \cite{2007ApJ...656.1121R}\\
  Gl 873             &  EV Lac   & M3.5 & 3.9     & \cite{1996ApJ...459L..95J, 2000ASPC..198..371J}\\
                     &           &      & 3.7     & \cite{1994IAUS..154..493S}\\
  Gl 729             &           & M3.5 & 2.0     & \cite{1996ApJ...459L..95J, 2000ASPC..198..371J}\\
                     &           &      & 2.2     & \cite{2007ApJ...656.1121R}\\
  Gl 87              &           & M3.5 & 3.9     & \cite{2007ApJ...656.1121R}\\
  Gl 388             &  AD Leo   & M3.5 & 2.8     & \cite{1985ApJ...299L..47S}\\
                     &           &      & 2.6     & \cite{1994IAUS..154..493S}\\
                     &           &      & 3.3     & \cite{2000ASPC..198..371J}\\
                     &           &      & 2.9     & \cite{2007ApJ...656.1121R}\\
                     &           &      & 3.2     & \cite{2009AIPC.1094..124K}\\
  GJ 3379            &           & M3.5 & 2.3     & \cite{2009ApJ...692..538R}\\
  GJ 2069~B          &           & M4.0 & 2.7     & \cite{2009ApJ...692..538R}\\
  Gl 876             &           & M4.0 & $<$~0.2 & \cite{2007ApJ...656.1121R}\\
  GJ 1005A	     &           & M4.0 & $<$~0.2  & \cite{2007ApJ...656.1121R}\\
  Gl 490 B           &  G 164-31 & M4.0 & 3.2     & \cite{2009ApJ...704.1721P}\\
  Gl 493.1           &           & M4.5 & 2.1     & \cite{2009ApJ...692..538R}\\
  GJ 4053            &  LHS 3376 & M4.5 & 2.0     & \cite{2009ApJ...692..538R}\\
  GJ 299             &           & M4.5 & 0.5     & \cite{2007ApJ...656.1121R}\\
  GJ 1227            &           & M4.5 & $<$~0.2 & \cite{2007ApJ...656.1121R}\\
  GJ 1224            &           & M4.5 & 2.7     & \cite{2007ApJ...656.1121R}\\
  Gl 285             &  YZ CMi   & M4.5 & 3.3     & \cite{2000ASPC..198..371J}\\
                     &           &      & $>$~3.9 & \cite{2007ApJ...656.1121R}\\
                     &           &      & 4.5     & \cite{2009AIPC.1094..124K}\\
  GJ 1154~A          &           & M5.0 & 2.1     & \cite{2009ApJ...692..538R}\\
  GJ 1156            &           & M5.0 & 2.1     & \cite{2009ApJ...692..538R}\\
  Gl 905             &           & M5.5 & $<$~0.1 & \cite{2007ApJ...656.1121R}\\
  GJ 1057            &           & M5.0 & $<$~0.2 & \cite{2007ApJ...656.1121R}\\
  Gl 905             &           & M5.5 & $<$~0.1 & \cite{2007ApJ...656.1121R}\\
  GJ 1245B	     &           & M5.5 & 1.7     & \cite{2007ApJ...656.1121R}\\
  GJ 1286            &           & M5.5 & 0.4     & \cite{2007ApJ...656.1121R}\\
  GJ 1002            &           & M5.5 & $<$~0.2 & \cite{2007ApJ...656.1121R}\\
  Gl 406             &           & M5.5 & 2.4     & \cite{2007ApJ...656.1121R}\\
                     &           &      & 2.1\,--\,2.4 & \cite{2007AA...466L..13R}\\
  Gl 412~B           &           & M6.0 & $>$~3.9 & \cite{2009ApJ...692..538R}\\
  GJ 1111            &           & M6.0 & 1.7  & \cite{2007ApJ...656.1121R}\\
  Gl 644 C           &  VB 8     & M7.0 & 2.3  & \cite{2007ApJ...656.1121R}\\
  GJ 3877 	     &  LHS 3003 & M7.0 & 1.5  & \cite{2007ApJ...656.1121R}\\
2M~0440232--053008   &           & M7.0 & 1.6  & \cite{2010ApJ...710..924R}\\
2M~0741068+173845    &           & M7.0 & 1.0  & \cite{2010ApJ...710..924R}\\
2M~0752239+161215    &           & M7.0 & 3.5  & \cite{2010ApJ...710..924R}\\
2M~0818580+233352    &           & M7.0 & 1.0  & \cite{2010ApJ...710..924R}\\
2M~1048126--112009   &  GJ 3622  & M7.0 & 0.6  & \cite{2010ApJ...710..924R}\\
2M~1356414+434258    &           & M7.0 & 2.7  & \cite{2010ApJ...710..924R}\\
2M~1456383--280947   &           & M7.0 & 1.2  & \cite{2010ApJ...710..924R}\\
2M~1534570--141848   &           & M7.0 & 2.0  & \cite{2010ApJ...710..924R}\\
  LHS~2645	     &           & M7.5 & 2.1  & \cite{2007ApJ...656.1121R}\\
2M~0331302--304238   &           & M7.5 & 2.0  & \cite{2010ApJ...710..924R}\\
2M~0351000--005244   &           & M7.5 & 1.4  & \cite{2010ApJ...710..924R}\\
2M~0417374--080000   &           & M7.5 & 1.8  & \cite{2010ApJ...710..924R}\\
2M~0429184--312356A  &           & M7.5 & 2.5  & \cite{2010ApJ...710..924R}\\
2M~1006319--165326   &           & M7.5 & 1.6  & \cite{2010ApJ...710..924R}\\
2M~1246517+314811    &           & M7.5 & $<$~0.4 & \cite{2010ApJ...710..924R}\\
2M~1253124+403403    &           & M7.5 & 1.6  & \cite{2010ApJ...710..924R}\\
2M~1332244--044112   &           & M7.5 & 1.6  & \cite{2010ApJ...710..924R}\\
2M~1546054+374946    &           & M7.5 & 2.7  & \cite{2010ApJ...710..924R}\\
  LP~412-31	     &           & M8.0 & $>$~3.9 & \cite{2007ApJ...656.1121R}\\
  VB~10              &           & M8.0 & 1.3  & \cite{2007ApJ...656.1121R}\\
2M~0248410--165121   &           & M8.0 & 1.4  & \cite{2010ApJ...710..924R}\\
2M~0320596+185423    &           & M8.0 & 3.7  & \cite{2010ApJ...710..924R}\\
2M~0517376--334902   &           & M8.0 & 1.6  & \cite{2010ApJ...710..924R}\\
2M~0544115--243301   &           & M8.0 & 1.2  & \cite{2010ApJ...710..924R}\\
2M~1016347+275149    &           & M8.0 & 2.1  & \cite{2010ApJ...710..924R}\\
2M~1024099+181553    &           & M8.0 & $<$~1.4 & \cite{2010ApJ...710..924R}\\
2M~1141440--223215   &           & M8.0 & 1.8  & \cite{2010ApJ...710..924R}\\
2M~1309218--233035   &           & M8.0 & 1.2  & \cite{2010ApJ...710..924R}\\
2M~1440229+133923    &           & M8.0 & $<$~0.6 & \cite{2010ApJ...710..924R}\\
2M~1843221+404021    &           & M8.0 & 1.2  & \cite{2010ApJ...710..924R}\\
2M~2037071--113756   &           & M8.0 & $<$~0.2 & \cite{2010ApJ...710..924R}\\
2M~2306292--050227   &           & M8.0 & 0.6  & \cite{2010ApJ...710..924R}\\
2M~2349489+122438    &           & M8.0 & 1.2  & \cite{2010ApJ...710..924R}\\
2M~0024442--270825B  &           & M8.5 & 2.1  & \cite{2010ApJ...710..924R}\\
2M~0306115--364753   &           & M8.5 & 1.6  & \cite{2010ApJ...710..924R}\\
2M~1124048+380805    &           & M8.5 & 2.0  & \cite{2010ApJ...710..924R}\\
2M~1403223+300754    &           & M8.5 & 2.1  & \cite{2010ApJ...710..924R}\\
2M~2226443--750342   &           & M8.5 & 1.8  & \cite{2010ApJ...710..924R}\\
2M~2331217--274949   &           & M8.5 & 3.1  & \cite{2010ApJ...710..924R}\\
2M~2353594--083331   &           & M8.5 & 2.0  & \cite{2010ApJ...710..924R}\\
  LHS~2924	     &           & M9.0 & 1.6  & \cite{2007ApJ...656.1121R}\\
  LHS~2065	     &           & M9.0 & $>$~3.9 & \cite{2007ApJ...656.1121R}\\
2M~0019457+521317    &           & M9.0 & 3.7  & \cite{2010ApJ...710..924R}\\
2M~0109511--034326   &           & M9.0 & 1.4  & \cite{2010ApJ...710..924R}\\
2M~0334114--495334   &           & M9.0 & 1.4  & \cite{2010ApJ...710..924R}\\
2M~0443376+000205    &           & M9.0 & $<$~1.0 & \cite{2010ApJ...710..924R}\\
2M~0853362--032932   &           & M9.0 & 2.9  & \cite{2010ApJ...710..924R}\\
2M~1048147--395606   &           & M9.0 & 2.3  & \cite{2010ApJ...710..924R}\\
2M~1224522--123835   &           & M9.0 & 1.4  & \cite{2010ApJ...710..924R}\\
2M~1438082+640836    &           & M9.5 & 1.2  & \cite{2010ApJ...710..924R}\\
2M~2237325+392239    &           & M9.5 & 1.0  & \cite{2010ApJ...710..924R}\\

  \bottomrule
\ifpdf
\end{longtable}}
\else
\end{tabular}}
\end{table}
\fi

\newpage

\begin{table}[htbp]
  \caption{Average magnetic fields from Stokes~I in pre-main sequence stars and young brown dwarfs.}
  \label{tab:StokesIPreMain}
   \centering
   {\small
   \begin{tabular}{lcrl}
     \toprule
     \textbf{Star} & \textbf{SpType} & \textbf{\emph{Bf} [kG]} & \textbf{Reference}\\
   \midrule

    TAP 10         & G5   & $<$~0.7 & \cite{1992ApJ...390..622B}\\
  GW Ori         & G5   & $<$~1.0 & \cite{1999AA...341..768G}\\
  T Tau	         & K0	& 2.4    & \cite{2007ApJ...664..975J}\\
                 &      & 2.5    & \cite{1999AA...341..768G}\\
  TAP 35         & K1   & 1.0    & \cite{1992ApJ...390..622B}\\
2MASS 05361049--0519449	& K3	& 2.31 & \cite{2011ApJ...729...83Y}\\
V1735 Orig	        & K4	& 2.08 & \cite{2011ApJ...729...83Y}\\
  LkCa 15               & K5    & 1.55 & \cite{1999AA...341..768G}\\
V1227 Ori	        & K5-K6	& 2.14 & \cite{2011ApJ...729...83Y}\\
OV Ori	                & K5-K6	& 1.85 & \cite{2011ApJ...729...83Y}\\
  GI Tau	 & K6	& 2.7    & \cite{2007ApJ...664..975J}\\
  TW Hya	 & K7	& 2.6    & \cite{2007ApJ...664..975J}\\
  GK Tau	 & K7	& 2.3    & \cite{2007ApJ...664..975J}\\
  GM Aur         & K7	& 1.0    & \cite{2007ApJ...664..975J}\\
  Hubble 4       & K7   & 2.5    & \cite{2004ApJ...617.1204J}\\
  AA Tau	 & K7	& 2.8    & \cite{2007ApJ...664..975J}\\
  BP Tau	 & K7	& 2.2    & \cite{2007ApJ...664..975J}\\
                 &      & 2.6    & \cite{1999ApJ...516..900J}\\
  DK Tau	 & K7	& 2.6    & \cite{2007ApJ...664..975J}\\
  GG TauA	 & K7	& 1.2    & \cite{2007ApJ...664..975J}\\
  TWA 9A         & K7   & 2.9    & \cite{2008AJ....136.2286Y}\\
  TW Hya         & K7   & 2.7    & \cite{2008AJ....136.2286Y}\\
V1568 Ori	 & K7	& 1.42   & \cite{2011ApJ...729...83Y}\\
  DG Tau	 & K7.5	& 2.6    & \cite{2007ApJ...664..975J}\\
2MASS 05353126--0518559	& K8	 & 2.84 & \cite{2011ApJ...729...83Y}\\
V1348 Ori	        & K8  	 & 3.14 & \cite{2011ApJ...729...83Y}\\
V1123 Ori	        & M0/K8	 & 2.51 & \cite{2011ApJ...729...83Y}\\
  DN Tau	 & M0	& 2.0    & \cite{2007ApJ...664..975J}\\
LO Ori	                & M0	 & 3.45 & \cite{2011ApJ...729...83Y}\\
LW Ori	                & M0.5	 & 1.30 & \cite{2011ApJ...729...83Y}\\
2MASS 05350475--0526380	& M0.5	 & 2.79 & \cite{2011ApJ...729...83Y}\\
V568 Ori	        & M1	 & 1.53 & \cite{2011ApJ...729...83Y}\\
2MASS 05351281--0520436	& M1	 & 1.70 & \cite{2011ApJ...729...83Y}\\
  CY Tau	 & M1	& 1.2    & \cite{2007ApJ...664..975J}\\
  DF Tau	 & M1	& 2.9    & \cite{2007ApJ...664..975J}\\
  TWA 9B         & M1   & 3.3    & \cite{2008AJ....136.2286Y}\\
  DH Tau	 & M1.5	& 2.7    & \cite{2007ApJ...664..975J}\\
  TWA 5A         & M1.5 & 4.2\super{a} & \cite{2008AJ....136.2286Y}\\    
V1124 Ori	        & M1.5   & 2.09 & \cite{2011ApJ...729...83Y}\\
  DE Tau	 & M2	& 1.1    & \cite{2007ApJ...664..975J}\\
  TWA 8A         & M2   & 2.7    & \cite{2008AJ....136.2286Y}\\
  TWA 7          & M3   & 2.0    & \cite{2008AJ....136.2286Y}\\  
UpSco 55         & M5.5 & 2.3    & \cite{2009ApJ...697..373R}\\
CFHT-BD-Tau 4    & M7   & $<$~1.8   & \cite{2009ApJ...697..373R}\\
UpSco-DENIS 160603 & M7.5 & $<$~0.4 & \cite{2009ApJ...697..373R}\\
2MASS~1207       & M8   & $<$~1.0   & \cite{2009ApJ...697..373R}\\
$\rho$-Oph-ISO 32& M8   & $<$~2.4   & \cite{2009ApJ...697..373R}\\

  \bottomrule
  \super{a}\,very high $v\,\sin{i}$
\end{tabular}}
\end{table}

\begin{table}[htbp]
  \caption[Magnetic field measurements not listed in 
    Tables~1, 2, and 3.]{Magnetic field measurements not listed in 
    Tables~\ref{tab:StokesIsunlike}, \ref{tab:StokesIMtype}, and
    \ref{tab:StokesIPreMain}.}
  \label{tab:StokesIpapers}
  \centering
  {\small
  \begin{tabular}{ll}
    \toprule
    Publication & Comment \\
    \midrule
    \cite{1980ApJ...236L.155R} & $\xi$~Boo~A, 70~Oph~A, 61~Vir\\
    \cite{1980ApJ...240..567V} & BY~Dra, HD~88230, 61~Cyg~A, HD~209813\\
    \cite{1981ApJ...245..624M} & $\xi$~Boo~A\\ 
    \cite{1983ApJ...268L.121G} & $\lambda$~And\\
    \cite{1984ApJ...276..286M} & 29 G- and K main sequence stars\\
    \cite{1984ApJ...281..286M} & 8 evolved stars\\
    \cite{1984ApJ...277..640G} & 18 F-, G-, and K-dwarfs\\
    \cite{1985ApJ...297..710G} & $\xi$~Boo~A, 61~UMa, $\lambda$~And\\
    \cite{1986ApJ...302..777S} & EQ~Vir\\
    \cite{1987LNP...291...36B} & 7 K- and M-dwarfs\\
    \cite{1989AA...208..189M} & Preliminary results, ``Stenflo--Lindegren'' technique\\
    \cite{1989ApJ...339.1059B} & VY Ari\\
    \cite{1990IAUS..138..427S} & Selection of 31 G\,--\,M star measurements; including unpublished data\\
    \bottomrule
  \end{tabular}}
\end{table}

\clearpage

\subsection{Longitudinal fields and Zeeman Doppler maps from Stokes~V}

\subsubsection{Dwarfs and subgiants}
\label{sect:VMapsDwarfs}

Advantages and caveats in searching for cool star magnetism through
polarization measurements were discussed in the Sections above.  For a
detection, the detailed line shape in the unpolarized case does not
have to be understood at very high level, which means that the signal
of a potential field will be relatively straightforward to detect
(given suitable instrumentation). The signal expected from magnetic
cool stars, however, may be extremely weak because of flux
cancellation and complicated field geometries. 

Early programs to search for longitudinal fields in late-type stars
were presented by \citet{1981ApJ...246..899B} and
\citet{1984ApJ...284..211B}. Both works show sophisticated methods to
search for circular polarization simultaneously in many spectral lines
and obtain encouraging results in hot stars with known strong magnetic
fields. Both programs, however, fail to detect polarization signals in
late-type stars confirming the suspicion that net longitudinal fields
are difficult to detect in these stars. The key idea in the two
mentioned programs is to co-add information from many spectral lines
in order to enhance the polarization signal. Masks transmitting only
the light in the vicinity of stellar absorption lines are used so that
the final signal is constructed to be something like an average signal
from many lines. The basic idea is very similar to the construction of
a mean line profile using the approach of Least Squares Deconvolution
(Section~\ref{sect:LSD}).

A detection of circular polarization in a low-mass star was successful
in an effort to create Zeeman Doppler Images in RS~CVn
binaries. \citet{1990AA...232L...1D} show signals in circular
polarization in three Fe\,{\sc i} lines of HR~1099, and more successful
detections in RS~CVn's are presented in \citet{1992AA...265..669D}.
These works obtained very high SNR data in order to detect
polarization signals in individual lines. Later, the technique of
Least Squares Deconvolution entered the domain of polarization
measurements and Zeeman Doppler Imaging (see
Section~\ref{Sect:ZeemanEffect}) and, since then, Stokes~V signatures
were investigated in many different stars
\citep[e.g.,][]{1997MNRAS.291..658D}. Table~\ref{tab:StokesVDwarfs}
gives a summary of Stokes~V measurements in cool dwarfs and
subgiants. It is sometimes difficult to compare the results from
different projects, because results are sometimes presented in the
form of surface maps and sometimes in the form of average magnetic
field strengths. Note that also in this context, average fields mean
the average value for the detected \emph{unsigned} magnetic field,
$|B|$, a definition similar to the value measured in Stokes~I; the
average value of the \emph{signed} magnetic field is zero by
construction.

A tremendous amount of work was put into the analysis of magnetic
geometries in stars through ZDI, and the possibility of reconstructing
magnetic fields on stellar surfaces is truly amazing. As laid out in
Section~\ref{Sect:ZeemanEffect}, however, the interpretation of the
field maps is very difficult, and conclusions have to be drawn with
great care. Typical average field values for sun-like stars of
spectral types F\,--\,K are of the order of ten Gauss, local field
strengths in Doppler maps go up to several hundred Gauss. Note that
most of the stars imaged with ZDI are rapid rotators that are much
more active than the slow-rotating Sun.

Much work was done also on maps of magnetic fields in M~stars.
Results derived from time-series of Stokes~V measurements are
presented in \citet{2008MNRAS.390..545D} and \citet{2008MNRAS.390..567M,
  2010MNRAS.407.2269M}. The typical average magnetic field strengths
found in Stokes~V measurements of M~dwarfs are significantly larger
than average fields from Stokes~V work found in hotter, sun-like
stars. Average fields up to 1.5~kG were detected, and the range of
$|B|$ in Doppler maps from Stokes~V extend up to 2~kG, a value at
which the weak-field approximation probably approaches its
limitations.

\enlargethispage{\baselineskip}
\cite{2011ApJ...732L..19K} presented very high quality measurements of
all four Stokes parameters in three sun-like stars. Using LSD, they
detect circular and linear polarization in the RS~CVn binary
HR~1099. For the first time, linear polarization is detected in
sun-like stars. Fields between 10 and 25~G are found in different
observations of HR~1099. The signal detected in linear polarization is
significantly more complex than the circular polarization signal and,
as expected, linear polarization is weaker than circular polarization
by roughly a factor 10.

\subsubsection{Giants}

Another class of stars in the focus of magnetic field research are
giants. Some of these evolved stars can be rather active, and they
possess large convection zones potentially allowing the operation of a
dynamo. Stokes~V observing campaigns are available in a handful of
giants providing information about their magnetism. For the active
FK~Com star HD~199178, \citet{2004MNRAS.351..826P} constructed Zeeman
Doppler maps with field strengths up to several hundred Gauss. The
other work summarized in Table~\ref{tab:StokesVGiants} find mean
fields that are on the order of 1~G or below, this means they are on
average much weaker than the fields found in dwarfs.

\subsubsection{Young stars}

Pre-main sequence stars are particularly interesting objects for
magnetic field measurements because fields may be of fossil origin or
generated through dynamo operation, and magnetism is required for
magnetospheric accretion (see
Section~\ref{sect:results_young_I}). The observational situation from
polarization measurements is similar to the one in main-sequence
stars: results are available in the form of magnetic Doppler maps
providing a range and geometry of the detected net field, and there
are results reporting average field values from multiple or single
polarization measurements.

A summary of magnetic field reports from circular polarization
measurements in pre-main sequence stars is given in
Table~\ref{tab:StokesVPreMS}. Field strengths up to several hundred
Gauss, and in the case of BP~Tau up to 3~kG are found in Zeeman
Doppler maps from photospheric lines. Average fields on the order of
ten to several hundred Gauss have been found in the analysis of
photospheric lines. A remarkable difference exists between the field
strengths found in polarization measurements carried out in
photospheric lines and those carried out in \emph{emission} lines that
are formed predominantly in the region of the accretion shock, usually
the He\,{\sc i} line at 5876~\AA. In the latter, field strengths
are on the order of a few kilo-Gauss similar to field strengths
detected in Stokes~I measurements.

Assuming the same flux values across the star, the net flux seen in
the accretion region alone may be higher than the net flux averaged
over the photosphere, which would mean that the difference is a purely
geometric effect. The accretion column could stem from a more or less
unipolar bundle of fluxtubes allowing to see the true (almost
uncanceled) flux, while the photospheric lines come from the entire
star and are subject to cancellation. Furthermore, field strengths may
actually be higher in the region of accretion, and the differences
between magnetic fields measurements from integrated light and results
from circular polarization in photospheric and accretion lines may
provide additional information on the magnetic field geometry in
pre-main sequence stars \citep[see e.g.,][and references
therein]{2007ApJ...664..975J, 2011MNRAS.417.1747D}. We come back to
this point in Section~\ref{sect:Geometry}.

\begin{table}[htbp]
  \caption{Longitudinal magnetic fields or Zeeman Doppler maps from
    Stokes~V for dwarfs and subgiants.}
  \label{tab:StokesVDwarfs}
  \centering
  {\small
  \begin{tabular}{llcccl}
  \toprule
  \textbf{Star} & \textbf{Other Name} & \textbf{SpType} & \textbf{\emph{\textless\,B\,\textgreater}} & \textbf{\emph{B} Range} & \textbf{Reference}\\
  ~             & ~                   & ~               & [G]  & [G] & ~ \\
  \midrule
  $\tau$~Boo	&         & F7V	  &     & 0\,--\,3    &   \cite{2007MNRAS.374L..42C}\\
  HR 1817       &         & F8V   & 13  & 0\,--\,250  &   \cite{2006ASPC..358..401M}\\
  \midrule
  HD 73350	&         & G0V	  & 12\super{b} & 0\,--\,20 &   \cite{2008MNRAS.388...80P}\\
  Sun\super{a}	&         & G2V	  & 4   &         &   \cite{2006AJ....131..520D}\\
  $\alpha$~Cen~A&         & G2V   & \textless~0.2 &        &   \cite{2011ApJ...732L..19K}\\
  HD 171488	&         & G2V	  &     & 0\,--\,500  &   \cite{2006MNRAS.370..468M}\\
                &         &       & 31  & 0\,--\,500  &   \cite{2006ASPC..358..401M}\\
  HD 146233	&         & G2V	  & 3.6 & 0\,--\,5    &   \cite{2008MNRAS.388...80P}\\
  HD 76151	&         & G3V	  & 5.6 & 0\,--\,10   &   \cite{2008MNRAS.388...80P}\\
  HD 190771	&         & G5IV  & 15\super{b} & 0\,--\,20 &   \cite{2008MNRAS.388...80P}\\
  \midrule
  LQ Hya	&         & K0V	  &     & 0\,--\,800  &   \cite{1999MNRAS.302..457D}\\
  AB Dor	&         & K0V	  &     & 0\,--\,800  &   \cite{1999MNRAS.302..437D}\\
  HD 46375	&         & K0V	  &     & 0\,--\,5    &   \cite{2010AA...524A..47G}\\
  II Peg	&         & K1IV  &     & 0\,--\,700  &   \cite{2007AN....328.1043C}\\
  HR 1099	&         & K1IV  &     & 0\,--\,800  &   \cite{1999MNRAS.302..457D}\\
                &         &       & 12\,--\,25 &        &   \cite{2011ApJ...732L..19K}\\
  HD 189733	&         & K2V	  &     & 0\,--\,40   &   \cite{2007AA...473..651M}\\
  $\epsilon$~Eri&         & K2V   & --6\,--\,5 &       &  \cite{2011ApJ...732L..19K}\\
  \midrule
  Gl 890   	&         & M0V	  & \textless~18 &       &   \cite{2009ApJ...704.1721P}\\
  Gl 410        & DS Leo  & M0V   & 100 &       &   \cite{2008MNRAS.390..545D}\\
  Gl 182        &         & M0.5V & 172 &       &   \cite{2008MNRAS.390..545D}\\
  Gl 494        & DT Vir  & M0.5V & 150 &       &   \cite{2008MNRAS.390..545D}\\
  Gl 49         &         & M1.5  &  27 &       &   \cite{2008MNRAS.390..545D}\\
  GJ 9520       & OT Ser  & M1.5  & 130 &       &   \cite{2008MNRAS.390..545D}\\
  Gl 569~A      & CE Boo  & M2.0V & 103 &       &   \cite{2008MNRAS.390..545D}\\
  Gl 752~A	& LHS 473 & M2.5V & 16  &       &   \cite{2009ApJ...704.1721P}\\
  GJ 3241	& KP Tau  & M3V	  & 100 &       &   \cite{2009ApJ...704.1721P}\\
  Gl 388        & AD Leo  & M3V   & 185 &       &   \cite{2008MNRAS.390..567M}\\
  Gl 896~A      & EQ Peg A& M3.5V & 480 &       &   \cite{2008MNRAS.390..567M}\\
  Gl 873	& EV Lac  & M3.5V & 18\,--\,40&       &   \cite{2006ApJ...646L..73P}\\
                &         &       & 530 &       &   \cite{2008MNRAS.390..567M}\\
  GJ 4247 	& V374 Peg& M4V	  &     & 0\,--\,2000&   \cite{2006Sci...311..633D}\\
                &         &       & 710 &       &   \cite{2008MNRAS.390..567M}\\
  Gl 490~B	& G 164-31& M4V	  & 680 & 0\,--\,1800 &   \cite{2009ApJ...704.1721P}\\	
  Gl 285        & YZ CMi  & M4.5V & 555 &       &   \cite{2008MNRAS.390..567M}\\
  Gl 896~B      & EQ Peg B& M4.5V & 450 &       &   \cite{2008MNRAS.390..567M}\\
          	&         &       & 290 &       &   \cite{2009ApJ...704.1721P}\\
  2E 4498	& 2E 4498 & M4.5V & 440 &       &   \cite{2009ApJ...704.1721P}\\
  Gl 51         &         & M5V   &1500 &       &   \cite{2010MNRAS.407.2269M}\\
  GJ 1156       &         & M5V   & 100 &       &   \cite{2010MNRAS.407.2269M}\\
  GJ 1245~B     &         & M5.5V & 150 &       &   \cite{2010MNRAS.407.2269M}\\
  Gl 905  	& HH And  & M5.5V & \textless~5&       &   \cite{2006ApJ...646L..73P}\\
  Gl 412~B      & WX UMa  & M6V   &1000 &       &   \cite{2010MNRAS.407.2269M}\\
  GJ 1111       & DX Cnc  & M6V   & 100 &       &   \cite{2010MNRAS.407.2269M}\\
  GJ 3622       &         & M6.5V &  55 &       &   \cite{2010MNRAS.407.2269M}\\
  \bottomrule
  \multicolumn{3}{l}{\super{a}Using observations of Vesta}\\
  \super{b}\,priv.\ comm.
  \end{tabular}}
\end{table}

\begin{table}[htbp]
  \caption{Longitudinal magnetic fields or Zeeman Doppler maps from
    Stokes V for giants.}
  \label{tab:StokesVGiants}
  \centering
  {\small
  \begin{tabular}{lcccl}
    \toprule
    \textbf{Star} & \textbf{SpType} & \textbf{\emph{\textless\,B\,\textgreater}} & \textbf{\emph{B} Range} & \textbf{Reference}\\
    ~ & ~            & [G] & [G] & ~ \\
    \midrule
    Betelgeuse	 & M2Iab   & 0.5\,--\,1.6 & ~          & \cite{2010AA...516L...2A}\\
    HD 199178	 & G5III   & ~            & 0\,--\,600 & \cite{2004MNRAS.351..826P}\\
    V390 Aur	 & G8III   & 5\,--\,15    & ~          & \cite{2008AA...480..475K}\\
    Pollux	 & K0III   & 0.1\,--\,1.4 & ~          & \cite{2009AA...504..231A}\\
    Arcturus     & K1.5III & 0.4\,--\,0.7 & ~          & \cite{2011AA...529A.100S}\\
    EK Boo	 & M5III   & 0.1\,--\,0.8 & ~          & \cite{2010AA...524A..57K}\\
    \bottomrule
  \end{tabular}}
\end{table}

\begin{table}[htbp]
  \caption{Longitudinal magnetic fields or Zeeman Doppler maps from
    circular polarization in Pre-main sequence stars.}
  \label{tab:StokesVPreMS}
  \centering
  {\small
  \begin{tabular}{lcccl}
    \toprule
    \textbf{Star} & \textbf{SpType} &
    \textbf{\emph{\textless\,B\,\textgreater}} & \textbf{\emph{B} Range} & \textbf{Reference}\\
    ~ & ~            & [G] & [G] & ~ \\
    \midrule
    HD 155555~A & G5    &       & 0\,--\,500            &\cite{2008MNRAS.387..481D}\\
    CV Cha	& G8	&       & 0\,--\,700		&\cite{2009MNRAS.398..189H}\\
    HD 155555~B & K0    &       & 0\,--\,300            &\cite{2008MNRAS.387..481D}\\
    T Tau	& K0	& 12    &	                &\cite{2006AJ....131..520D}\\
    CR Cha	& K2	&       & 0\,--\,400		&\cite{2009MNRAS.398..189H}\\
    V410 Tau    & K4    &       & 0\,--\,1000           &\cite{2010MNRAS.403..159S}\\
    V2129 Oph	& K5	&       & 0\,--\,800		&\cite{2007MNRAS.380.1297D}\\
                &       &       & 0\,--\,2000           &\cite{2011MNRAS.412.2454D}\\ 
    Tw Hya	& K6	& 150   &		        &\cite{2007AJ....133...73Y}\\
   		&	& 2000\super{a} &		&\cite{2007AJ....133...73Y}\\
                &       &       & 0\,--\,3000           &\cite{2011MNRAS.417..472D}\\
    BP Tau	& K7	& 500   & 0\,--\,3000           &\cite{2008MNRAS.386.1234D}\\
                &       & 2750\super{a} &	        &\cite{2000ASPC..198..371J}\\
		&	& 200   &                       &\cite{1999ApJ...510L..41J}\\
		&	& 2500\super{a} &               &\cite{1999ApJ...510L..41J}\\
    DK Tau	& M0	& 1450\super{a} &		&\cite{2000ASPC..198..371J}\\
    AA Tau	& M0	& 2900\super{a} &		&\cite{2000ASPC..198..371J}\\
    DF Tau	& M2	& 1000\super{a} &	        &\cite{2000ASPC..198..371J}\\
    V2247 Oph   & M2.5  &       & 0\,--\,800            &\cite{2010MNRAS.402.1426D}\\
    \bottomrule
    \super{a}\,from accretion lines\\
  \end{tabular}}
\end{table}

\newpage

\section{The Rotation--Magnetic Field--Activity Relation}
\label{sect:rotmagnact}

The generation of stellar magnetic fields is the result of complex
mechanisms acting in the moving plasma of the stellar interior. The
variety of magnetic field-related phenomena observed on the Sun should
be explainable by a theory of the solar dynamo, but this dynamo
continues to pose serious challenges to both observers and
theoreticians. Although our knowledge about the solar magnetic
properties and its time-dependence is rich and growing with the
growing fleet of instrumentation observing the magnetic Sun, this
single star can only exhibit magnetic features according to its own
properties, and the investigation of the Sun alone will not lead to a
full understanding of stellar dynamos in general. It is, therefore, of
large interest for a deep understanding of stellar dynamos and, in
particular, of the solar dynamo to understand the dependence of
magnetic field generation on stellar properties.

The driving force of dynamos operating in the Sun and low-mass stars
is the interplay between convective plasma motion, density and
temperature stratification, and stellar rotation. Differential
rotation and shear play particularly important roles in the most
favored versions of sun-like dynamos. For overviews on the solar
dynamo and theoretical backgrounds, I refer to the many reviews on
this topic, for example \citet{2003AARv..11..287O} and
\citet{2010LRSP....7....3C}. One often-mentioned expectation from
stellar dynamo models is the relation between the magnetic field
strength and the rotation of a star. The relation is expected because
the efficiency of a dynamo can be described by the ``dynamo number'',
$D$, that is related to the so-called $\alpha$-effect that itself
depends on rotation. The exact functional dependence between dynamo
efficiency and rotation is difficult to assess, but one can argue that
a dynamo can only exist if the rotational influence on convection,
expressed as the Coriolis number $Co$, exceeds a certain value in
order to create the \emph{differential} rotation that is required for
the dynamo process \citep[see, e.g.,][]{1978GApFD...9..241D}. The
Coriolis number is proportional to the Rossby number that is often
used in work on stellar activity, $Ro = P_{\mathrm{rot}}/\tau_{\mathrm{conv}}$,
with $P_{\mathrm{rot}}$ the rotational period and $\tau_{\mathrm{conv}}$ the
convective overturn time.

The definition of the convective overturn time is not exactly well
defined particularly in very low-mass stars
\citep[e.g.,][]{1986ApJ...300..339G, 1996ApJ...461..499K}. It has
therefore been attempted to derive ``empirical'' turnover times
assuming a relation between magnetic activity and stellar rotation. A
first and very successful approach was presented by
\citet{1984ApJ...279..763N} who connected observations of stellar
chromospheric activity and rotation. Investigating a rich sample of
X-ray, coronal, activity measurements, \citet{2003AA...397..147P}
were able to show a well-defined rotation-activity relation connecting
normalized X-ray luminosity and Rossby number. The latter is a
mass-dependent function chosen to minimize the scatter in the
rotation-activity relation (Figure~\ref{fig:Pizzolato}). Empirical
convective overturn times in low-mass stars were derived by
\citet{2007AcA....57..149K}, and \citet{2010ApJ...721..675B} cover a
wide range of masses.

\epubtkImage{img138-img143.png}{%
\begin{figure}[htbp]
  \centerline{
    \includegraphics[width=0.46\textwidth]{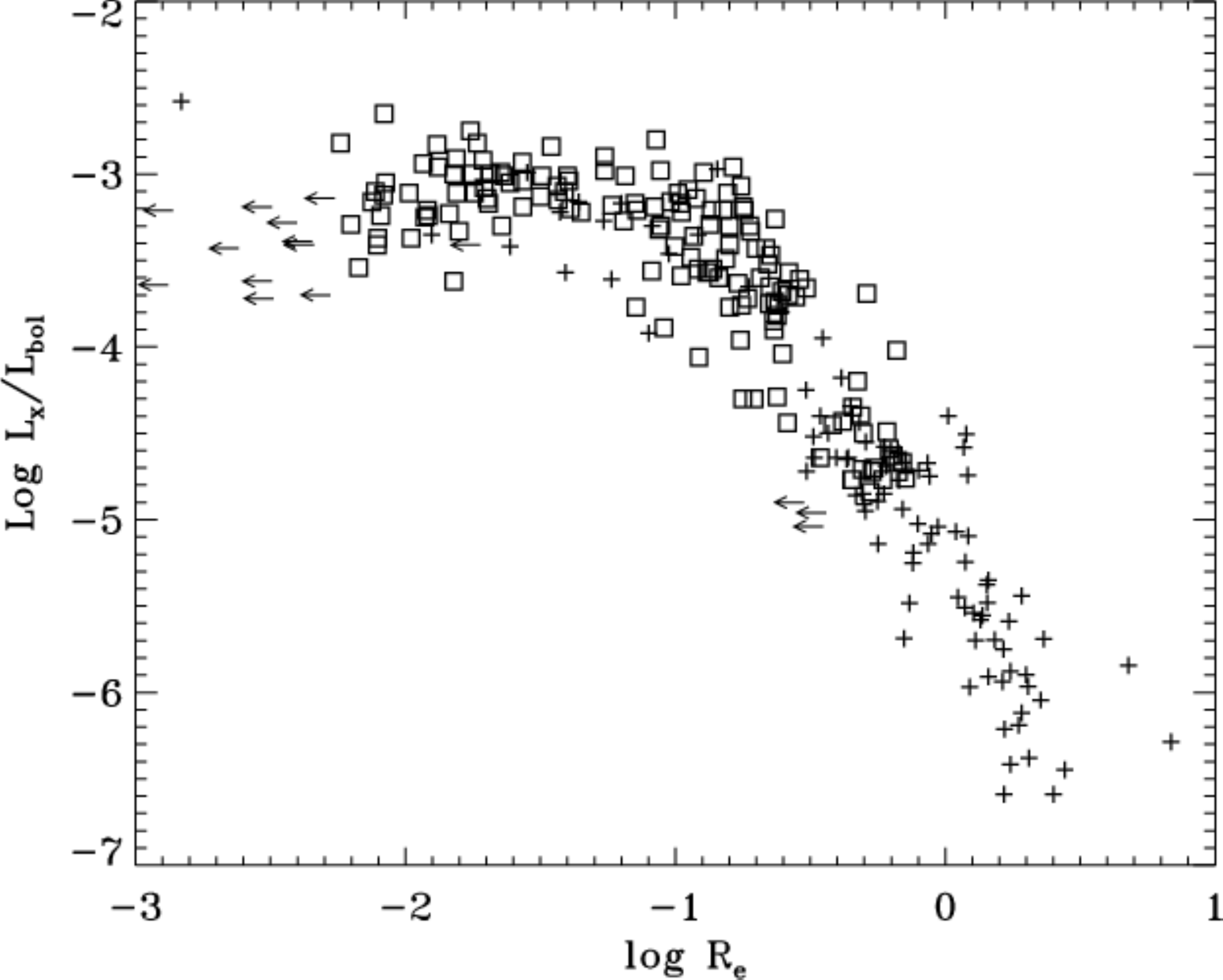}\qquad
    \includegraphics[width=0.46\textwidth]{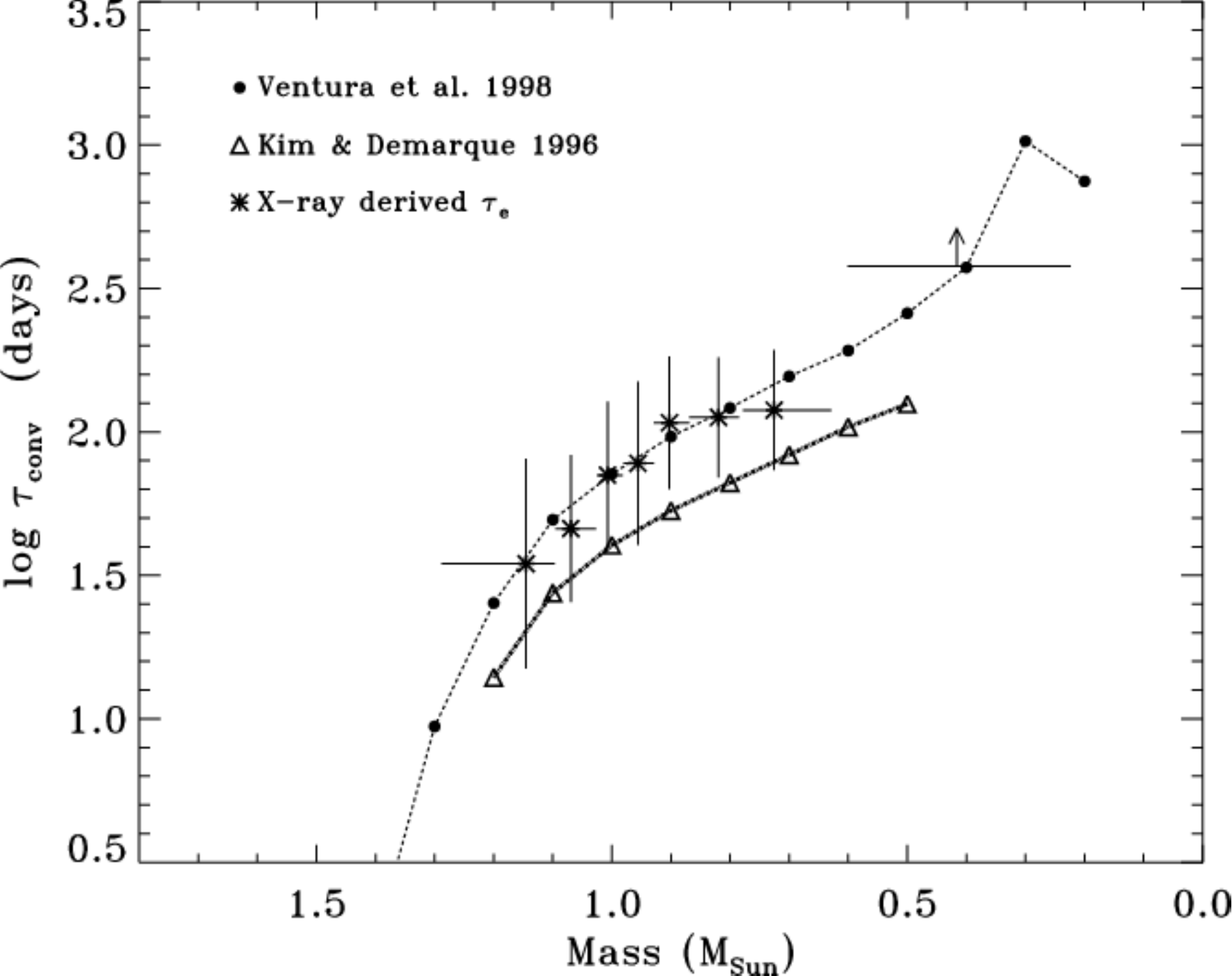}
  }
  \caption{\emph{Left panel:} Rotation-activity
    relation showing the normalized X-ray luminosity as a function of
    Rossby number. \emph{Right panel}: Empirical turnover time chosen
    to minimize the scatter in the rotation-activity relation
    \citep[from][]{2003AA...397..147P}.}
  \label{fig:Pizzolato}
\end{figure}}

It has been argued that the construction of a mass-dependent empirical
convective overturn time in order to minimize scatter in the
rotation-activity relation is in principle nothing else than a
compensation for a dependence of activity on stellar luminosity
\citep{1986LNP...254..184B, 2003AA...397..147P}. In other words,
while normalized activity seems to scale with Rossby number, total
(unnormalized) flux seems to scale with rotation period. It is argued
that the reason for this is the approximate scaling of
$\tau_{\mathrm{conv}}$ with $(L_{\mathrm{bol}})^{-0.5}$. Nevertheless,
a tight dependence between Rossby number and normalized activity is
clearly observed in sun-like and early- to mid-M~type stars. The
result is that magnetic activity is rising with decreasing Rossby
number as long as $Ro \ge 0.1$. At $Ro \approx 0.1$, activity
saturates and does not grow further with decreasing Rossby
number. This behavior is interpreted as increasing dynamo efficiency
with faster rotation in the regime where rotation is not yet
dominating convection ($Ro \ge 0.1$). This is what may be expected
from the dynamo models introduced above. At fast rotation ($Ro \approx
0.1$) the dynamo reaches a level of saturation that cannot be exceeded
even if the star is spinning much faster.

We know from the Sun that activity is caused by magnetic
fields. Together with expectations of the relation between rotation
and magnetic field strength from dynamo theory, it is straightforward
to conclude that the reason for the observed rotation-activity
relation is a rotation-dependence of magnetic field generation, i.e.,
what we observe is a direct consequence of the magnetic dynamo
efficiency. From an observational standpoint, this is not entirely
clear because all we have discussed so far is that \emph{activity}
scales with rotation (or Rossby number), but this may also be due to a
constant magnetic field translating to observable activity in a
fashion that depends on rotation. A direct link between rotation and
magnetic field observations was shown by \citet{1996IAUS..176..237S,
  2001ASPC..223..292S}, the observational basis of this work was
discussed in Section~\ref{sect:results_I}. In sun-like and young
stars it is found that magnetic field strengths indeed are a function
of rotational period: \Bf follows a relation that is proportional to
some power of \Ro consistent with expectations. In sun-like stars,
however, magnetic fields can not be measured at very low Rossby
numbers (saturated regime) because spectral line widths are too broad
due to the rotational broadening at the corresponding rotation
rates. Mainly because of the smaller radii, and perhaps also because
of longer convective overturn times, the relation between Rossby
number and equatorial velocity favors the detection of Zeeman
splitting at low Rossby numbers (in the saturated regime) in low-mass
stars \citep[see, e.g.,][]{2007AA...467..259R}; M~dwarfs have very
small radii (and long overturn times) so that for small Rossby numbers
the corresponding surface velocities are relatively low. This allows
measuring magnetic fields of M~stars well within the saturated
regime. For M~dwarfs of spectral type M6 and earlier,
\citet{2009ApJ...692..538R} found that average magnetic fields indeed
show evidence for saturation at low Rossby numbers. This can be
interpreted as evidence that saturation of activity at high rotation
rates is a consequence of saturation of the average magnetic field and
that \B itself is limited (in contrast to a limit of the filling
factor \f or of the coupling between magnetic fields and non-thermal
heating). Unfortunately, it was not yet possible to separate \B from
\f in the measurements of magnetic flux density \Bf so that the true
range and variation of both field strength \B and filling factor \f
remains unknown. Although there is good evidence for a firm upper
limit, there is still room for some variation of \B as a function of
\Ro in the saturated regime. Further observations, and especially
information about values of both \B and \f are highly desired.

\epubtkImage{Bf_Ro.png}{%
\begin{figure}[htbp]
  \centerline{\includegraphics[width=0.6\textwidth]{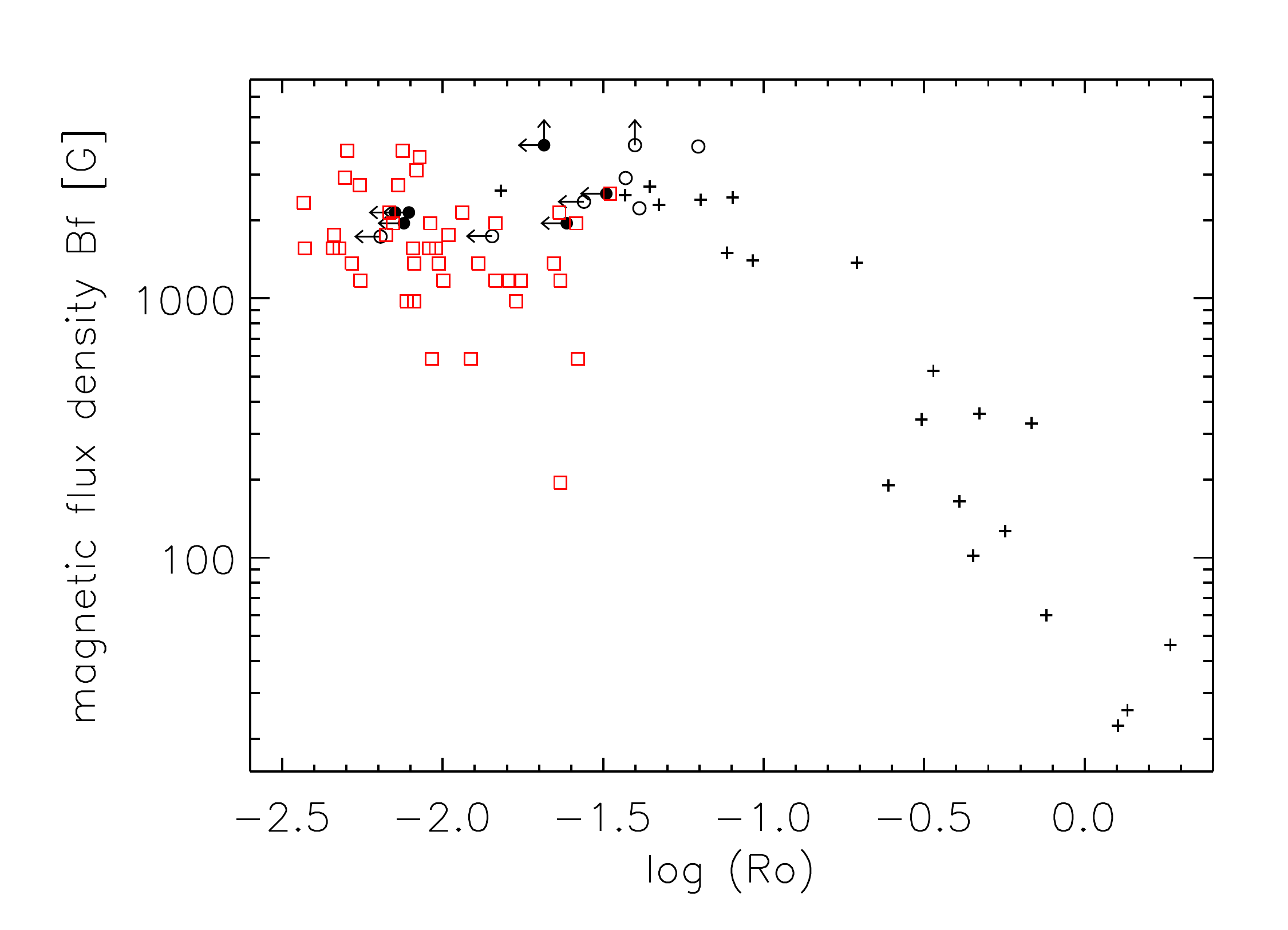}}
  \caption{Magnetic fields as a function of Rossby
    number. Crosses are sun-like stars \citet{1996IAUS..176..237S,
      2001ASPC..223..292S}, circles are M-type of spectral class M6
    and earlier \citep[see][]{2009ApJ...692..538R}. For the latter, no
    period measurements are available and Rossby numbers are upper
    limits (they may shift to the left hand side in the figure). The
    black crosses and circles follow the rotation-activity relation
    known from activity indicators. Red squares are objects of
    spectral type M7\,--\,M9 \citep{2010ApJ...710..924R} that do not seem
    to follow this trend ($\tau_{\mathrm{conv}}$~=~70~d was assumed for
    this sample).}
  \label{fig:Bf_Ro}
\end{figure}}

The relation between magnetic flux density and Rossby number is shown
in Figure~\ref{fig:Bf_Ro}. Crosses are from sun-like stars and define
the rising, unsaturated part of the rotation-activity relation, and
circles are M-type stars defining saturation at a few kilo-Gauss
average field strength. At least for sun-like stars and early-M~stars,
the rotation-magnetic field relation seems to be rather well
defined. Looking back to the discussion on detectability of magnetic
fields, at least in sun-like stars, some cautious doubt may be allowed
as to the relation between Rossby number and \Bf in the ``low-field''
regime (\Bf~\textless~1~kG). First, in principle, the detection of magnetic
fields in this region and, in particular, from optical data, is
extremely difficult and the significance of the data points is
difficult to assess (see Section~\ref{sect:posteriori}). Second,
assuming that the Sun has a Rossby number somewhere between 2 and 0.5
($P_{\mathrm{rot}} \sim$~26~d, $\tau_{\mathrm{conv}} \sim$~12\,--\,50~d), the
average magnetic flux for the Sun is on the order of 20\,--\,100~G, which
is significantly above the value detectable in the Sun if it is
observed as a star.

One interesting and relatively firm conclusion from M~dwarf magnetic
field measurements is that the typical upper limit for average
magnetic fields is of the order of a few kilo-Gauss, average fields of
10~kG are not observed, and the upper limit does not seem to
significantly depend on temperature. This contradicts the prediction
of a close correlation between (maximum) magnetic field strength and
spectral type introduced by assuming a limiting influence of buoyancy
forces on the dynamo efficiency \citep{1982ApJ...253..290D}. However,
this conclusion is only valid for M-type main sequence stars because
we have no good estimate of maximum field strengths in F\,--\,K-type
stars, and magnetic fields in pre-main sequence stars may follow
different rules.

\subsection{The dynamo at very low masses}

Stars of spectral type G\,--\,K are considered sun-like stars, their
interior structure with an outer convective envelope and an inner
radiative core, and the general observational evidence for similar
evolutionary paths lead to the conclusion that this group of stars
follows physical principles that are very much alike. Early-M~type
stars can be sorted into the same category. However, at mid- and
late-M spectral types, serious changes occur to stellar structure that
are predicted from theory and observed in different aspects of stellar
evolution. The first important change in stellar structure occurs at
spectral types M3/M4 (in field dwarfs). Stars hotter than M3 have a
radiative core like the Sun, and cooler sun-like stars have convective
envelopes that extend deeper into the interior of the star. On the other
hand, stars cooler than M4 are believed to be fully convective without
a radiative core, and without a transition region between the outer
convective envelope and the inner radiative core. Because in solar
dynamo models this transition region, the \emph{tachocline}, is
believed to be the locus where (at least the cyclic part of) the
stellar dynamo is most efficient, a change in dynamo efficiency has
been expected at spectral type M3/M4. So far, measurements of average
field strengths show no evidence for such a break in dynamo behavior
(but see Section~\ref{sect:Geometry}); rapidly rotating stars on both
sides of the convective boundary can produce magnetic fields of
kG-strength.

Following stellar evolution to even lower temperatures and later
spectral types leads into the regime of very-low-mass objects. At
spectral type $\sim$~M6 and later, there is no longer a unique relation
between effective temperature and the mass of an object, because this
is the regime where field stars and young brown dwarfs co-exist. Brown
dwarfs are objects with mass lower than $\approx0.08\,M_{\odot}$
that cannot burn hydrogen into helium for significant fractions of
their lifetime. The difference from stars to brown dwarfs is dramatic
if long-term evolution is concerned. For a potential dynamo operating
in objects at the cool end of the main-sequence, however, the source
of energy is not necessarily expected to make any
difference. Therefore, no fundamental difference should exist between
magnetic fields on very-low mass stars and massive brown
dwarfs. However, we have seen in Section~\ref{sect:results_young_I}
that so far no field could be detected in a brown dwarf.

Another effect that probably bears some importance for magnetic field
generation in very-low mass stars is growing atmospheric neutrality
\citep{1999AA...341L..23M, 2002ApJ...571..469M}. Since magnetic fields
can only couple to ions and electrons, the lack of ionization in
atmospheres below a few thousand K should play a role in the
generation of magnetic fields. The growth of ionisation fraction with
depth may still allow field coupling sufficiently deep within the
stellar interior and atmospheric ionisation may be provided by dust
ionisation \citep[see][]{2011ApJ...737...38H}. Growing atmospheric
neutrality certainly is important for the coupling between magnetic
fields and the stellar wind. Evidence for the latter is found in
observations of high rotation velocities interpreted as very weak
angular momentum loss in very-low mass stars and brown dwarfs
\citep{2008ApJ...684.1390R, 2010ApJ...723..684B}.

Recently, magnetic fields measurements in a sample of very-low mass
stars of spectral types M7\,--\,M9 were reported in
\citet{2010ApJ...710..924R}. Stars in this regime are probably all
fully convective and their atmospheres are significantly cooler than
atmospheres in sun-like stars. The overall distribution of field
strengths in very-low-mass stars does not seem to differ from
higher-mass, earlier M~dwarfs; average fields of up to $\sim$~4~kG are
detected. However, there is evidence for a change in the relation
between rotation and magnetic field strength.
\citet{2010ApJ...710..924R} show that the correlation between
projected surface velocity $v\,\sin{i}$ gradually weakens from
spectral type M7 to M9 showing virtually no relation at lowest
temperatures. We can estimate the Rossby number of the M7\,--\,M9 stars
and plot the average magnetic fields of the M7\,--\,M9 stars as a function
of Rossby number (red squares in Figure~\ref{fig:Bf_Ro}). This plot
shows that the Rossby numbers for the sample stars are much lower than
those for earlier stars. For given Rossby number, the distribution of
average field strength \Bf extends to lower values than seen in
hotter stars. This also may be interpreted as evidence for the
breakdown of the rotation-magnetic field relation in very-low mass
stars.

\subsection{Magnetism and H$\alpha$ activity}

The possibility to measure the surface magnetic fields in M~stars from
molecular FeH lines opens the opportunity to study the generation and
consequences of magnetic fields in a large sample (see
Table~\ref{tab:StokesIMtype}). The most frequently used FeH lines
around 1~\mum can be observed with near-infrared instrumentation,
but also with spectrographs like HIRES or UVES that operate at optical
wavelengths. With optical spectrographs, it is often possible to
observe molecular FeH and simultaneously cover the most frequently
used activity indicator in cool stars, the emission line of
H$\alpha$. Chromospheric emission is known to be variable on
timescales of minutes and strictly simultaneous observation of
H$\alpha$ together with the magnetic field is, therefore, particularly
useful. Figure~\ref{fig:HaBfM} shows the relation between normalized
H$\alpha$ luminosity $\log{L_{\mathrm{H\alpha}}/L_{\mathrm{bol}}}$ and the
average magnetic field \Bf for the M~dwarf measurements given in
Table~\ref{tab:StokesIMtype}. A deeper discussion of this relation
can be found in \citet{2007ApJ...656.1121R, 2010ApJ...710..924R}.

\epubtkImage{Halpha_Bf_M.png}{%
\begin{figure}[htbp]
  \centerline{\includegraphics[width=0.6\textwidth]{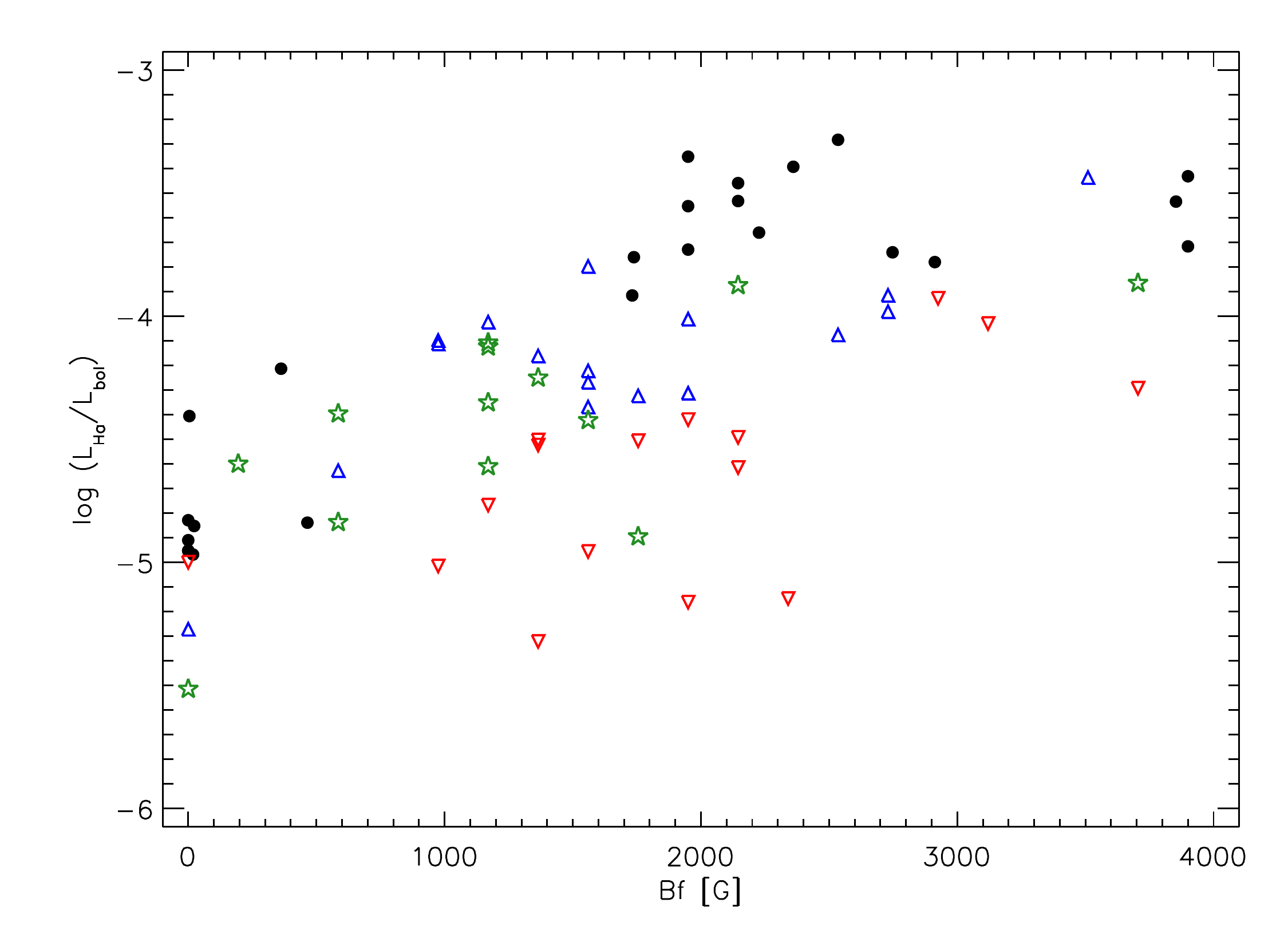}}
  \caption{Normalized H$\alpha$ luminosity as a function of
    \Bf. Filled black circles: Early-/mid-M type stars of spectral
    type M0\,--\,M6; blue triangles: spectral type M7; green stars:
    spectral type M8; red upside down triangles: spectral type M9.}
  \label{fig:HaBfM}
\end{figure}}

One can draw two interesting conclusions from the data shown in
Figure~\ref{fig:HaBfM}. First, in early-M~dwarfs ($\le$M6), the relation
between magnetic field and chromospheric activity follows a curve similar to
the rotation-activity relation; chromospheric activity grows with average
field strength in the low-field regime ($Bf \le 2000$~G) but saturates at a
critical field strength. The critical field strength in early M~stars seems to
be close to 2000~G. At this point, however, the data sample is rather sparse
and uncertainties are high so that such conclusions can only be
preliminary. This saturation -- if existent -- is different from the
saturation of the field itself (at $Ro \sim 0.1$). The rotation-activity
relation implies that fields cannot be stronger than 3\,--\,4~kG in general. A
saturation of chromospheric activity at 2000~G would mean that additionally,
H$\alpha$ emission saturates at even lower rotation rates when the field is
sufficiently strong.

Figure~\ref{fig:HaBfM} also contains data for cooler stars of spectral
type M7\,--\,M9. In these stars, the level of
$\log{L_{\mathrm{H\alpha}}/L_{\mathrm{bol}}}$ is lower on average than
in hotter stars, a reason may be the growing atmospheric neutrality
that weakens the coupling between the ionized atmosphere and magnetic
fields hence rendering magnetic heating ineffective. Interestingly,
there is a hint in Figure~\ref{fig:HaBfM} that the field strength at
which saturation occurs may grow to larger fields in cooler stars. In
other words, cooler stars need stronger field strengths to generate
the same level of activity than hotter stars, and the saturation is
not limited by a fixed field strength but by a maximum level of
chromospheric emission ($\log{L_{\mathrm{H\alpha}}/L_{\mathrm{bol}}}$).

\subsection{A posteriori knowledge about detectability of magnetic fields}
\label{sect:posteriori}

I have argued in Section~\ref{sect:results_I} and
\ref{sect:rotmagnact} that the detection of magnetic fields from
optical lines is extremely difficult, and that our picture of the
non-saturated part of the rotation-magnetic field relation may be
biased by influences of activity on line profiles other than the
Zeeman effect. In other words, the relation between rotation period
and measured fields may be driven by the presence of starspots that
are generated when stars rotate more rapidly, and they influence the
line profiles in a way that may mimic the presence of Zeeman
splitting. It is currently very difficult to assess the influence of
this effect. The rotation-magnetic field relation would probably look
similar to the one shown in Figure~\ref{fig:Bf_Ro}, but its absolute
values may differ significantly. One consequence of this might be the
mismatch between the solar average field and the value predicted
according to its rotation period.

\epubtkImage{Field_veq.png}{%
\begin{figure}[htbp]
  \centerline{\includegraphics[width=0.6\textwidth]{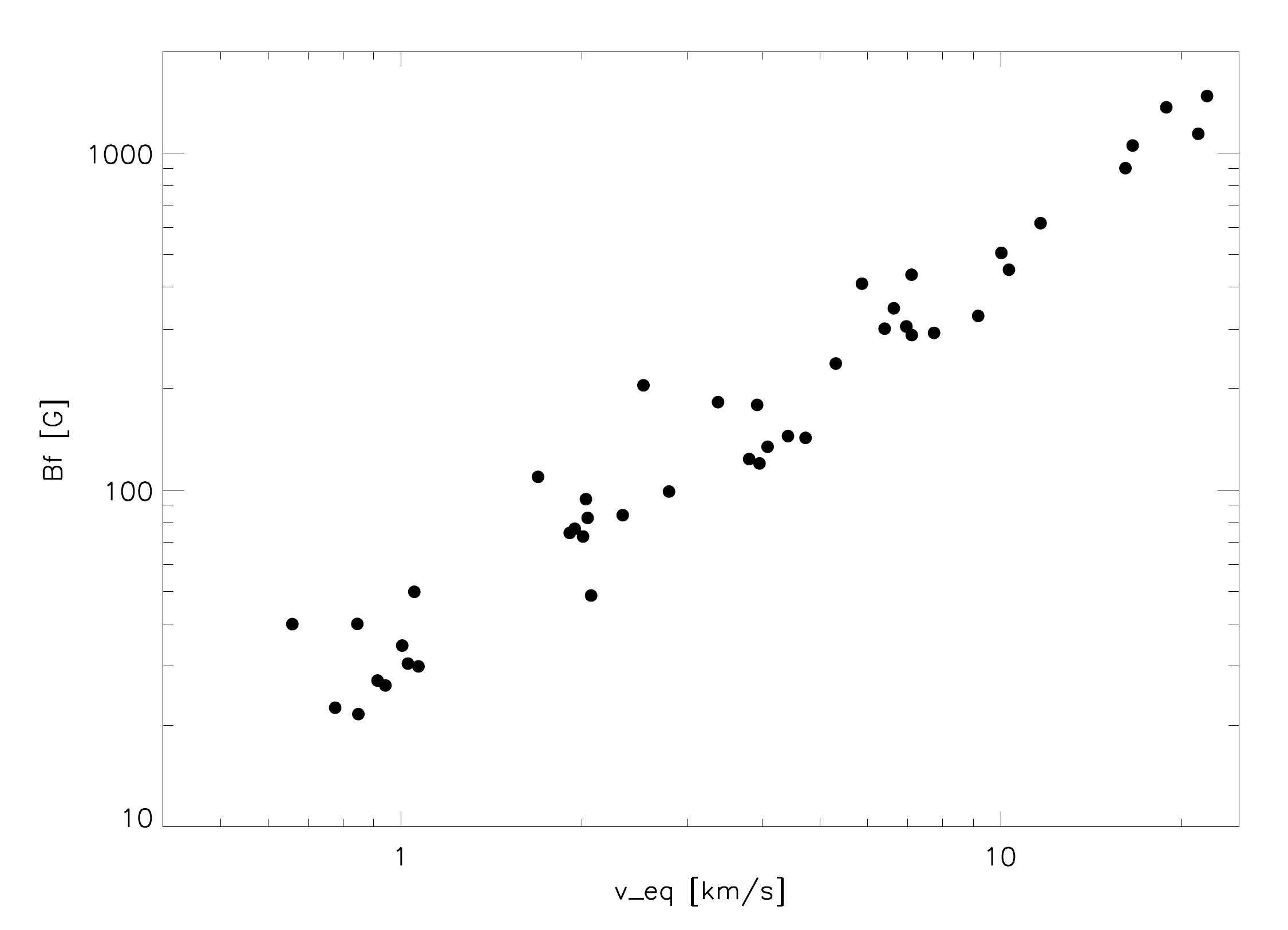}}
  \caption{Estimated average magnetic fields as a
    function of equatorial rotational velocity. Equatorial velocities
    are calculated for stars with measured rotation periods in
    \citet{1984ApJ...279..763N} and \citet{1996ApJ...466..384D}.
    Average fields are estimated from the relation given in the text.}
  \label{fig:Bf_veq}
\end{figure}}

As an interesting exercise, we can take the rotation-magnetic field
relation in the unsaturated part of Figure~\ref{fig:Bf_Ro} and estimate
magnetic field strengths for a sample of stars with measured rotation
periods. The relation in Figure~\ref{fig:Bf_Ro} can be approximated by
\Bf~=~70~Ro\super{-1.5}. We can then take empirical Rossby numbers (using
measured rotation periods) from the work of
\citet{1984ApJ...279..763N} and \citet{1996ApJ...466..384D}, convert
rotation period into approximate surface rotation velocity for each
star, and estimate the average magnetic field strength from the
relation between \Bf and \Ro. The result of this exercise is shown
in Figure~\ref{fig:Bf_veq}. We can conclude that according to the
relation in Figure~\ref{fig:Bf_Ro}, sun-like stars with surface
rotation velocities of 5~\kms generate average magnetic
fields of approximately \Bf~=~200~G. Kilo-Gauss field strengths are
generated in stars that rotate as rapidly as
$v_{\mathrm{eq}}$~=~15~\kms. A rough approximation to the relation
shown in Figure~\ref{fig:Bf_veq} is $Bf \approx 50\,v_{\mathrm{eq}}$,
with \Bf in Gauss and $v_{\mathrm{eq}}$ in \kms. We can insert this
relation into Equation~(\ref{Eq:Zeemanv}) to achieve a very rough
estimate of the ratio between rotational broadening and Zeeman
splitting in sun-like stars. The resulting ratio is
\begin{equation}
  \frac{\Delta v_{\mathrm{Zeeman}}}{v_{\mathrm{eq}}} \approx 0.07 \, \lambda_0 \, g,
\end{equation}
with $\lambda_0$ in \mum. Thus, at optical wavelengths, the
approximate Zeeman shift according to the rotation-magnetic field
relation in the non-saturated dynamo regime is usually well below
10\% of rotational broadening. Given the limitations and systematic
uncertainties of detailed spectral synthesis, this is a very
challenging problem for Zeeman observations. So far, conclusive
investigations of Zeeman splitting at infrared wavelengths are
lacking, but there is certainly a great need to verify the
rotation-magnetic field relation at longer wavelengths $\lambda_0$.

\newpage

\section{Equipartition}
\label{sect:Equipartition}

In stellar atmospheres, a magnetic field is a source of magnetic
pressure, $P = B^2/8\pi$. In hydrostatic equilibrium, this pressure
must be balanced by the gas pressure so that magnetic flux is limited
by the total gas pressure. Such stability considerations lead to the
expectation that stellar magnetic fields in general are limited by
atmospheric structure
\citep[e.g.,][]{1979SoPh...62...15S}. \citet{1985ApJ...299L..47S} and
\citet{1990IAUS..138..427S} estimate scaling relations of
equipartition magnetic fields in field stars, and
\citet{2007ApJ...664..975J} provides a detailed comparison between
magnetic fields in pre-main sequence stars and corresponding
equipartition field strengths. 

The values of equipartition fields and, therefore, the expected maximum
field strengths in main sequence stars are obviously a function of
surface temperature and gravity. They turn out to be of the order of
1\,--\,2~kG for spectral type G, 2\,--\,3~kG for spectral type K, and
3.5\,--\,4.0~kG for spectral type early- to mid-M
\citep{1990IAUS..138..427S}. From a first glance, this is in remarkable
agreement with the maximum average field strengths observed in M
dwarfs. However, the \emph{average} field of a star with flux tubes
satisfying equipartition would probably be much lower than the maximum
field because the gas pressure available to balance magnetic pressure
must be available in some non-magnetic regions (in other words, \f
cannot be 1).

As discussed above, we have no conclusive information about maximum
field strengths in hotter, sun-like stars. We can, therefore, not
exclude that maximum field strengths in sun-like stars are
significantly lower than $\sim$~4~kG. Even if the local field
strengths in M~dwarfs are probably above our estimate of equipartition
field strengths, there still may be a scaling of maximum field
strengths with gas pressure, and only the absolute value of the
limiting field differs somewhat from our approximations. Again,
infrared observations of Zeeman splitting are required in active
sun-like stars to shed light on this fundamental question.

Equipartition field strengths are predicted to be much lower than 4~kG
in pre-main sequence stars. \citet{2007ApJ...664..975J} has found a
significant mismatch between observed field strengths and predictions
from equipartition. Average magnetic fields in pre-main sequence stars
appear to have typical values of 2.5~kG, but predicted field strengths
for the young stars of the sample in \citet{2007ApJ...664..975J} are
between 0.5 and 3~kG, and the observed field values do not show a
correlation with equipartition estimates. Therefore, at least in
pre-main sequence stars, magnetic fields exist with magnetic pressure
dominating gas pressure. Such stars probably have no field-free
regions on their surface. \citet{2007ApJ...664..975J} also provide
evidence that fields in these young stars may be of fossil origin and
not generated by a dynamo, although \cite{2006AA...446.1027C}
estimate survival times for fossil fields in fully convective stars
well below 1000~yr. Assuming that such fields survive much longer,
a tentative conclusion consistent with observations is that dynamo
generated fields are in agreement with pressure balance, while fossil
fields can exceed this boundary. An alternative conclusion from the
available data, however, is that the limit for stellar average fields
is simply on the order of 3\,--\,4~kG in all stars, and pressure balance
has a minor effect on the generated field strength. A physical
motivation for such a minimalistic approach, however, is missing.

\newpage

\section{Geometries of Stellar Magnetic Fields}
\label{sect:Geometry}

Stellar magnetic fields are vector fields. The total strength and energy
contained in a stellar magnetic field are probably characteristic of the
overall dynamo efficiency and resulting activity, and probably determine the
rules of magnetic braking. The geometry of stellar fields adds information
that is crucial for our understanding of these effects. The difficulties
measuring both, total field strength and geometry, were discussed in the
sections above.

As a starting point, again, we can take a look at the Sun. The
surface-averaged flux density on the Sun is much lower than on many
other stars, but we have a better view on
it. Figure~\ref{fig:SchrijverTitle} show a recent visualization of the
Sun's magnetic field during the eruption happening on August~1,
2010. As in stellar magnetic field reconstructions, this visualization
rests on model assumptions and leaves some room for fields not
captured by the applied methods. Nevertheless, the picture is
tremendously rich in details revealing an enormously complex structure
of the solar magnetic field.

\epubtkImage{locations.jpg}{%
\begin{figure}[htbp]
  \centerline{\includegraphics[width=0.8\textwidth]{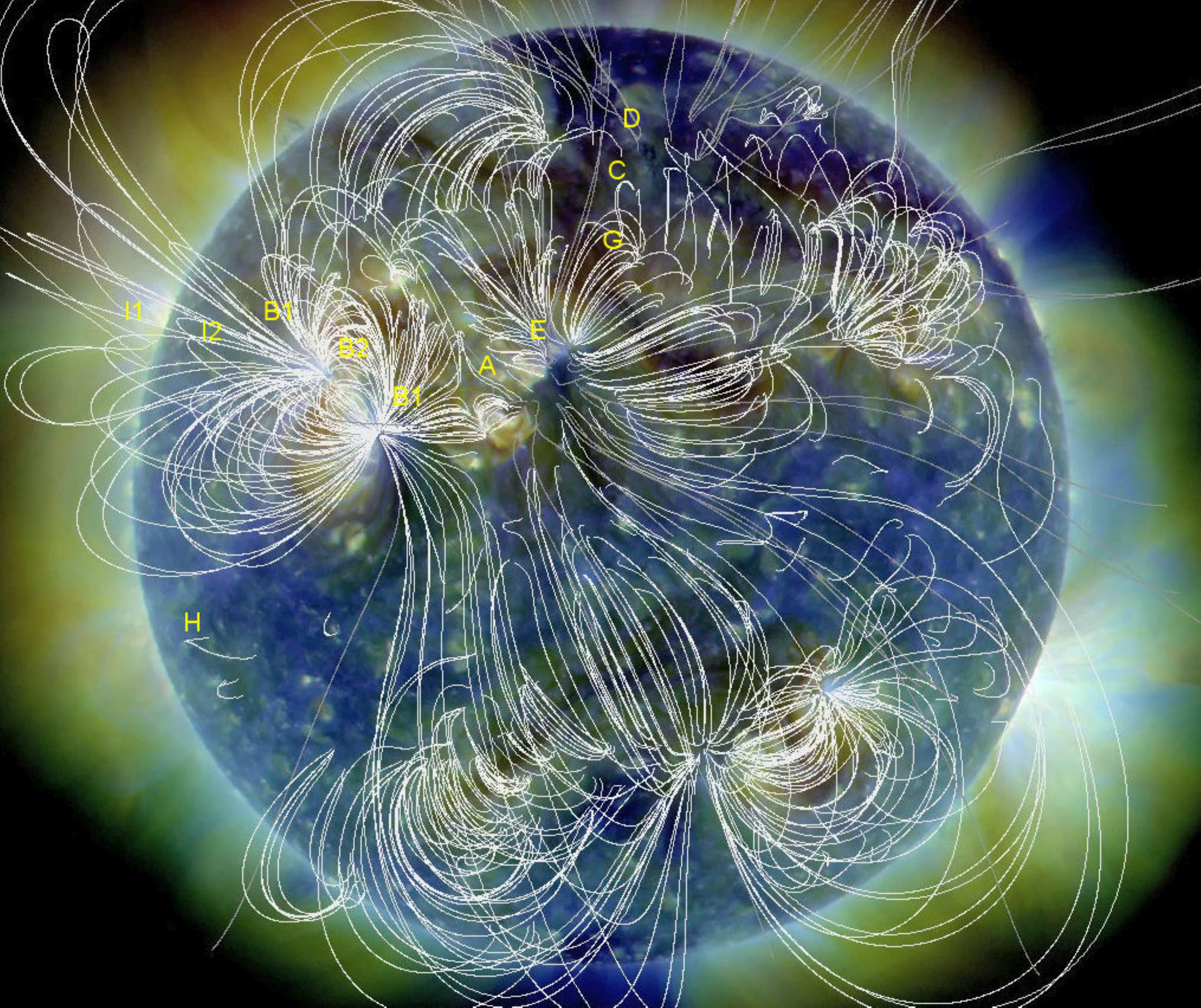}}
  \caption{Three-color composite of EUV
    images of the Sun obtained by the Solar Dynamics Observatory
    during the Great Eruption of August~1, 2010. White lines show a
    model of the Sun's complex magnetic field based on an
    extrapolation for the full-sphere magnetic field
    \citep[from][]{2011JGRA..11604108S}.}
  \label{fig:SchrijverTitle}
\end{figure}}

A remarkable feature of the Sun's magnetic field are the large
magnetic loops visible in Figure~\ref{fig:SchrijverTitle}. Such loops
also occur in images of the solar upper atmosphere where the plasma
seems to follow the magnetic field. If the Sun was observed as a star,
what part of its magnetic field would we be able to see? It was
mentioned before that the Sun's magnetic field would probably be too
small to be detected at any rate. In integrated light, Zeeman
broadening would be too small by one or two orders of magnitude to
produce any detectable signal. Linear polarization would probably
remain undetectable, too. In circular polarization, the situation is
more difficult. Measurements of polarization from net magnetic fields
on the order of one or even a tenth of a Gauss were reported for some
stars, and this may be within the range of an observable net field of
the Sun at a given moment. However, the information about field
geometry from such a measurement alone would certainly be very
limited.

More information is available if a Doppler Image from a star with a
stronger field can be obtained. Such an image takes into account all
the net field snapshots visible at different rotational phases, which
greatly enhances the detectability of tangled fields. An overview
about the current picture of magnetic field geometries in low-mass
stars, in particular among stars of spectral type M, was given by
\citet{2009ARAA..47..333D}. The powerful methods of Least Squares
Deconvolution and Zeeman Doppler Imaging have provided a wealth of
Doppler Images showing very different pictures of stellar magnetic
field geometries. A particularly interesting example are low-mass
stars of spectral class M; not only are there many Doppler Images of M
stars, this spectral range is also of particular interest for our
understanding of the solar and stellar dynamos as pointed out earlier
in this review.

\epubtkImage{Donati11.png}{%
\begin{figure}[htbp]
  \centerline{\includegraphics[width=0.9\textwidth]{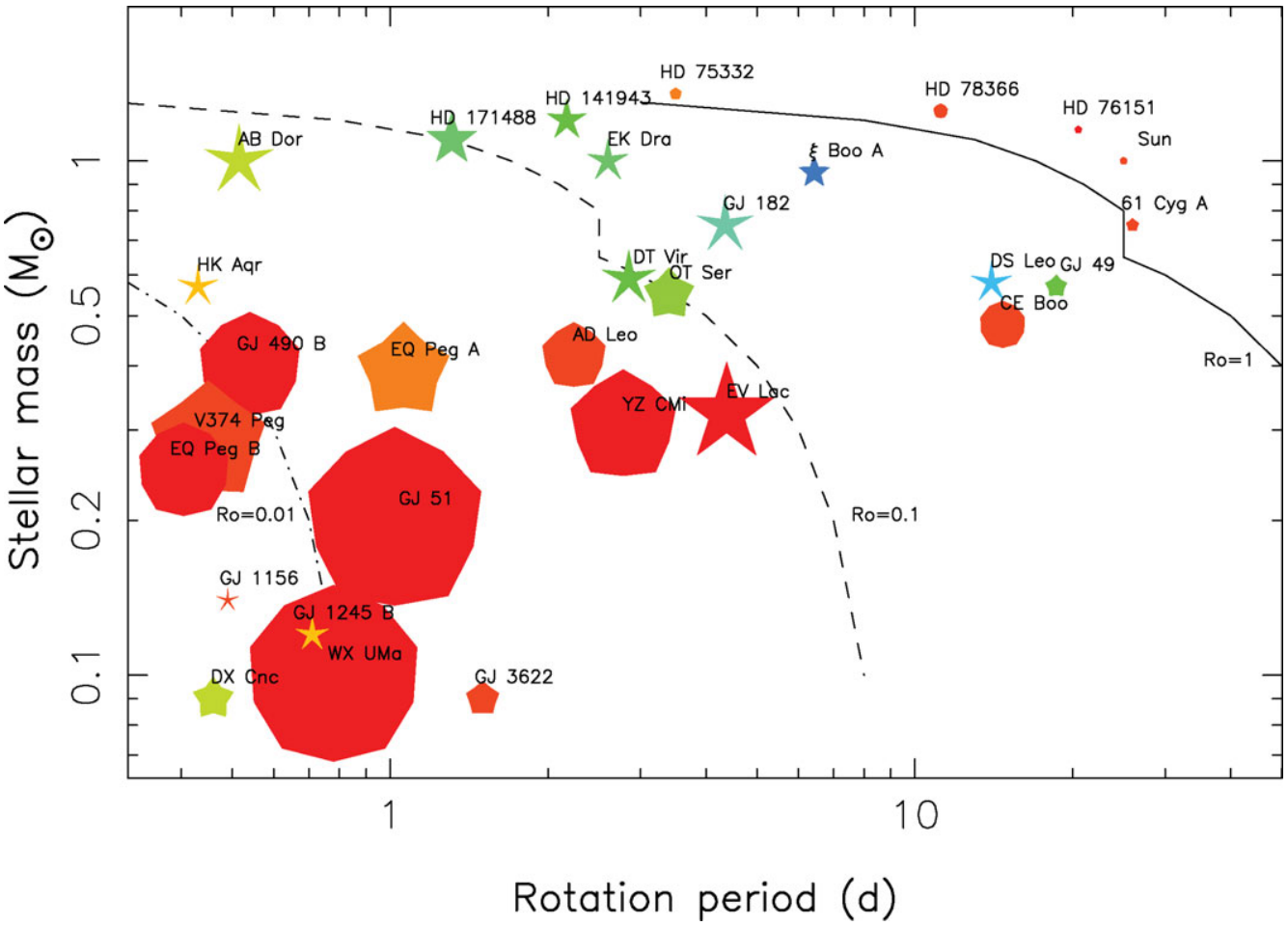}}
  \caption{Properties of the large-scale magnetic
    geometries of cool stars \citep{2011IAUS..271...23D} as a function
    of rotation period and stellar mass. Symbol size indicates
    magnetic densities with the smallest symbols corresponding to mean
    large-scale field strengths of 3~G and the largest symbols to
    1.5~kG. Symbol shapes depict different degrees of axisymmetry of
    the reconstructed magnetic field (from decagons for purely
    axisymmetric fields to sharp stars for purely non-axisymmetric
    fields). Colors illustrate field configuration (dark blue for
    purely toroidal fields, dark red for purely poloidal fields,
    intermediate colors for intermediate configurations). Full,
    dashed, and dash-dot lines trace lines of equal Rossby number
    \Ro~=~1, 0.1, and 0.01, respectively \citep[from][reproduced by
    permission of Cambridge University Press]{2011IAUS..271...23D}.}
  \label{fig:Morin2010}
\end{figure}}

\citet{2010MNRAS.407.2269M} summarized the results from Zeeman Doppler
Imaging currently available in M-type stars. Including the results of
\citet{2010MNRAS.407.2269M}, Figure~\ref{fig:Morin2010} shows
properties of the large-scale magnetic geometries of cool stars from
\citet{2011IAUS..271...23D} in a visualization of magnetic field
geometries as a function of mass and rotation period. Many of the
stars follow the trend of stronger average fields in less massive and
more rapidly rotating stars \citep{2009ARAA..47..333D}. These more
active stars have field geometries that seem to be more axisymmetric
and predominantly poloidal. This leads to the suggestion that rapidly
rotating low-mass stars tend to produce strong, axisymmetric, and poloidal
fields. Whether the reason for such a trend is due to rotation, mass
(radius), or structural differences in the interior of the stars, is
unknown. In any case, a more axisymmetric and poloidal field geometry
is not what one expects from the general picture of magnetic dynamos;
distributed dynamos in fully convective stars should not be able to
produce strong fields that are \emph{more} symmetric and poloidal than
fields in sun-like stars in which the dynamo operating at the
tachocline is believed to produce a rather organized global field. The
trend towards stronger and more organized fields in low-mass stars is
challenged by a number of very-low mass ($M \sim 0.1\,M_{\odot}$)
rapid rotators ($P \sim 1\mathrm{\ d}$) exhibiting rather weak fields
and geometries with a low degree of axisymmetry: a number of
very-low-mass stars produce fields with entirely different geometries
and field strengths (lower left in Figure~\ref{fig:Morin2010}).

It is well known that early-M~dwarfs (M0\,--\,M3) in the field are
generally much less active and slower-rotating than later, fully
convective M~stars \citep[e.g.,][]{1998AA...331..581D,
  2008ApJ...684.1390R, 2012ReinersJoshiGoldman}. Early-M~dwarfs appear
to suffer much more severe rotational braking so that their activity
lifetime is shorter than in later M~dwarfs
\citep{2008AJ....135..785W}, and this can be explained by the severe
change in radius and its consequences to angular momentum loss
\citep{2012ApJ...746...43R}.  Do magnetic fields suffer significant
change around spectral type M3/M4? In total field strength, visible to
Stokes~I, no change is detected; differences between field strengths
are consistent with the assumption that flux generation is ruled by
Rossby number (or rotation period) on both sides of the threshold to
fully convective stars. On the other hand, Doppler Images show that
differences between sun-like and early-M~type stars on one side and
very-low-mass stars on the other are enormous. If we assume that these
differences are real, low-mass stars must be able to somehow generate
fields of structure radically different from fields in sun-like
stars. This would probably imply either a small scale dynamo mechanism
capable of generating fields with very different global properties, or
the co-existence of different dynamo mechanisms in fully convective
stars.

It is important to realize that at spectral type M3/M4, severe changes
happen also in more basic parameters of these stars, and the
reason for a change for example in braking timescales seems to be
much more fundamental than magnetic field geometry. For example, from
spectral type M2 to M5, radius and mass diminish by more than a factor
of two, which is enough to cause the observed differences in rotation
and activity \citep{2012ApJ...746...43R}. In other words, less
effective magnetic braking in fully convective stars does not require
a change in field geometry. To what extent such changes may also
influence the detectability of magnetic fields, in particular of
small-scale magnetic structures, is a question that is important for
our understanding of stellar dynamos and, in particular, for the
differences between dynamos in fully and partially convective stars.

We have seen in Section~\ref{sect:methodology} that the fraction of
the magnetic flux detected in the currently available Zeeman Doppler
Images (from Stokes~V) may be substantially lower than one, due to
cancellation effects or the weak-field approximation. This fraction
can be determined if the field is also visible in Stokes~I, where the
full field is measurable. The typical average field strength of a few
hundred Gauss, as detected in Doppler Images, is much lower than
average field strengths of magnetically active stars observed in
Stokes~I that are a typically few kG. We can compare the field
measurements in Stokes~I and V for stars contained in
Tables~\ref{tab:StokesIMtype} and \ref{tab:StokesVDwarfs}. This
comparison is shown in Figure~\ref{fig:BigPlot}, which is an update of
Figure~2 in \citet{2009AA...496..787R}.

\epubtkImage{BigPlot.png}{%
\begin{figure}[htb]
  \centerline{\includegraphics[width=0.9\textwidth]{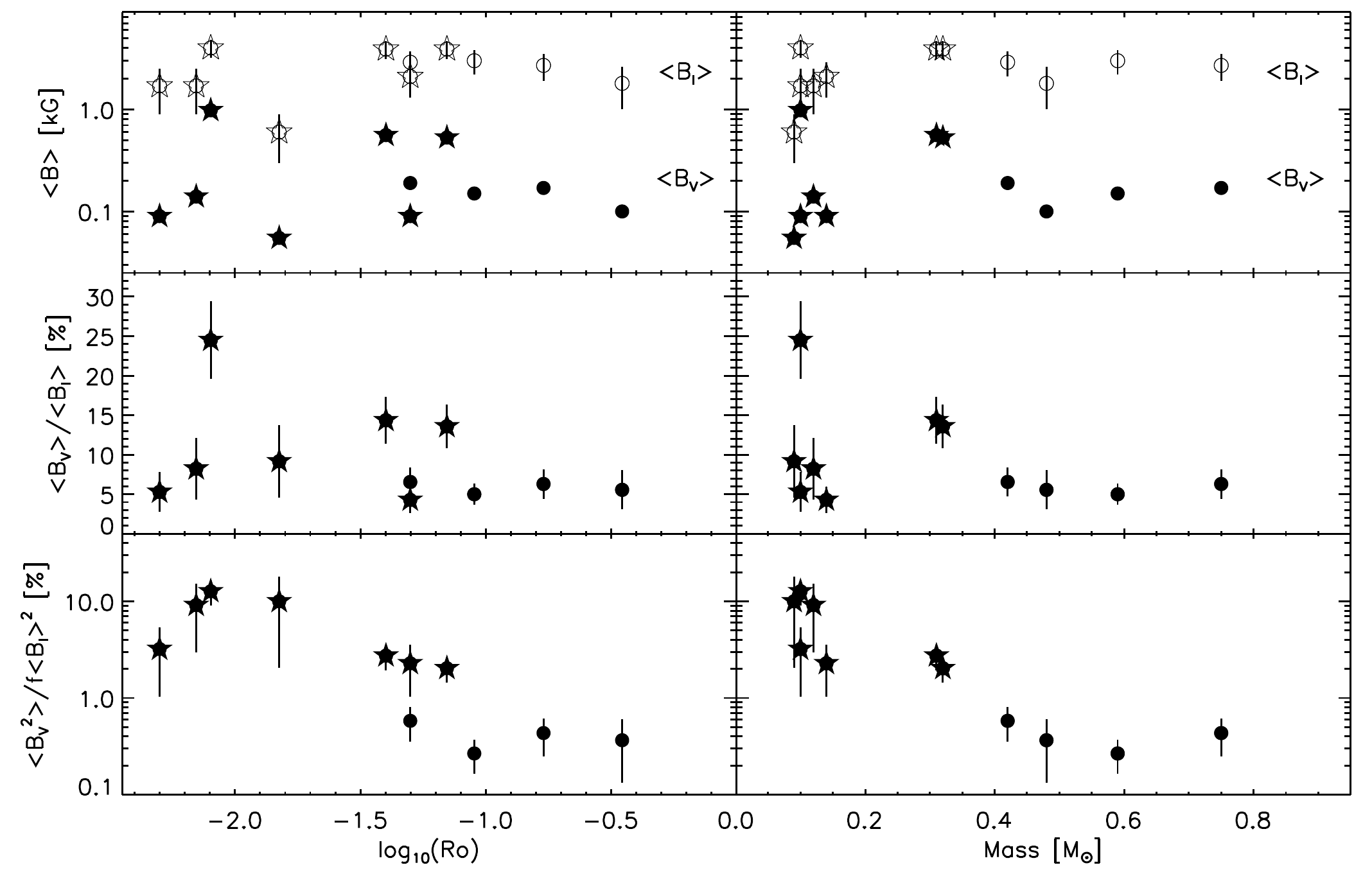}}
  \caption{Measurements of M~dwarf magnetic fields
    from Stokes~I and Stokes~V. \emph{Top panel:} Average magnetic
    field -- Open symbols: measurements from Stokes~I; Filled symbols:
    measurements from Stokes~V. \emph{Center panel:} Ratio between
    Stokes~V and Stokes~I measurements. \emph{Bottom panel:} Ratio
    between magnetic energies detected in Stokes~V and Stokes~I.
    Circles show objects more massive than $0.4\,M_{\odot}$, stars
    show objects less massive than that.}
  \label{fig:BigPlot}
\end{figure}}

Figure~\ref{fig:BigPlot} shows the average magnetic fields from
Stokes~I and V, their ratios, and the ratios of magnetic energies as a
function of Rossby number and stellar mass. In the top panel, the
measurements are shown directly, the center panel shows the ratio
between the average magnetic fields $\langle B_{V} \rangle / \langle
B_{I} \rangle$. For the majority of stars, the ratio is on the order
of ten percent or less, which means that \textless~10\% of the full
magnetic field is detected in the Stokes~V map. In other words, more
than 90\% of the field detected in Stokes~I is invisible to this
method. As discussed above, this is probably a consequence of
cancellation between field components of different polarity. One very
interesting case with a very high value of $\langle B_{V} \rangle /
\langle B_{I} \rangle$ is the M6 star WX~Uma, which has an average
field of approximately 1~kG in Stokes~V (Gl~51 shows an even higher
field but has not yet been investigated with the Stokes~I method).

A second observable that comes with the Stokes~V maps is average
\emph{squared} magnetic field, $\langle B^{2} \rangle$, which is
proportional to the magnetic energy of the star. Under some basic
assumptions, this value can be approximated from the Stokes~I
measurement, too \citep[see][]{2009AA...496..787R}. The ratio between
approximate magnetic energies detected in Stokes~V and I is shown in
the bottom panel of Figure~\ref{fig:BigPlot}, it is between 0.3 and
15\% for the stars considered. In contrast to the conclusions
suggested in \citet{2008MNRAS.390..545D} and
\citet{2009AA...496..787R}, evidence for a change in magnetic
geometries at the boundary between partial and complete convection is
not very obvious when the latest results are included. Four of the
late-M~dwarfs have ratios $\langle B_{V} \rangle / \langle B_{I}
\rangle$ below 10\% while earlier results suggested that more flux is
detectably in Stokes~V in fully convective stars. On the other hand,
the ratio of detectable magnetic energies stays rather high in this
regime ($\ge$~2\%), which may reflect an influence of the convective
nature of the star. An important question, however, is why the five
low-mass stars with $<0.2\,M_{\odot}$ show a relatively high fraction
of detected magnetic energy, while they show such a low fraction of
detected field strength? This may well be an effect of different
magnetic geometries but cannot be clearly identified at this point.

In a typical Zeeman Doppler Image of a low-mass star, about 90\% of
the magnetic field and much more than 90\% of the magnetic energy
remains undetected. It is a challenging task to derive global
properties for a field of which only a small fraction is
visible. Small-scaled structures like sunspots are in principle
difficult to detect in Stokes~V measurements, but it is believed that
the method can well reproduce the large-scale
structures. Nevertheless, in field geometries as complex as shown in
Figure~\ref{fig:SchrijverTitle}, it is not immediately clear which part
can be reconstructed by a given observation, and which part
cannot. Our understanding of magnetic field geometries and magnetic
dynamos, both in the Sun and other stars, will therefore depend on
whether it will be possible to characterize the properties of the
remaining 90\% magnetic flux on stars other than the Sun.

\newpage

\section{Beyond Rotation}

Most of our knowledge about the occurrence and strength of magnetic
fields in cool stars can be explained by a close relation between
rotation and magnetic field generation. While geometries are more
difficult to interpret -- partly because the observational picture is
not unambiguous -- the general idea of the rotation-induced dynamo
seems to be valid over a rather large scale of physical objects with
outer convection zones. There are a few topics beyond this relation
between rotation and field generation that should be discussed briefly
in this section.

\subsection{A scaling law for saturated planetary and stellar dynamos}
\label{sect:scaling}

The surface magnetic field of a star is controlled by its rotation
rate, but magnetism saturates when rotation is faster than a critical
velocity, perhaps associated to the rate where $Ro \approx 0.1$. The
value of the field strength at this saturation level, however, may
vary between objects and depend on additional
parameters. \citet{2009Natur.457..167C} suggested a scaling law based
on energy-flux consistent with the maximum magnetic fields found in
rapidly rotating low-mass stars, and some planetary fields like Earth
and Jupiter (Figure~\ref{fig:nature}). In this picture, the available
heat flux in the convection zone is converted into magnetic energy. A
few assumptions are necessary to explain fields of other planets, but
the general idea is that a single scaling relation may hold in objects
of vastly different dimensions like planets, brown dwarfs, and stars
\citep[for a deeper discussion of flux scaling relations,
see][]{2010SSRv..152..565C}.

\epubtkImage{nature07626-f2-2.png}{%
\begin{figure}[htbp]
  \centerline{\includegraphics[width=0.6\textwidth]{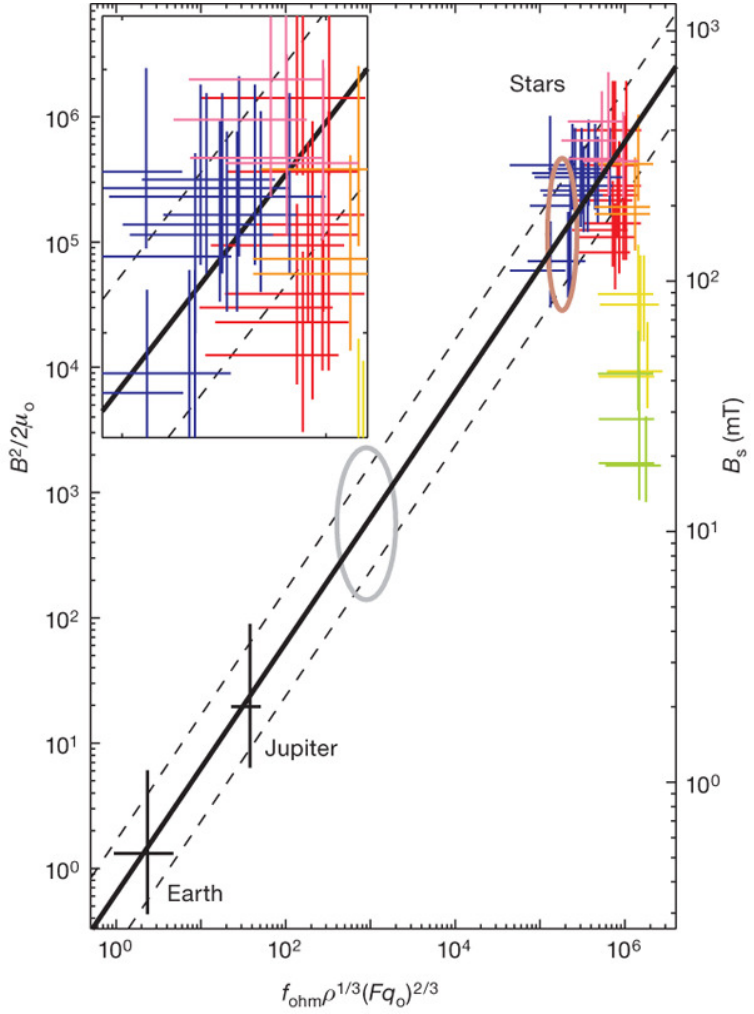}}
  \caption{Magnetic energy density in the dynamo
    vs.\ a function of density and bolometric flux (both in units of
    J m\super{-3}) according to \citet{2009Natur.457..167C}. The scale on the
    right shows r.m.s.\ field strength at the dynamo
    surface. \emph{Blue:} T Tauri stars; \emph{red:} old M
    dwarfs. Black lines show the rescaled fit from
    \citet{2009Natur.457..167C} with 3$\sigma$ uncertainties (solid
    and dashed lines, respectively). The stellar field is enlarged in
    the inset. Brown and grey ellipses indicate predicted locations of
    a brown dwarf with 1500 K surface temperature and an extrasolar
    planet with seven Jupiter masses, respectively
    \citep[from][]{2009Natur.457..167C}.}
  \label{fig:nature}
\end{figure}}

In order to test a relation like this one, magnetic fields must be
measured in stars in the saturated dynamo regime, i.e., at fast
rotation. So far, this was possible only in pre-MS stars and M~dwarfs,
and fields on the order of a few kG were found here. The two species
have available a comparable amount of heat flux so that the scaling
relation yields comparable results for both groups, which is
consistent with the empirical results from the field measurements. In
order to further test the applicability of the scaling relation, it
would be necessary to observe the saturation level in stars of very
different nature, brown dwarfs, and finally exoplanets. The scaling
law provides a prediction for magnetic fields in brown dwarfs;
according to the relation, magnetic fields on the order of several
hundred Gauss up to kG-strength should exist on rapidly rotating,
evolved brown dwarfs.

\clearpage
\subsection{Brown dwarfs}

The sample of objects investigated for magnetic fields by
\citet{2009ApJ...697..373R} and \citet{2010ApJ...710..924R} includes a
number of brown dwarfs. All of them are of spectral type late-M
implying that they are young, a few of them probably only ten Myr or
less. So far, no significant magnetic field could be detected in a
brown dwarf using Zeeman splitting although radio observations in
brown dwarfs provide evidence for kG-strength magnetic fields (see
Section~\ref{sect:radio}). Interestingly, all young brown dwarfs
investigated for Zeeman splitting so far are known accretors. Thus,
they harbor a disk, which is probably the reason why they are not
rotating beyond the critical rate at which Zeeman broadening becomes
undetectable. Upper limits for the magnetic fields in young, accreting
brown dwarfs are on the order of a few hundred Gauss, significantly
lower than fields found in higher-mass young T~Tauri stars or in older
stars of the same spectral type. The field limits are also
significantly below the predictions from the scaling model introduced
in Section~\ref{sect:scaling}. It is currently unknown whether the
low fields in accreting brown dwarfs are due to a less effective
dynamo, the influence of accretion, or some other effect.

More evolved brown dwarfs at ages higher than typical disk lifetimes
of ten Myr are usually rotating too fast for a successful detection
of Zeeman broadening, and they are often also too faint for current
spectroscopic instrumentation. An observation of a magnetic field (or
its upper limit) in non-accreting brown dwarfs, regardless of age,
would be important to make progress in this field.

\subsection{Fossil or dynamo fields in young low-mass stars}
\label{sect:Flux}

One remarkable result from the observation of magnetic fields in young
low-mass stars is the relatively small scatter in the fields detected
in Stokes~I. Most of the fields reported for 14 classical T~Tauri
stars investigated by \citet{2007ApJ...664..975J} have average flux
densities on the order of 2~kG and do not follow predictions according
to equipartition. This sample of stars was augmented with 14 other
T~Tauri stars by \citet{2011ApJ...729...83Y} still finding similar
results. A potential explanation put forward by
\citet{2011ApJ...729...83Y} is that the field strength in these young
stars is not actually maintained through a magnetic dynamo that would
be affected by the scatter of physical properties in the rather
diverse sample of young stars. Instead, \citet{2011ApJ...729...83Y}
propose that the weak scatter in magnetic field measurements is the
result of the gradually weakening magnetic flux from a fossil
field. Because the stars are still contracting, the younger stars with
more magnetic flux, $4 \pi R_{\star}^2 B$, are the stars that are
larger than the older stars with less magnetic flux. According to
pre-MS evolutionary models, the observed average field strengths are
consistent with magnetic flux decreasing by about a factor of ten
between 1 and 10~Myr according to the evolution of available
convective energy. However, estimates of survival times of fossil
fields predict much faster decay well below 1000~yr
\citep{2006AA...446.1027C}. Furthermore, the weakening of total flux with age shown by
\citet{2011ApJ...729...83Y} is based on age estimates derived from comparison to
evolutionary models. However, \citet{2009ApJ...702L..27B} show that episodic
accretion can lead to a luminosity spread that can be misinterpreted
as an age spread on the order of 10\,Myr, hence age and radius estimates have
large uncertainties in this regime. More Stokes~I observations
of average magnetic fields in stars of several 10~Myr age would be
helpful to see whether the total flux is indeed weakening on this
timescale.

The obvious alternative to fossil fields are dynamo generated fields in young
low-mass stars, and convective timescales are supporting this scenario (see
above). Furthermore, Zeeman Doppler maps suggest that large-scale fields in
classical T~Tauri stars can undergo secular variations on timescales of a few
years \citep[e.g.,][]{2009ARAA..47..333D, 2011MNRAS.412.2454D}. If the
variability in these magnetic field maps is indeed due to a change in magnetic
geometry of the star, this readily implies that fields cannot be of fossil
origin but must be generated by a dynamo mechanism.

Nevertheless, an idea similar to the decay of fossil magnetic flux may be
useful to answer the
question why young (accreting) brown dwarfs have such small
fields. Effective dynamos may not yet be operating in these young
objects. The radii of young brown dwarfs are much larger than radii in
older brown dwarfs and low-mass field stars, implying that average
fields may be very low if the available magnetic flux (not a generated
field) is the limiting factor. Observational evidence for any of these
scenarios among brown dwarfs, however, demands conclusive magnetic
field observations.


\section{Summary}

Cool-star magnetic fields are in the center of interest for a large
number of reasons. Their influence on star-formation, solar and
stellar activity, habitability of planets, magnetic dynamos, stellar
structure, angular momentum evolution, and many other topics, put
magnetic fields in the focus of vastly different research areas. Our
vision of magnetic fields in different astronomical contexts, however,
is motivated by only a very limited number of empirical facts, but
influenced by an almost unlimited number of assumptions.  An arsenal
of different methods is available to search for magnetic fields in
cool stars, but usually the different methods provide insights into
very different aspects of magnetism.

Together with observations of stellar activity and rotation, direct
magnetic fields measurements and reconstruction of field geometries
provide an empirical picture of stellar dynamos. With
observations of young stars on one side, and detailed information
about the solar magnetic field on the other, we have boundary
conditions that allow an investigation of an evolution of stellar
magnetism. Our observations are crossing the physically important
boundaries between partially and fully convective stars, between
saturated activity and unsaturated stars, and between stars and brown
dwarfs.

Magnetic fields in cool stars are just on the edge of detectability
for our methods. The interpretation of results from the various
methods opens a parameter space that certainly contains deep
information about the fields, their geometry, and the underlying
physical principles. Unfortunately, it is not always clear how to
interpret our observational results, and how measurements from
different analyses can be compared. Therefore, it is not only
important to improve the general observational picture by collecting
more observational material, but it is perhaps of even greater
importance to understand the assumptions in and limitations of the
methodology we are applying. Technology and computer power is
improving rapidly and will at some point make more and more details
available to direct observation. Nevertheless, the design of future
instrumentation is driven by expectations and interpretation of
earlier results. Given the fact that still a large fraction of the
magnetic universe is unknown, we should be prepared to find a richness
of magnetic phenomena that does not yet exist in our imagination of
stellar magnetic fields -- and we must ensure that our observations
can find it.


\section{Acknowledgments}

I am very thankful to Steve Saar for his careful reading and
thoughtful comments on the manuscript, and to Karel Schrijver and
anomymous referees for very useful comments and their help to shape
this article. Financial support is acknowledged from the Deutsche
Forschungsgemeinschaft through the Emmy Noether programme,
RE~1664/4-1, and the Heisenberg Programme, RE~1664/9-1.

\newpage



\bibliography{refs}

\end{document}